\documentstyle[worldsci]{article}
\input epsf
\def \beq{\begin{equation}}
\def \eep{e^+ e^-}
\def \eeq{\end{equation}}

\def \ga{\gamma \gamma}
\def \I{{\rm Im~}}
\def \ite{{\it et al.}}
\def \jape{J/\psi}
\def \k{K^0}
\def \bk{\bar K^0}
\def \m{{\cal M}}
\def \pipe{\pi^+ \pi^-}
\def \poop{\pi^0 \pi^0}
\def \R{{\rm Re~}}

\begin{document}
\pagestyle{plain}

\title{\bf PRESENT AND FUTURE ASPECTS OF CP VIOLATION}

\author{Jonathan L. Rosner \\
Enrico Fermi Institute and Department of Physics \\
University of Chicago, Chicago, IL 60637}

\maketitle

\begin{abstract}
This series of five lectures describes aspects of CP violation, emphasizing its
description within the standard electroweak model. After discussing the kaon
system, the only place in which CP violation has been seen so far, we turn to
the leading contender for the effect, complex phases in the
Cabibbo-Kobayashi-Maskawa (CKM) matrix.  A number of suggestions are made for
improved tests of the standard picture.  Hadrons containing $b$ quarks play a
key role in this program, and are discussed separately.  We also mention a
number of non-standard and speculative aspects of CP violation, including
alternatives to the CKM description, direct tests of time reversal invariance,
and baryogenesis.
\end{abstract}

%%%%%%% For reproduction comment out the next six lines %%%%%%%%%%%%%%%%%
\vspace{-3.1in}
\centerline{Lectures presented at the VIII J. A. Swieca Summer School}
\centerline{Rio de Janeiro, Brazil, February 7 -- 11, 1995}
\centerline{Proceedings to be published by World Scientific}
\centerline{EFI 95-36; hep-ph/9506364}
\vspace{2.4in}
%%%%%%%%%%%%%%%%%%%%%%%%%%%%%%%%%%%%%%%%%%%%%%%%%%%%%%%%%%%%%%%%%%%%%%%%%

\leftline{1. INTRODUCTION}
\bigskip

A. ~~ Overview of CP violation
\medskip

For many years, it was widely believed that the laws of physics were invariant
under the separate discrete symmetry operations of spatial reflection ({\it
parity,} or P), time reversal (T), and charge conjugation (C).  However, it was
only possible to prove, under the assumptions of locality and
Lorentz-invariance, that a quantum field theory must respect the product
CPT.\cite{CPT} In 1956 it was realized that no tests of P or C invariance of
the weak interactions had yet been performed.\cite{LY} Experiments soon showed
that both P and C were separately violated in weak decays.\cite{Pviol}

The theory of weak interactions as formulated\cite{VA} in 1957 did conserve the
product CP.  Seven years later, the discovery of the two-pion decay of the
neutral kaon\cite{CCFT} showed that even the product CP was violated.  Since
1964, although no new CP-violating phenomena have been observed, we have a
theory for this effect and the prospect of many experimental tests. These
lectures describe our present understanding of CP violation, some of the
tests which are likely to bear fruit in the near future, and other ways in
which CP violation can manifest itself.
\bigskip

B. ~~ Plan of these lectures
\medskip

Section 2 is devoted to a description of CP violation in the kaon system,
without regard to its fundamental origin.  A candidate theory for CP violation
based on complex phases in the Cabibbo-Kobayashi-Maskawa (CKM) matrix is
described in Section 3.  Prospects for improving information about that theory
(or disproving it altogether by exposing inconsistencies) are mentioned in
Section 4.  A crucial role is played by $B$ mesons (particularly by CP
violation in their decays), to which Section 5 is addressed. Non-standard and
speculative aspects of CP violation are the subject of Section 6, while
Section 7 concludes.
\bigskip

C. ~~ A short bibliography
\medskip

A good overall reference on quantum field theory, with an excellent section on
CP violation, is the textbook by Cheng and Li.\cite{CL}  There are many fine
articles on CP violation in a recent World Scientific publication edited by
Cecilia Jarlskog.\cite{CJ}

For historical perspectives I recommend the proposals for a short- and
long-lived neutral kaon;\cite{GP,KCP} the original report of the discovery of
CP violation;\cite{CCFT} early and later reviews of CP violation in the neutral
kaon system;\cite{Revs} and original articles on the Cabibbo theory,\cite{Cab}
charmed quarks,\cite{Charm} and the third family of quarks.\cite{KM} An article
which was of tremendous use in anticipating properties of the charmed quark is
the study of kaon decays by Gaillard and Lee.\cite{GL}

We shall make frequent use of a parametrization of quark couplings,\cite{WP}
updating earlier analyses\cite{CKMans} in which data constrain these
parameters. More general possibilities for the study of CP violation have
arisen since the suggestion\cite{BCP} that these effects may be large for
mesons containing $b$ quarks.  The reader is invited to consult reviews of
earlier developments in this area.\cite{Brevs}

For aspects of CP violation not confined to the kaon and $B$ meson systems,
one may consult a review\cite{PecR} of the strong CP problem\cite{PQ} and one
proposed solution of it,\cite{AX} as well as the paper of Sakharov\cite{Sak}
which proposed that CP violation was one of the ingredients responsible for the
observed baryon asymmetry of the Universe.

The present lectures are based in part on earlier
treatments,\cite{TASI,JRCKM,PASCOS,JRCharm,DPF,Fest} updated to take account of
recent developments.  In some cases these references may be consulted for
greater detail.
\bigskip

\leftline{2. CP VIOLATION IN THE KAON SYSTEM}
\bigskip

A. ~~ CP Eigenstates of neutral kaons
\medskip

In order to make sense of a class of ``strange'' particles produced strongly
but decaying weakly, Gell-Mann and Nishijima\cite{GN} proposed an additive
quantum number, ``strangeness,'' conserved in the strong interactions but not
necessarily in the weak interactions.  The reaction $\pi^- p \to K^0 \Lambda$,
for example, would conserve strangeness $S$ if $S(\pi) = S(p) = 0$, $S(\Lambda)
= -1$, and $S(K^0) = +1$.  But then the kaon could not be its own antiparticle;
there would have to also exist a $\bar K^0$ with $S(\bar K^0) = -1$.  It could
be produced, for example, in the reaction $\pi^- p \to K^0 \bar K^0 n$.

The states $K^0$ and $\bar K^0$ would be degenerate in the absence of coupling
to final states (or to one another).  However, both states can decay to the $2
\pi$ final state in an S-wave (orbital angular momentum $\ell = 0$). Gell-Mann
and Pais\cite{GP} noted that since $C(\pi^+ \pi^-)_{\ell = 0} = +$, $C(K^0) =
\bar K^0$, and $C(\bar K^0) = K^0$, the linear combination of $K^0$ and $\bar
K^0$ which decayed to $\pi^+ \pi^-$ had to be $K_1 \equiv (K^0 + \bar
K^0)/\sqrt{2}$. Then there should be another state $K_2 \equiv (K^0 - \bar
K^0)/\sqrt{2}$ forbidden to decay to $\pi^+ \pi^-$, and thus long-lived.  (It
should be able to decay to $3 \pi$, for example.) This state was looked for
and found.\cite{KL} Its lifetime was measured to be about 600 times that of
$K_1$.

The $K^0 - \bar K^0$ system resembles many coupled degenerate problems in
physics.  For example, a drum-head in its first excited state possesses a line
of nodes. A degenerate state exists with the line of nodes perpendicular to the
first, but nothing specifies the absolute orientations of the two lines of
nodes.  However, if a fly alights off-center on the drum, it will define the
two lines of nodes.  One mode will couple to the fly, thereby changing
in frequency, and one mode will not.

Gell-Mann and Pais assumed that C was conserved in the weak decay process.  In
a CP-conserving weak interaction theory in which C and P are individually
violated, the above argument can be recovered by replacing C with CP.\cite{KCP}

The discovery\cite{CCFT} in 1964 that both the short-lived ``$K_1$'' and
long-lived ``$K_2$'' states decayed to $\pi \pi$ upset this tidy picture and
signified that not even CP symmetry was valid in Nature.  Henceforth the states
of definite mass and lifetime would be known as $K_S$ (for ``short'') and
$K_L$ (for ``long'').  Before describing the manifestations of CP violation
in neutral kaon decays, we discuss briefly the CP properties of final states
containing pions.
\bigskip

B. ~~ CP properties of $2 \pi$ and $3 \pi$ states
\medskip

1.  {\em Two-pion states.}
A $\pipe$ state of relative orbital angular momentum $\ell$ has charge
conjugation eigenvalue $C = (-1)^\ell$ and parity $ P = (-1)^\ell$, so it is
always an eigenstate of CP with eigenvalue $ +1$. For a $\poop$ state one must
have $C = P = +$, and only even $\ell$ are allowed.

2.  {\em Three-pion states.}
We shall concentrate on neutral states with total angular momentum $J = 0$.
Consider first the $ \pipe \pi^0$ state, with the relative orbital singular
momentum of $\pi^+$ and $\pi^-$ defined to be $\ell_{12}$, and that of the
$\pi^0$ with respect to the $\pipe$ system taken to be $\ell_3$. Since $J = 0$,
one must have $\ell_3 = \ell_{12}$. The intrinsic CP of the $\pipe$ system is
+, and that of $\pi^0$ is $-$, since $C(\pi^0) = +, ~P(\pi^0) = -$. Then
\beq
CP (\pipe \pi^0 ) |_{J=0} = - (-1)^{\ell_3}
= ({\rm odd, ~ even, ~ odd}, \ldots ) ~{\rm for} ~ \ell_3 = ( 0, 1, 2,\ldots)
\eeq
The corresponding isospins allowed for the $\pipe$ system are (even, odd, even,
$\ldots$). (Recall $\ell_{12} = \ell_3$). The states with $\ell_{12} = \ell_3
> 0$ are highly suppressed in kaon decays by centrifugal barrier effects. Thus
in $J = 0$ states of $\pipe \pi^0$, the CP-odd state predominates.
The situation is simpler for the $ 3 \pi^0$ states. Here $\ell_{12}$ (and hence
$\ell_3$) must be even, so {\em only} the CP-odd states can occur.
The $3 \pi$ states are the dominant hadronic decay mode of the $K_L$.
\bigskip

C. ~~ The $\Delta I = 1/2$ rule
\medskip

The nonleptonic decays of strange particles are governed by the quark
subprocess $s \to u d \bar u$, which produces three $I = 1/2$ quarks in the
final state and thus can lead to $\Delta I = 1/2$ and $\Delta I = 3/2$. In
almost all processes, $\Delta I = 1/2$ dominates. The reason for this involves
a combination of several effects, as we shall see for $K \to 2 \pi$ decays.

The isospin states $|I \rangle $ of two pions with momenta $p_1$ and $p_2$ may
be expressed in terms of the charge states $|Q (\pi [p_1 ] ) Q (\pi
[p_2])\rangle$, where $Q = + , 0$, or $-$, as
\beq
|0 \rangle = [ | + - \rangle - | 0~0 \rangle + | - + \rangle ] /\sqrt{3}
{}~~~ (J = ~{\rm even} ) ~~~,
\eeq
\beq
|1 \rangle = [ | + - \rangle - | - + \rangle ] / \sqrt{2} ~~~
 (J = ~{\rm odd} ) ~~~,
\eeq
\beq
|2 \rangle = [ | + - \rangle + 2| 0~0 \rangle + | - + \rangle ] / \sqrt{6}
{}~~~ (J = ~{\rm even} ) ~~~.
\eeq
Inverting these relations, we find, for even $J$ (the case of interest in
neutral kaon decay)
\beq
\langle \pm \mp | T | K \rangle = \langle 0 | T | K \rangle / \sqrt{3}
+ \langle 2 | T | K \rangle / \sqrt{6}~~~,
\eeq
\beq
\langle 0~0 | T | K \rangle = - \langle 0 | T | K \rangle / \sqrt{3}
+ \langle 2 | T | K \rangle \sqrt{2/3}~~~.
\eeq
To take into account identical particles we integrate over phase space with
$\hat{p}_1$ in one hemisphere and $\hat{p}_2$ in the other, taking both $
\langle \pi^+ (p_1) \pi^- (p_2) |$ and $\langle \pi^- (p_1) \pi^+ (p_2)|$ into
account. For the $\pipe$ final state, this avoids double counting in $\langle
\pi^0 (p_1) \pi^0 (p_2)|$. We then find that if $I_{\pi \pi} = 0$ is dominant,
$\Gamma (K^0 \to \pipe ) = 2 \Gamma (K^0 \to \poop)$, while if $I_{\pi \pi} =
2$, $\Gamma (K^0 \to \poop) = 2 \Gamma (K^0 \to \pipe)$. Experimentally the
ratio\cite{PDG} is much closer to that for $I_{\pi \pi} = 0$. The small
deviation from a $2:1$ ratio for $\pi^\pm \pi^\mp : \poop$ is due to Coulomb
corrections and to a small $I_{\pi \pi} = 2$ admixture in the amplitude, whose
magnitude may be estimated by comparing the rate for $ K^+ \to \pi^+ \pi^0$
(which must involve $I_{\pi \pi} = 2$; $I_{\pi \pi} = 1$ is forbidden by Bose
statistics) with that for $K_S^0 \to \pi \pi$:
\beq
\Gamma (K^+ \to \pi^+ \pi^0) = B (K^+ \to \pi^+ \pi^0 ) / \tau_{K^+} =
1.71 \times 10^7 s^{-1}
\eeq
\beq
\Gamma (K_S^0 \to \pi \pi ) \simeq 1 /\tau_{K_S} = 1.12 \times 10^{10} s^{-1}
\eeq
The $K_S \to \pi \pi$ decay rate is more than 600 times as large as the $K^+
\to \pi^+ \pi^0$ rate.

Several mechanisms probably contribute to the enhancement of the $\Delta I =
1/2$ transition (leading to $I_{\pi \pi} = 0$) with respect to the $\Delta I =
3/2$ transition (leading to $I_{\pi \pi} = 2$). They include the following:

{\it 1. The ``penguin graph''}\cite{pen} involves the transition $s \to d$
with emission of at least one gluon.  (The transition without gluon emission
corresponds to a term which can be removed from the Lagrangian by a small
redefinition of quark fields.\cite{CG}) Since a single $d$ quark is created,
the penguin graph automatically has $\Delta I = 1/2$.

{\it 2. The ``exchange graph''}, corresponding to the process $s \bar d \to u
\bar u$ mediated by $W$ exchange, would lead to a $u \bar u$ final state if
applied to an initial neutral kaon. A $u \bar u$ state cannot have $I = 2$, so
this graph could also play a role in enhancement of $K \to (2 \pi )_{I=0}$.
However, it is thought to be an unlikely contributor to decays involving light
quarks in a state of $J = 0$. In the limit of vanishing quark mass, the $s$
quark should participate in the process with left-handed helicity, while the
$\bar d$ should participate with right-handed helicity. The total $z$ component
of spin of the $s$ and $\bar d$ quarks would then be $-1$ along the direction
of the $s$ quark in the $s \bar d$ c.m.s., which is impossible for a $J=0$
state like the kaon. Perturbative QCD effects appear inadequate to overcome the
expected suppression.\cite{BSS}

{\it 3. Final-state interactions} may well contribute to the enhancement of
$I_{\pi \pi} = 0$. The $ \pi \pi$ phase shift $\delta_I$ is large and positive
$(\approx 45^0)$ for $I_{\pi \pi} = 0$, but small and negative $( \approx -
7^0)$ for $I_{\pi \pi} = 2$. Resonant behavior would correspond to $\delta_I =
\pi/2$.

{\it 4. Perturbative QCD effects} undoubtedly enhance $\Delta I = 1/2$
transitions. The effective weak Hamiltonian for $s \to u d \bar u$ can be
decomposed into pieces symmetric $(I_{ud} = 1)$ and antisymmetric $(I_{ud} =
0)$ under the interchange of the final $u$ and $d$ quarks. The $I_{ud} = 0$
piece receives an enhancement from perturbative QCD, while the $I_{ud} = 1$
piece is slightly suppressed.\cite{PQCD} The $I_{ud} = 0$ piece contributes
only to $\Delta I = 1/2$ amplitudes, since the only other quark carrying
isospin in the process is a $\bar u$ quark.

The actual enhancement of $\Delta I = 1/2$ transitions in $K$ decays appears to
be due to a combination of several of the effects mentioned above.\cite{DIH}
\bigskip

D. ~~ Mass eigenstates of short- and long-lived kaons
\medskip

In order to discuss neutral $K$ decays (and, later, neutral $B$ decays), we
need information on the matrix ${\cal M}$ whose eigenstates correspond to
particles of definite mass and lifetime.\cite{CL,Revs} In the kaon rest frame,
the time evolution of basis states $\k$ and $\bk$ can be written\cite{Sachs} as
\beq
i \frac{\partial}{\partial t}
\left[ \begin{array}{c} \k \\ \bk \end{array} \right]
= {\cal M} \left [ \begin{array}{c} \k \\ \bk \end{array} \right]~~~;
{}~~ {\cal M} = M - i \Gamma /2~~~.
\eeq
An arbitrary matrix ${\cal M}$ can be written in terms of Hermitian matrices
$M$ and $\Gamma$.  CPT invariance can be shown to imply the restriction ${\cal
M}_{11} = {\cal M}_{22}$ and hence $M_{11} = M_{22}, ~ \Gamma_{11} =
\Gamma_{22}$. We shall adopt this limitation in all our subsequent discussions;
one may consult several reviews\cite{Revs} for the case of CPT violation.

We denote the eigenstates of ${\cal M}$ by
\beq
|S \rangle + p_S | \k \rangle + q_S | \bk \rangle ,
\eeq
\beq
| L \rangle + p_L | \k \rangle + q_L | \bk \rangle ,
\eeq
with $|p_{S,L} |^2 + | q_{S,L} |^2 = 1$, and the corresponding eigenvalues
by $ \mu_{S,L} \equiv m_{S,L} - {i \over 2} \Gamma_{S,L}$, where
$m_{S,L}$ and $\Gamma_{S,L}$ are real.  Then
\beq \label{eqn:meq}
{\cal M} \left[ \begin{array}{c c}
p_S  & p_L \\ q_S  & q_L \end{array} \right ] =
\left[ \begin{array}{c c} p_S  & p_L \\ q_S  & q_L \end{array} \right]
\left[ \begin{array}{c c} \mu_S  & 0 \\ 0  & \mu_L \end{array} \right ]~~~ ,
\eeq
implying
\beq \label{eqn:pqs}
\left( \frac{p_S}{q_S} \right )^2 = \frac{{\cal M}_{12}}{{\cal M}_{21}} =
\left( \frac{p_L}{q_L} \right )^2
\eeq
when the condition ${\cal M}_{11} = {\cal M}_{22}$ is taken into account. Thus
there are two solutions for $p$'s and $q$'s. We choose $p_S = p_L \equiv p$;
then $q \equiv q_S = - q_L$, and the eigenstates are
\beq
|S \rangle = p | \k \rangle + q | \bk \rangle ,
\eeq
\beq
|L \rangle = p | \k \rangle - q | \bk \rangle ,
\eeq
or with $\epsilon \equiv (p-q) /( p+q)$,
\beq
|S \rangle = \frac{1}{\sqrt{2(1+| \epsilon |^2)}}
\left[ (1 + \epsilon ) | \k \rangle + (1 - \epsilon )| \bk \rangle \right]~~~,
\eeq
\beq
| L \rangle = \frac{1}{\sqrt{2(1+| \epsilon |^2)}}
\left[ (1 + \epsilon ) | \k \rangle - (1 - \epsilon )| \bk \rangle \right]~~~.
\eeq
In a CPT-invariant theory, a single complex parameter $\epsilon$ specifies
the eigenstates.

One can relate $\epsilon$ more directly to the properties of the mass matrix
and mass eigenvalues.
Making a phase choice when taking the square root of (\ref{eqn:pqs}), we write
\beq
{q \over p} = \sqrt{\frac{{\cal M}_{21}}{{\cal M}_{12}}}
\eeq
and note from (\ref{eqn:meq}) that
\beq
\mu_S = \m_{11} + \sqrt{\m_{12} \m_{21}}~~; ~~~
\mu_L = \m_{11} - \sqrt{\m_{12} \m_{21}}~~~,
\eeq
so
\beq \label{eqn:mudiff}
\mu_S - \mu_L = 2 \sqrt{\m_{12} \m_{21}}~~~.
\eeq
Then
\beq
\epsilon = \frac{p-q}{p+q} =
\frac{\sqrt{\m_{12}} - \sqrt{\m_{21}}}{\sqrt{\m_{12}} + \sqrt{\m_{21}}}
\simeq
\frac{\m_{12}-\m_{21}}{4 \sqrt{\m_{12} \m_{21}}}~~~,
\eeq
where the smallness of $\epsilon$ has been used.
With the definition of ${\cal M}$ and (\ref{eqn:mudiff}) we can then write
\beq
\epsilon \simeq
\frac{{\rm Im} (\Gamma_{12}/2) + i~{\rm Im} M_{12}}{\mu_S - \mu_L} ,
\eeq
so that the CP-violation parameter $\epsilon$ arises from imaginary parts of
off-diagonal terms in the mass matrix.

The matrices $\Gamma$ and $M$ may be expressed\cite{Kabir} in terms of sums
over states connected to $\k$ and $\bk$ by the weak Hamiltonian $H_W$:
\beq \label{eqn:g12}
\Gamma_{12} = 2 \pi \sum_F \rho_F \langle \k | H_W | F \rangle \langle F | H_W
| \bk \rangle,
\eeq
where $\rho_F$ denotes the density of final states $F$, and
\beq
M_{12} = \langle \k | H_W | \bk \rangle + \sum_n
\frac{\langle \k | H_W | n \rangle \langle n |H_W | \bk \rangle}{m_{K^0} - m_n}
\eeq
By considering specific $2 \pi,~3 \pi,~\pi l \nu$, and other final states, one
can show that $|{\rm Im} \Gamma_{12}/2| \ll |{\rm Im} M_{12}|$, and so
\beq
\epsilon \simeq \frac{i~{\rm Im}M_{12}}{\mu_S - \mu_L}~~~.
\eeq
This result implies a specific phase of $\epsilon$:
\beq
{\rm Arg}~\epsilon \approx \left\{ \begin{array}{c} 90^0 \\ 270^0 \end{array}
\right \} - ~{\rm Arg} (\mu_S - \mu_L ) ~~ {\rm for} ~~
\left \{ \begin{array}{c} {\rm Im}M_{12} > 0 \\ {\rm Im}M_{12} < 0 \end{array}
\right \}
\eeq
Given the measurements:\cite{PDG}
\beq
m_S - m_L = - 0.476 ~ \Gamma_S ; ~~
\Gamma_S - \Gamma_L = 0.998 ~ \Gamma_S ,
\eeq
we have $\mu_S - \mu_L = - (0.476 + 0.499i) \Gamma_S$, or Arg $(\mu_S -
\mu_L) = (3 \pi /2) - $  arctan $(0.476/0.499) = (3 \pi/2) - 43.6^0$.
Thus
$$
{\rm Arg} ~ \epsilon = (43.6 \pm 0.2)^0 ~~ ({\rm Im}~ M_{12} < 0 )~~~,
$$
\beq \label{eqn:epsph}
{\rm Arg }~ \epsilon = \pi + (43.6 \pm 0.2)^0 ~~ ({\rm Im}~ M_{12} > 0 )~~~.
\eeq
\bigskip

E. ~~ CP-violating observables and expected phases
\medskip

The parameter $\epsilon$ actually depends on the phase convention used to
relate $\k$ to $\bk$. Observable convention-independent quantities can be
defined in terms of ratios of decay amplitudes. We define
\beq \label{eqn:epsdef}
\epsilon_0 \equiv
\frac{\langle0 | T | L \rangle }{\langle 0 | T | S \rangle }
{}~~~;~~~ \epsilon_2 \equiv \frac{1}{\sqrt{2}}
\frac{\langle2 | T | L \rangle }{\langle 0 | T | S \rangle },
\eeq
which relate CP-violating decays of $K_L$ to $2 \pi$ in $I = 0$ and $I = 2$
final states to the CP-conserving decay $K_S \to (2 \pi )_{I=0}$, and the
ratio of CP-conserving $I = 2$ and $I = 0$ amplitudes
\beq \label{eqn:omdef}
\omega \equiv
\frac{\langle 2 | T | S \rangle }{\langle 0 | T | S \rangle } ~.
\eeq
The ratios for specific charge states corresponding to CP-forbidden and
CP-allowed $2 \pi$ decays are defined as
\beq
\eta_{+-} \equiv
\frac{\langle + - | T | L \rangle }{\langle +- | T | S \rangle } ~~;
{}~~ \eta_{00} \equiv
\frac{\langle 00 | T | L \rangle }{\langle 00 | T | S \rangle } ~~~.
\eeq
Recalling the expressions for $|+- \rangle$ and $|00 \rangle$ in terms of
isospin states, and substituting, we find
$$
\eta_{+-} ~~=~~
\frac{\langle2 | T | L \rangle + \sqrt{2} \langle 0 | T | L \rangle}{\langle2 |
T | S \rangle + \sqrt{2} \langle 0 | T | S \rangle}~~~,
$$
\beq \label{eqn:etas}
\eta_{00}~~=~~ \frac{ \sqrt{2}\langle2 | T | L \rangle - \langle 0 | T | L
\rangle}{\sqrt{2}\langle2 | T | S \rangle - \langle 0 | T | S \rangle}~~~,
\eeq
or
\beq
\eta_{+-} =
\frac{\epsilon_0 + \epsilon_2}{1 + \omega / \sqrt{2}}
\approx \epsilon_0 + \epsilon '
\eeq
\beq
\eta_{00} =
\frac{\epsilon_0 -2 \epsilon_2}{1 - \omega \sqrt{2}}
\approx \epsilon_0 - 2 \epsilon'
\eeq
where
\beq \label{eqn:epspdef}
\epsilon' \equiv \epsilon_2 - \frac{\omega \epsilon_0}{\sqrt{2}} ~~.
\eeq

In order to relate these results to an expression involving $\epsilon$, we must
first discuss phases of amplitudes. We may factor out the final state phase
shift $\delta_I$ in the amplitude $\langle I|T|\k \rangle$ to write
\beq
\langle I | T | \k \rangle \equiv A_I e^{i \delta_I} ~~.
\eeq
Applying CPT to this result, we shall now show that
\beq
\langle I | T | \bk \rangle = A_I^* e^{i \delta_I} ~~,
\eeq
so that the same final state phase appears, accompanied by the complex
conjugate amplitude. The proof goes as follows.\cite{WolfCPT} Let a final state
$F_{\rm out}$ be related to a ``standing-wave'' state by $F_{\rm out} = e^{i
\delta} F_0$. Then, since CPT is an antiunitary operator, CPT $F_{\rm out} = e
^{-i \delta} F_0$. But also CPT $F_{\rm out} = F_{\rm in}$, so we have shown
that
\beq
F_{\rm in} = e^{-2i\delta} F_{\rm out} ~~~.
\eeq
This relation holds for eigenstates of the $S$-matrix (such as $2 \pi$ states
with $\m_{\pi \pi} = \m_k$ and $J = 0$). Let us start with
$\langle F_{\rm out}  | T | \k \rangle = A_I e^{i \delta_I}$.
Now consider the desired matrix element:
\beq
\langle F_{\rm out} | T | \bar K \rangle = \langle CPT ~ F_{\rm out} | T |
CPT ~ \bar K \rangle^*
\eeq
(CPT is antiunitary)
\beq
= \langle F_{\rm in} | T | K \rangle^*
\eeq
(by definition of $F_{\rm in}$ and $K$)
\beq
= \langle F_{\rm out} | T | K \rangle^* e^{2 i \delta_I} = (A_I e^{i \delta_I})
^* e^{2 i \delta_I} = A_I^* e^{i \delta_I}~~~,
\eeq
which was to be proved.

We may now substitute the expressions for $|S \rangle$ and $|L\rangle$ into the
definitions (\ref{eqn:epsdef}) and (\ref{eqn:omdef}). First of all, we have
\beq
\epsilon_0 =
\frac{\langle0 | T | \k \rangle ( 1 + \epsilon ) - \langle0 | T | \bk \rangle
(1 - \epsilon ) }{\langle0 | T | \k \rangle ( 1 + \epsilon ) + \langle0 | T |
\bk \rangle (1 - \epsilon ) } = \frac{i~\I A_0 + \epsilon~\R A_0}
{\R A_0 + i~\epsilon~\I A_0 }~~~.
\eeq
The relation between $\epsilon$ and $\epsilon_0$ depends on the phase of $A_0$.
For now, we shall adopt a definition of $\k$ and $\bk$ such that ${\rm Im}A_0 =
0$ (the Wu-Yang\cite{WY} convention), in which case $\epsilon_0 = \epsilon$.
If, instead, we were to take $A_0 = \rho_0 e^{i \phi_0}$ ($\rho_0$ real), then
\beq
\epsilon_0 =
\frac{i \sin \phi_0 + \epsilon \cos \phi_0}{\cos \phi_0 + i~\epsilon \sin
\phi_0}  \approx
\epsilon ( 1 - i \phi_0 ) + i \phi_0
\eeq
to lowest order in small quantities. This result implies that
\beq
\R \epsilon_0 \simeq \R \epsilon + \phi_0~\I \epsilon
\eeq
Here the second term is small in comparison with the first if $\phi_0$ is
small, which it usually is in most conventions differing from the Wu-Yang
convention.\cite{NonWY}

The corresponding results for $\epsilon_2$ and $\omega$ involve the difference
of final state phases $\delta_2$ and $\delta_0$ in $I_{\pi \pi} = 2$ and
$I_{\pi \pi} = 0$ channels:
\beq
\epsilon_2 = \frac{1}{\sqrt{2}}
\frac{i~\I A_2 + \epsilon~\R A_2}{\R A_0 + i~\epsilon~\I A_0 }~
e^{i(\delta_2 - \delta_0 )}~~~,
\eeq
\beq
\omega =
\frac{\R A_2 + i~\epsilon~\I A_2}{\R A_0 + i~\epsilon~\I A_0}~
e^{i (\delta_2 - \delta_0)}~~~.
\eeq
When $\I A_0 = 0$, taking account of (\ref{eqn:epspdef}), we have the
simple result
\beq
\epsilon' = \frac{i}{\sqrt{2}} \frac{\I A_2}{A_0} ~e^{i (\delta_2 -
\delta_0
)}
\eeq
The phase of $\epsilon'$ is then governed entirely by the final state $\pi \pi$
phases. We shall show that, just as for $\epsilon$, the phase of $\epsilon'$
should be close to 45$^\circ$.

Present information on $\pi \pi$ phase shifts\cite{phs} for $I_{\pi \pi} = 0$
is based at very low $\m_{\pi \pi}$ on $K_{e4}$ decay,\cite{Ke4} and at higher
$\m_{\pi \pi}$ on data from the reaction $\pi^- p \to \pipe n$ using pion
exchange.\cite{OPE} It is estimated\cite{phs} that $\delta_0 - \delta_2 = (42
\pm 4)^\circ$, and
\beq \label{eqn:epspph}
{\rm Arg} ~ \epsilon' = 48 \pm 4^\circ ~~,
\eeq
which implies [cf. Eq.~(\ref{eqn:epsph})] that $\epsilon$ and $\epsilon'$ have
nearly the same phase.

Combining $\epsilon_0 = \epsilon$ with Eqs.~(\ref{eqn:etas}), we have
\beq
\eta_{+-} = \epsilon + \epsilon'~~~;~~~\eta_{00} = \epsilon - 2 \epsilon'~~~.
\eeq
In view of the near-equality of the phases of $\epsilon$ and $\epsilon'$,
we find
$$
| \eta_{+-} | \simeq | \epsilon | [ 1 + {\rm Re} (\epsilon '/\epsilon) ] ~~~,
$$
\beq
| \eta_{00} | \simeq | \epsilon | [ 1 - 2 {\rm Re} (\epsilon'/\epsilon ] ~~~,
\eeq
so
\beq
\left | \frac{\eta_{00}}{\eta_{+-}} \right |^2 =
\frac{\Gamma (K_L \to 2 \pi^0)}{\Gamma ( K_S \to 2 \pi^0 )} /
\frac{\Gamma (K_L \to \pipe )}{\Gamma (K_S \to \pipe )} = 1 - 6~{\rm Re}
\frac{\epsilon '}{\epsilon} ~~~.
\eeq
This double ratio is measurable with considerably less systematic error than
any individual ratios involving a single type of decaying particle or single
type of final state. Two experiments, one at Fermilab (E731)\cite{E731} and one
at CERN (NA31),\cite{NA31} have provided the most recent values, implying
\beq
{\rm NA31}: ~~ {\rm Re} (\epsilon '/\epsilon) = (23.0 \pm 6.5) \times 10^{-4}
{}~~~,
\eeq
\beq
{\rm E731}: ~~ {\rm Re} (\epsilon '/ \epsilon) = (7.4 \pm 6.0) \times 10^{-4}
{}~~~.
\eeq

The central values obtained by the two experiments are very different in their
implications for fundamental theories of CP violation. If $ \epsilon' \neq 0$,
as suggested by the NA31 result, CP violation must be occurring in a decay
amplitude (specifically, via ${\rm Im} A_2 \neq 0$) as well as through the mass
matrix.  If $\epsilon' = 0$ and only $\epsilon$ is nonzero, as suggested by the
E731 result, CP violation could arise from any of a number of sources,
including a special $\Delta S = 2$ ``superweak'' interaction concocted
especially to affect the mass matrix,\cite{sw} many other types of interactions
to be discussed in Section 6, or even an accidental cancellation of effects
within the context of CKM physics. We shall discuss this possibility in Section
4. More data will be forthcoming from both groups.

The results just mentioned suggest that $0 \leq {\rm Re}(\epsilon'/\epsilon)
\leq 2 \times 10^{-3}$. With Arg$(\epsilon'/\epsilon) = (4.3 \pm 4)^0$, one
then finds that $\phi_{00} \equiv {\rm Arg} (\eta_{00})$ and $\phi_{+-} \equiv
{\rm Arg} (\eta_{+-})$ should differ by no more than $0.1^0$. Anything more
implies a breakdown in the logic leading to Eqs.~(\ref{eqn:epsph}) and
(\ref{eqn:epspph}), which would be most naturally blamed on a violation of CPT
invariance.

Recently, Fermilab experiment E773\cite{Schwing} has obtained the result
$\phi_{00} - \phi_{+-} = (0.62 \pm 1.03)^{\circ}$, compatible with CPT
invariance.  A combined fit to E731 and E773 data also finds $\phi_{+-}
= (43.53 \pm 0.97)^{\circ}$, consistent with the value (\ref{eqn:epsph})
expected for the phase of $\epsilon$ and with the world average\cite{PDG}
$\phi_{+-} = (44.3 \pm 0.8)^{\circ}$.
\bigskip

F. ~~ Semileptonic decays
\medskip

The charge asymmetry in the semileptonic decays of $K_L$ provides further
evidence that it is not a CP eigenstate. Let us define
\beq
\delta_l \equiv
\frac{\Gamma(K_L \to \pi^- l^+ \nu ) - \Gamma (K_L \to \pi^+ l^- \bar \nu
)}{\Gamma(K_L \to \pi^- l^+ \nu ) + \Gamma (K_L \to \pi^+ l^- \bar \nu )}
{}~~~.
\eeq
In the standard model of weak charge-changing interactions, only the $\k$
component of the $K_L$ leads to $\pi^- l^+ \nu$, since the corresponding quark
subprocess is $\bar s \to \bar u l^+ \nu$. Similarly, only the $\bk$ component
leads to $\pi^+ l^- \bar \nu$; at the quark level this decay is $s \to u l^-
\bar \nu$.  These processes have $\Delta Q$ (the change in charge of the
hadron) equal to $\Delta S$ (the change in its strangeness). Processes with
$\Delta Q = - \Delta S$, such as $\bk \to \pi^- l^+ \nu$, have not been
observed and are not expected in the standard model. Neglecting their
contribution (one may also include them in a more complete
treatment\cite{dsdq}), we may use the expression for $K_L$ in terms of $\k$ and
$\bk$ to find
\beq
\delta_l \simeq 2 ~ {\rm Re} ~ \epsilon ~~~.
\eeq

Experimental averages\cite{PDG} for this quantity are
$$
\delta_e = (3.33\pm 0.14) \times 10^{-3} ~~~;~~~
\delta_\mu = (3.04 \pm 0.25) \times 10^{-3} ~~~,
$$
leading to an overall average
\beq \label{eqn:deltal}
\delta_l = (3.27 \pm 0.12) \times 10^{-3} ~~~.
\eeq
Using $|\epsilon| \simeq |\eta_{+-}|[1 - {\rm Re}~(\epsilon'/\epsilon)] = (2.26
\pm 0.02) \times 10^{-3}$ and Arg $\epsilon = (44.8 \pm 0.8)^{\circ}$ from the
experimental world average,\cite{PDG} one obtains $ 2~{\rm Re}~\epsilon = (3.24
\pm 0.07) \times 10^{-3}$, in excellent accord with Eq.~(\ref{eqn:deltal}).
\bigskip

G. ~~ $K_S$ decays
\medskip

{\it 1. The decay $K_S \to 3 \pi^0$} is purely CP violating. The CP of the
initial state is positive, that of each pion is negative, and the zero spin of
$K_S$ and each pion leads to positive parity of the final $3 \pi$ spatial wave
function.

By searching for interference between $K_L \to 3 \pi^0$ and $K_S \to 3 \pi^0$,
it has been found possible\cite{KSbd} to place the indirect bound $B(K_S \to 3
\pi^0) < 3.7 \times 10^{-5}$.  If this decay were to occur via mixing alone,
one would expect
\beq
B(K_S \to 3 \pi^0) = \frac{\tau_S}{\tau_L} |\epsilon|^2 B(K_L \to 3 \pi^0)
\simeq 2 \times 10^{-9}~~~.
\eeq

{\it 2.  The decay $K_S \to \pi^+ \pi^- \pi^0)$} has been studied in the
CPLEAR\cite{CPLEAR} and Fermilab E621\cite{E621} experiments. CPLEAR can
directly ``tag'' the flavor of the produced neutral kaon in the reaction $\bar
p p \to [\k K^- \pi^+~{\rm or}~\bk K^+ \pi^-]$, thereby measuring both
$\eta_{+-}$ (though not, so far, with smaller errors than other experiments)
and $\eta_{+-0} \equiv A(K_S \to [\pi^+ \pi^- \pi^0]|{\rm CP = -}) /A(K_L \to
\pi^+ \pi^- \pi^0)$.  By searching for $K_S - K_L$ interference in the $\pi^+
\pi^- \pi^0$ final state, the bound $B(K_S \to \pi^+ \pi^- \pi^0) < 8 \times
10^{-7}$ has been set. If the decay were to occur via mixing, one would expect
\beq
B(K_S \to [\pi^+ \pi^- \pi^0]|_{\rm CP = -}) = \frac{\tau_S}{\tau_L}
|\epsilon|^2 B(K_L \to \pi^+ \pi^- \pi^0) \simeq 1.1 \times 10^{-9}~~~.
\eeq
The decay $K_S \to [\pi^+ \pi^- \pi^0]|_{\rm CP = +}$ has been identified at
CPLEAR.\cite{CPLEAR}  This involves orbital angular momenta between the
pions, as mentioned in Section 2 B.

{\it 3.  One application of a $\phi$ factory}, in which $K_S - K_L$ pairs are
produced via the reaction $\eep \to \phi \to K_S K_L$, is to select a
$K_L$ by means of some prominent decay mode and to look for (e.g.) the
$K_S \to 3 \pi^0$ decay.  At least $10^{10}~\phi$ mesons are necessary to
make a useful measurment; such numbers are envisioned for a machine currently
under construction at Frascati near Rome.\cite{phifact}  The absence of
$K_L K_L$ and $K_S K_S$ pairs in such a reaction follows from Bose statistics
since the $\phi$ has $J = 1$.
\newpage

\leftline{3. PHYSICS OF THE CKM MATRIX}
\bigskip

A. ~~ Quark masses and transitions
\medskip

Our understanding of the weak interactions has undergone tremendous progress in
the past century.  It is almost 100 years since beta-decay electrons were first
identified by J.~J.~Thomson in 1897.  Fermi first formulated the theory of
beta-decay as a charge-changing process. The space-time properties of such
processes were finally settled in terms of $V-A$ interaction\cite{VA} nearly a
quarter of a century later. We now understand the charge-changing weak
interactions as part of a unified structure,\cite{GWS} encompassing also
electromagnetism and charge-{\em preserving} weak interactions.

Both leptons and quarks participate in the charge-changing weak interactions.
The patterns, however, appear to be radically different.

At present, for lack of better information,  we view each charged lepton as
undergoing charge-changing transitions to or from its own neutrino, as
illustrated on the right-hand side of Fig.~1. On the other hand, the quarks as
shown on the left-hand side of Fig.~1, participate in a rich pattern of
charge-changing transitions. This pattern is summarized in a $3 \times 3$
unitary matrix, the Cabibbo-Kobayashi-Maskawa (CKM) matrix.\cite{Cab,KM}

% This is Figure 1
\begin{figure}
% \vspace{3in}
\centerline{\epsfysize = 3 in \epsffile {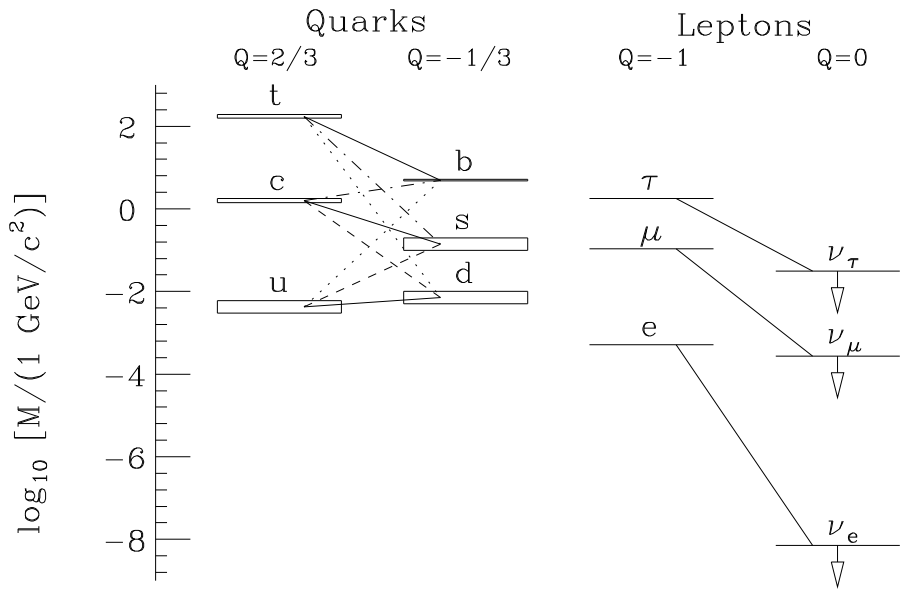}}
\caption{Patterns of charge-changing weak transitions among quarks and leptons.
Direct evidence for $\nu_\tau$ does not yet exist. The strongest inter-quark
transitions correspond to the solid lines, with dashed, dot-dashed, and dotted
lines corresponding to successively weaker transitions.}
\end{figure}

% This is Table 1
\begin{table}
\caption{Relative strengths of charge-changing weak transitions.}
\begin{center}
\begin{tabular}{||c|c|l||} \hline
Relative & Transition & Source of information \\
amplitude & & ~~~~~~~~(example) \\ \hline
$\sim$ 1 & $u \leftrightarrow d$ & Nuclear $\beta$-decay \\
$\sim$ 1 & $c \leftrightarrow s$ & Charmed particle decays \\
$\sim 0.22$ & $u \leftrightarrow s$ & Strange particle decays \\
$\sim 0.22$ & $c \leftrightarrow d$ & Neutrino prod. of charm \\
$\sim 0.04$ & $c \leftrightarrow b$ & $b$ decays \\
$\sim 0.003$ & $u \leftrightarrow b$ & Charmless $b$ decays \\
$\sim$ 1 & $t \leftrightarrow b$ & Only indirect evidence \\
$\sim 0.04$ & $t \leftrightarrow s$ & Only indirect evidence \\
$\sim 0.01$ & $t \leftrightarrow d$ & Only indirect evidence \\ \hline
\end{tabular}
\end{center}
\end{table}

We summarize the approximate relative strengths of the charge-changing weak
transitions in Table 1. The relative phases of these amplitudes are also
of importance. We shall describe how all this information is obtained after
a discussion of the way in which the CKM matrix arises.
\bigskip

B. ~~ Origin of the CKM matrix
\medskip

The electroweak Lagrangian, before electroweak symmetry breaking, may be
written in flavor-diagonal form as
\beq
{\cal L}_{\rm int} = - \frac{g}{\sqrt{2} }[ \overline{U '}_L
\gamma^\mu W_\mu^{(+)} {D'}_L + H.c.]~~~,
\eeq
where $U' \equiv (u',c',t')$ and $D' \equiv (d',s',b')$ are column vectors
decribing {\em weak eigenstates}. Here $g$ is the weak $SU(2)_L$ coupling
constant, and $\psi_L \equiv (1 - \gamma_5) \psi /2$ is the left-handed
projection of the fermion field $\psi = U$ or $D$.

Quark mixings arise because mass terms in the Lagrangian are permitted to
connect weak eigenstates with one another. Thus, the matrices ${\cal M}_{U,~D}$
in
\beq
{\cal L}_m = - [\overline{U '}_R {\cal M}_U {U'}_L + \overline {D '}_R {\cal
M}_D {D'}_L + H.c.]
\eeq
may contain off-diagonal terms. One may diagonalize these matrices by separate
unitary transformations on left-handed and right-handed quark fields:
\beq
R_{Q}^+ {\cal M}_{Q} L_{Q} = L_{Q}^+ {\cal M}_{Q}^+ R_Q = \Lambda_Q ~~~.
\eeq
where
\beq
{Q'}_L = L_Q Q_L ; ~~ {Q'}_R = R_Q Q_R ~~~ (Q = U, D)~~~.
\eeq
Using the relation between weak eigenstates and mass eigenstates:
${U'}_L = L_U U_L , ~ {D'}_L = L_D D_L$, we find
\beq
{\cal L}_{\rm int} = - \frac{g}{\sqrt{2}} [ \overline{U}_L \gamma^\mu W_\mu
V D_L + H.c.] ~~~,
\eeq
where $U \equiv (u,c,t)$ and $D \equiv (d,s,b)$ are the mass eigenstates, and
$V \equiv L_U^+ L_D$. The matrix $V$ is just the Cabibbo-Kobayashi-Maskawa
matrix. By construction, it is unitary: $V^+V = VV^+ = 1$. It carries no
information about $R_U$ or $R_D$. More information would be forthcoming from
interactions sensitive to right\--handed quarks or from a genuine theory of
quark masses.
\bigskip

C. ~~ Parameter counting
\medskip

For $n~u$-type quarks and $n~d$-type quarks, $V$ is $n \times n$ and
unitary. An arbitrary $n \times n$ matrix has $2n^2$ real parameters, but
unitarity $(V^+ V = 1)$ provides $n^2$ constraints, so only $n^2$ real
parameters remain. We may remove $2n-1$ of these by appropriate redefinitions
of relative quark phases. The number of remaining parameters is then $n^2 -
(2n-1) = (n-1)^2$. Of these, $n (n - 1)/2$ (the number of independent rotations
in $n$ dimensions) correspond to angles, while the rest, $(n-1)(n-2)/2$,
correspond to phases. We summarize these results in Table 2.

% This is Table 2
\begin{table}
\caption{Parameters of KM matrices for $n$ doublets of quarks.}
\begin{center}
\renewcommand{\arraystretch}{1.2}
\begin{tabular}{| c | c c c c |} \hline
 & $n =$  & 2~~~ &~~~ 3~~~ &~~~ 4 \\ \hline
Number of parameters & $(n-1)^2 $ & 1~~~ &~~~ 4~~~ &~~~ 9 \\
Number of angles & $n(n-1)/2$  & 1~~~ &~~~ 3~~~  &~~~ 6 \\
Number of  phases& $ (n-1)(n-2)/2$ & 0~~~ &~~~ 1~~~ &~~~ 3 \\ \hline
\end{tabular}
\end{center}
\renewcommand{\arraystretch}{1.0}
\end{table}

For $n=2$, we have one angle and no phases. The matrix $V$ then can always be
chosen as orthogonal.\cite{Cab,Charm} For $n=3$, we have three angles and one
phase, which in general cannot be eliminated by arbitrary choices of phases in
the quark fields. It was this phase that motivated Kobayashi and
Maskawa\cite{KM} to introduce a third quark doublet.  It provides a potential
source of CP violation, serving as the leading contender for the observed
CP-violating effects in the kaon system and suggesting substantial CP
asymmetries in the decays of mesons containing $b$ quarks.

The CKM matrix $V$ is then, explicitly,
\beq \label{eqn:CKM}
V = \left( \begin{array}{c c c}
V_{ud} & V_{us} & V_{ub} \\
V_{cd} & V_{cs} & V_{cb} \\
V_{td} & V_{ts} & V_{tb}
\end{array} \right) ~~~.
\eeq
We now parametrize its elements.

It is convenient to choose quark phases\cite{QP} so that the $n$ diagonal
elements and the $n-1$ elements just above the diagonal are real and positive.
Since all the angles $\theta_{ij}$ are small, the $V_{cs}$ element in
(\ref{eqn:CKM}) is nearly real, so only a small change in quark phases is
needed to accomplish this. The parametrization we shall employ is one suggested
by Wolfenstein.\cite{WP}

The diagonal elements of $V$ are nearly $1$, while the dominant off-diagonal
elements are $V_{us} \simeq - V_{cd} \equiv \lambda \simeq 0.22$. Thus to order
$\lambda^2$, the upper $2 \times 2$ submatrix of $V$ is already known from the
four-quark pattern. The empirical observation that $V_{cb} \simeq 0.04$ allows
one to express it as $A \lambda^2$, where $A = {\cal O}(1)$. Unitarity then
requires $V_{ts} \simeq - A \lambda^2$ as long as $V_{td}$ and $V_{ub}$ are
small enough (which they are).  Finally, $V_{ub}$ appears to be of order $A
\lambda^3 \times {\cal O}(1)$. Here one must allow for a phase, so one must
write $V_{ub} = A \lambda^3 (\rho - i \eta )$. Finally, unitarity specifies
uniquely the form $V_{td} = A \lambda^3 (1 - \rho - i \eta )$. To summarize,
the CKM matrix may
be written
\beq \label{eqn:CKMwp}
V \approx \left[ \matrix{1 - \lambda^2/2 & \lambda & A \lambda^3 (\rho -
i \eta) \cr
- \lambda & 1 - \lambda^2 /2 & A \lambda^2 \cr
A \lambda^3 (1 - \rho - i \eta) & - A \lambda^2 & 1 \cr } \right]~~~.
\eeq

We shall anticipate further results to note that $V_{cb} = 0.038 \pm 0.003$.
This enables us to write $A = 0.79 \pm 0.06$. The measurement of semileptonic
charmless $B$ decays gives $|V_{ub}/V_{cb}|$ in the range from 0.06 to 0.10,
where most of the uncertainty is associated with the spread in models for the
lepton spectra. Taking $0.08 \pm 0.02$ for this ratio, we find that the
corresponding constraint on $\rho$ and $\eta$ is $(\rho^2 + \eta^2 )^{1/2} =
0.36 \pm 0.09$. The main indeterminacy in the CKM matrix concerns the magnitude
of $V_{td}$, for which only indirect evidence exists.

The form (\ref{eqn:CKMwp}) is only correct to order $\lambda^3$ in the matrix
elements. For certain purposes it may be necessary to exhibit corrections of
higher order to the elements. This can be done using the unitarity of the
matrix.

Unitarity implies that $V_{ij}^* V_{ik} = \delta_{jk}$ and $V_{ij}^* V_{kj} =
\delta_{ik}$, where summation over repeated indices is understood. For example,
we have
\beq \label{eqn:ur}
V_{ud}^* V_{td} + V_{us}^* V_{ts} + V_{ub}^* V_{tb} = 0 ~~~.
\eeq
Since $V_{ud}^* \approx 1 , ~ V_{us}^* \approx \lambda , ~ V_{ts} \approx - A
\lambda^2$, and $V_{tb} \approx 1$ we have $V_{td} + V_{ub}^* = A \lambda^3$, a
useful relation expressing the least-known CKM elements in terms of relatively
well-known parameters.  This result can be visualized as a triangle\cite{UT} in
the complex plane [Fig.~2(a)]. In this figure the angles $\alpha, \beta$, and
$\gamma$ are defined as in the review by Nir and Quinn.\cite{NQ}

% This is Figure 2
\begin{figure}
% \vspace{1.2in}
\centerline{\epsfysize = 1.2 in \epsffile {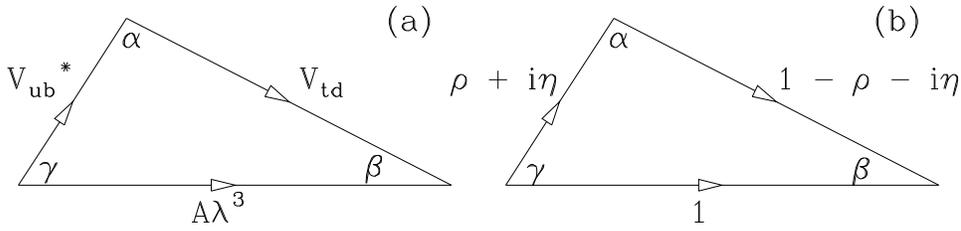}}
\caption{Unitarity triangle for CKM elements.
(a) The relation (\ref{eqn:ur}) in the complex plane;
(b) Eq.~(\ref{eqn:ur}) divided by the normalizing factor $A \lambda^3$.}
\end{figure}

Dividing (\ref{eqn:ur}) by $A \lambda^3$, since $V_{ub}^*/ A \lambda^3 = \rho +
i \eta, ~~ V_{td} / A \lambda^3 = 1 - \rho - i \eta$, one obtains a triangle of
the form shown in Fig.~2(b). The value of $V_{ub}^* / A \lambda^3$ may then be
depicted as a point in the $(\rho , \eta)$ plane. The major ambiguity which
still remains in the determination of the CKM matrix elements concerns the
shape of the unitarity triangle. The answer depends on the magnitude of
$V_{td}$. As we shall see, decays alone will not provide the answer. One
resorts to indirect means, which involve loop diagrams.
\bigskip

D. ~~ Direct measurements of elements
\medskip

{\it 1. The matrix element $|V_{ud}|^2$}
may be obtained by comparing the strengths of certain beta-decay transitions
involving vector transitions with that of muon decay. One can also measure the
neutron decay rate (which involves both vector and axial vector transitions),
and extract the vector coupling strength by finding $g_A$ from decay
asymmetries. This vector coupling strength may be compared with that obtained
in muon decay to learn $|V_{ud} |^2$. Finally, one can study the decay $\pi^+
\to \pi^0 e^+ \nu_e$.

A nuclear beta-decay process typically involves unknown matrix elements of
both vector and axial currents. In transitions between two $0^+$ states which
belong to the same isospin multiplet, however, the conserved vector current
hypothesis\cite{VA,GZ} serves to normalize the vector current. These
transitions are known as {\em superallowed Fermi transitions}.

In order to extract a fundamental coupling strength from beta-decay lifetimes
one must apply a phase space and Coulomb correction factor, independent of the
decaying nucleus, and further nuclear-dependent radiative corrections.  The
results for eight nuclei:  $^{14}$O, $^{26}$Al, $^{34}$Cl, $^{38}$K, $^{42}$Sc,
$^{46}$V, $^{50}$Mn, and $^{54}$Co lead to an average $|V_{ud}| = 0.9740 \pm
0.0006$. However, the possibility of systematic effects which become more
important for high-$Z$ nuclei have led Marciano\cite{WMVUD} to place most
reliance in the $^{14}$O value, with the error reflecting systematic as well
as statistical uncertainty: $|V_{ud}| = 0.9748 \pm 0.0010$.

The measurements of the neutron lifetime and $g_A$ have now become precise
enough that one can extract a value of $V_{ud}$:\cite{JRCKM} $|V_{ud}| = 0.9804
\pm 0.0005 ~ ({\rm rad.~corrs.}) ~ \pm 0.0010 ~(\tau_n ) ~ \pm 0.0020 ~ (g_A)$,
a bit above the values implied by superallowed Fermi beta-decay transitions.

The decay rate $\Gamma(\pi^+ \to \pi^0 e^+ \nu_e)$ is governed by a form
factor which is nearly at zero momentum transfer, and radiative corrections
appear to be well in hand. From present measurements one obtains $|V_{ud}| =
0.968 \pm 0.018$.\cite{pib} Improvement by a factor of 8 would match the
precision of the neutron experiments.

{\it 2.  The matrix element $V_{us}$} is probed by the semileptonic decays of
strange particles.  The vector current matrix elements provide the most
reliable information, since they are affected only to second order by SU(3)
symmetry-breaking effects.\cite{AG} Two main sources of information on the
vector $\Delta S = 1$ (strangeness-changing) current are the decays $K \to \pi
\ell \nu$ (known as $K_{\ell 3}$ decays) and the semileptonic decays of
hyperons.\cite{VUSfit} The result obtained from $K_{\ell 3}$ decays is $V_{us}
= 0.2196 \pm 0.0023$, while that obtained in a recent fit to semileptonic
hyperon decays is $|V_{us}| = 0.220 \pm 0.001 \pm 0.003$. In what follows we
shall use an average of the $K_{e3}$ and hyperon values:\cite{WMVUD} $|V_{us}|
= 0.220 \pm 0.002$.

{\it 3.  The quantity $|V_{cd}|$} can be learned from deep inelastic neutrino
excitation of charm via reactions such as $\nu_\mu d \to \mu^- c$ and
$\bar{\nu}_\mu \bar{d} \to \mu^+ \bar{c}$ and from semileptonic decays of
charmed mesons to nonstrange final states. Both are associated with some
model-dependence. The average of results from deep inelastic
scattering\cite{VCDdis} is $|V_{cd} | = 0.205 \pm 0.011$.  This value is
compatible with the Cabibbo-Glashow-Iliopoulos-Maiani\cite{Cab,Charm}
four-quark picture, which would imply $V_{cd} = - V_{us}$. The semileptonic
decays of charmed particles to nonstrange final states also lead to values
compatible with this hypothesis.\cite{VCDslc}

{\it 4.  The matrix element $V_{cs}$} is provided by neutrino interactions and
charmed particle semileptonic decays. The neutrino reactions $\nu_\mu s \to
\mu^- c $ and $\bar{\nu}_\mu \bar{s} \to \mu^+ \bar c$ both must proceed off
strange quarks or antiquarks in the nucleon sea. Some uncertainty is associated
with modeling this component of the proton. One then looks for semileptonic
decays of the $c$ or $\bar{c}$, leading (as above) to dimuon events. This
method\cite{VCDdis} leads to a lower bound $|V_{cs}| \ge 0.59$. The
semileptonic charmed particle decays providing information on $V_{cs}$ are $D^0
\to K^- e^+ \nu_e$ and $D^+ \to \overline{K}^0 e^+ \nu_e$. With assumptions
about the form factor,\cite{VCS} one finds $|V_{cs}| = 1.07 \pm 0.16$.

{\it 5. The measurement of $|V_{cb}|$} constituted the first evidence for a
structure of the $\Delta Q = 1$ weak transitions extending beyond the
four-quark picture. It is now known that the dominant decays of $b$
quarks involve charmed quarks in the final state, so that $|V_{cb}| >
|V_{ub}|$, but the rather small value of $|V_{cb}|$ proved to be a surprise.

The underlying physics in extracting $V_{cb}$ from data involves measurement of
the rate for the semileptonic decay of $B$ mesons (mesons containing the $b$
quark), and careful accounting of strong interaction effects relating this
process to the underlying $b \to c \ell \nu$ transition strength.  A free-quark
estimate\cite{JRCKM} suffices to illustrate the main features. For the decay of
a free $b$ quark to a light quark $q$ and an additional fermion pair $A
\overline{B}$, the predicted rate is
\beq
\Gamma (b \to q \overline{A} B) =
\frac{G_{F}^2 m_{b}^5 N_c}{192 \pi^3} | V_{qb} |^2 \Phi
\left( \frac{m_q}{m_b} ,\frac{m_A}{m_b}, \frac{m_B}{m_b} \right) ~~~,
\eeq
where $N_c = 1$ when $\overline{A} B$ is a lepton pair and $3$ when it is a
quark pair (in which case one should also multiply the result by $|V_{AB}|^2$).
The kinematic factor $\Phi$ is\cite{kin}
$$
\Phi (x, z, y) = 12 \int_{(x+y)^2}^{(1-z)^2} \frac{ds}{s}
(s - x^2 - y^2) (1 + z^2 - s)
$$
\beq
\times \left\{ [s-(x-y)^2][s-(x+y)^2][(1+z)^2-s][(1-z)^2-s]\right\}^{1/2}~~~.
\eeq
Specific limits include $\Phi (0,0,0) = 1$ and
\beq
\Phi (x,0,0) = 1 - 8 x^2 + 8 x^6 - x^8 - 24 x^4 \ln x~~~.
\eeq

For $q=c$ and $\overline{A} B = \bar{\nu}_\ell \ell$ the partial
rate is (neglecting $m_\ell$)
\beq
\Gamma (b \to c \bar{\nu}_\ell \ell^- ) = 7.14 \times 10^{-11} ~{\rm GeV}
\left( \frac{m_b}{5~{\rm GeV}} \right)^5 |V_{cb}|^2
\Phi \left( \frac{m_c}{m_b},0,0 \right) ~~~.
\eeq
This is to be compared with the measured semileptonic $b$ decay rate
\beq
\Gamma (B \to {\rm charm} + \bar{\nu}_\ell + \ell^- ) =
\frac{\hbar B (B \to {\rm charm} + \bar{\nu}_\ell + \ell^- )}{\tau_B}
\eeq
For $B(B \to {\rm charm}+\bar{\nu}_\ell+ \ell^-) = 10.5\%$ and $\tau_B = 1.49$
ps it was then concluded,\cite{JRCKM} based on the ranges $m_b = 5.0 \pm 0.3$
GeV/$c^2$ and $m_b - m_c = 3.37 \pm 0.03$ GeV/$c^2$ in typical descriptions of
hadron spectra, that
\beq \label{eqn:vcb}
|V_{cb}| = 0.038 \pm 0.003~~~;~~~ A \equiv |V_{cb}|/|V_{us}|^2 = 0.785 \pm
0.062~~~.
\eeq
More recent data\cite{BH} imply a slightly longer $B$ meson lifetime (the
average of $\tau(B^+)$ and $\tau(B^0)$ is $1.63 \pm 0.05$ ps), and an inclusive
semileptonic branching ratio of $10.98 \pm 0.28\%$ (with some additional
systematic error associated with choice of model).  The result (\ref{eqn:vcb})
nonetheless continues to represent the range of values obtained under various
assumptions\cite{HQS} about the way in which free quarks are incorporated into
hadrons, and is consistent with a recent determination\cite{PBall} with
slightly smaller quoted errors.

{\it 6.  The magnitude of $V_{ub}$} may be estimated by studying the spectra of
charged leptons in the decays $b \to q \bar \nu_\ell \ell^-$, which are
sensitive to the relative contributions of $b \to u \bar \nu_e e^-$ and $b \to
c \bar \nu_e e^-$ and hence to $|V_{ub}/V_{cb}|$. Defining $x \equiv 2E_e
/m_b$, where $E_e$ is the electron energy in the $b$ rest frame, $\Gamma_0
\equiv G_{F}^2 m_{b}^5/(192 \pi^3)$, and $\zeta \equiv m_{q}^2/ m_{b}^2$, we
have\cite{spect}
\beq
\frac{1}{\Gamma_0} \frac{d \Gamma}{d x} = 2x^2 | V_{qb} |^2
\left ( \frac{ 1 - \zeta - x}{1 - x} \right )^2
\left [ 1 - \zeta - x + \frac{2-x}{1-x} (1 + 2 \zeta - x) \right ] ~~~.
\eeq
When $\zeta = 0$ one obtains the Michel spectrum $\sim 2 x^2 (3-2x)$, which
attains its maximum at $x = 1$. When $\zeta \neq 0$ the spectrum falls to zero
at the kinematic limit $x = 1 - \zeta$. Corrections to the shape from
perturbative and nonperturbative strong interactions are important,\cite{corrs}
but the main points already can be visualized without them.

The integrals over the spectra for $ b \to u \ell^- \bar \nu_\ell$ and $b \to c
\ell^- \bar \nu_\ell$ lead to about a factor of two kinematic suppression of $b
\to c \ell^- \bar \nu_\ell$ relative to $b \to u \ell^- \bar \nu_\ell$. The
spectrum of leptons emitted in semileptonic $B$ decays is consistent on the
whole with the $b \to c \ell \bar \nu_\ell$ prediction after corrections for
QCD effects and detector resolution are taken into account. The decay $b \to
c\ell \bar \nu_\ell$ is far more prevalent than $b \to u \ell \bar \nu_\ell$.
However, by studying leptons with energies beyond the endpoint for $ b \to c
\ell \bar \nu_\ell$, both ARGUS\cite{ARGVub} and CLEO\cite{CLVub} have
concluded that $b \to u \ell \bar \nu_\ell$ does occur, at a rate about 2\%
(ARGUS) or 1\% (CLEO) of that for $ b \to c \ell \bar \nu_\ell$.  When combined
with the spread in various models\cite{Vubmods} for deviations from the
free-quark predictions, these measurements imply that
\beq
|V_{ub}/V_{cb}| = 0.08 \pm 0.02~~~;~~~(\rho^2 + \eta^2)^{1/2} = 0.36 \pm
0.09~~~.
\eeq

The use of exclusive decays such as $B \to (\pi,~\rho,{\rm or}~\omega) \ell
\nu$ may lead in the future to more restrictive bounds on $|V_{ub}|$. Some
recent progress in identifying $B \to \pi \ell \nu$ decays has been reported by
the CLEO Collaboration.\cite{pilnu}

A measurement of the branching ratio for $B \to \tau \nu$ or $B \to \mu \nu$,
though challenging, will be helpful in extracting the product $f_B |V_{ub}|$,
where the $B$ decay constant $f_B$ is defined by
\beq \label{eqn:fbdef}
\langle 0 | J^\mu | B (q) \rangle = i q^\mu f_B ~~~.
\eeq
The predicted decay rates are
\beq
\Gamma (B^+ \to l^+ \nu_l) = \frac{G_F^2 f_B^2 m_l^2 m_B}{8 \pi}
\left[ 1 - \frac{m_l^2}{m_B^2} \right]^2 |V_{ub} |^2 ~~~.
\eeq
Here we have chosen a normalization such that the analogous decay constant for
pions\cite{WMVUD} is $f_\pi = 132$ MeV. The expected branching ratios are
$$
B (B^+ \to \mu^+ \nu_\mu ) = (1.1 \times 10^{-7})
(f_B/f_\pi)^2 |V_{ub}/0.003|^2 ~~~;
$$
\beq
B (B^+ \to \tau^+ \nu_\tau ) = (2.5 \times 10^{-5})
(f_B/f_\pi)^2 |V_{ub}/0.003|^2 ~~~.
\eeq

Recent estimates\cite{FBL,FBQ} suggest $f_B \approx (1.4 \pm 0.2) f_\pi$. The
hadronic form factors for decays like $ B \to \pi l \nu , ~ B \to \rho l
\nu$, or $B \to \omega l \nu$ may be related to those of $ D \to \pi l \nu ,~
\rho l \nu $ or $\omega l \nu$ using heavy-quark symmetries,\cite{HQS} making
it possible to learn $|V_{ub}|$ from exclusive processes once the Dalitz plots
of these $D$ decays have been studied.

{\it 7. Elements involving the top quark} have not yet been measured directly.
{}From the unitarity of the CKM matrix, we expect $|V_{tb}| \simeq 1, |V_{ts}|
\simeq 0.04$, and $|V_{td}| \simeq 0.005$ to $0.012$. For $m_t > M_W + m_b,~t$
will decay to $W + (s ~{\rm or}~ d)$ with a branching ratio less than $ 2
\times 10^{-3}$, with the dominant decay being to $W + b$. The decay $t \to W
b$ is expected to have a partial width of $1.8 \pm 0.4$ GeV for $m_t = 180 \pm
12$ GeV/$c^2$ and $|V_{tb}| \approx 1$.

We shall now discuss indirect measurements of $|V_{ts}|$ and $|V_{td}|$ by
means of their contributions to box diagrams.
\bigskip

E. ~~ Box diagrams
\medskip

Indirect information on the CKM matrix is provided by $B^0 - \bar B^0$ mixing
and CP-violating $K^0 - \bar K^0$ mixing, through the contributions of box
diagrams involving two charged $W$ bosons and two quarks of charge 2/3
$(u,~c,~t)$ on the intermediate lines.  These calculations have acquired new
precision as a result of evidence for the top quark with a mass of $m_t = 180
\pm 12$ GeV/$c^2$, where we have averaged values of $176 \pm 8 \pm 10$
GeV/$c^2$ from CDF\cite{CDFt} and $199^{~+19}_{~-21} \pm 22$ GeV/$c^2$ from
D0.\cite{D0t}

Details of calculations of box diagrams have been given in several
places,\cite{CL,TASI,IL} so we shall give only the main results.  A key
feature is the cancellation of the highest powers of momenta in the loop
integrals as a result of the unitarity of the CKM matrix:
$V_{ud}^* V_{us} + V_{cd}^* V_{cs} + V_{td}^* V_{ts} = 0$ or
$V_{ud}^* V_{ub} + V_{cd}^* V_{cb} + V_{td}^* V_{tb} = 0$.
An approximation whose validity must be checked by explicit calculation
of hadronic matrix elements (e.g., in lattice gauge theory) is the
assumption of ``vacuum saturation,'' in which one takes the matrix element
of the box diagram between $\k$ and $\bk$ by inserting the vacuum in all
possible ways.  One writes\cite{CL}
\beq \label{eqn:bkdef}
\langle \k | (\bar d{La} \gamma^\mu s_L^a)(\bar d{Lb} \gamma_\mu s_L^b) |
\bk \rangle = - 2 m_K^2 f_K^2 B_K /3~~~,
\eeq
where $a$ and $b$ are color indices, $f_K = 161$ MeV is the kaon decay
constant, and $B_K = 1$ corresponds to the assumption of vacuum saturation.
\bigskip

F. ~~ CP-violating mixing of $K$ mesons
\medskip

A mass term in the Lagrangian is of the form ${\cal L}_m = - m^2 \phi_K^{\dag}
\phi_K$ where $\phi_K$ denotes the kaon field.  To make contact with what is
calculated from Feynman rules, we note that the electromagnetic interaction
term in the Lagrangian, ${\cal L}_{\rm int}^{\rm e.m.} = - e \bar \psi
\gamma^\mu A_\mu \psi$, corresponds to the Feynman rule $-ie \gamma_\mu$. Thus,
if we denote by $A_{\rm eff}$ the amplitude calculated according to Feynman
rules for $K-\bar K$ mixing, we have the correspondence
\beq
{1 \over i} \langle K^0|A_{\rm eff}|\bar K^0 \rangle = \delta m^2 =
2m_K{\cal M}_{12}~~~.
\eeq

% This is Figure 3
\begin{figure}
% \vspace{1.3in}
\centerline{\epsfysize = 1.3 in \epsffile {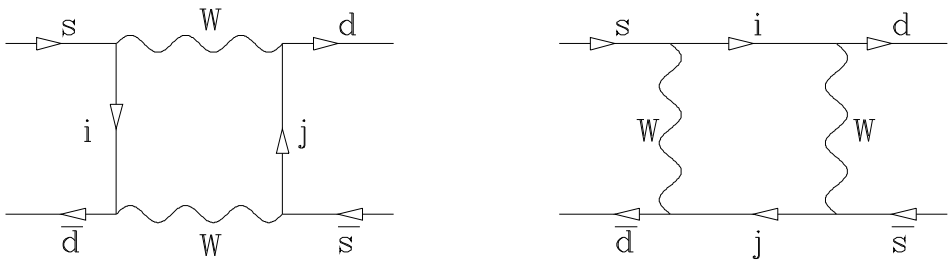}}
\caption{Box diagrams for mixing of a neutral kaon and its antiparticle.}
\end{figure}

The loop-diagram contribution to $A_{\rm eff}$ from the exchange of a pair of
$W$ bosons is illustrated in Fig.~3.  Here $i,j=u,c,t$ denote the quarks in the
intermediate states.  One finds in the limit in which all quark masses are
small compared to $M_W$ that the matrix element between a $\k$ and a $\bk$ is
\beq \label{eqn:kmixlo}
M_{12} = {-G_F^2 m_K f_K^2 B_K \over 12 \pi^2} \left[ m_c^2 \xi_c^2 + m_t^2
\xi_t^2 + 2 m_c^2 \xi_c \xi_t \ln {m_t^2 \over m_c^2} \right]~~~,
\eeq
where $\xi_i \equiv V_{id}^* V_{is}$. The first, second, and third terms
correspond to loop diagrams with two charmed quarks, two top quarks, and one of
each, respectively.

The top quark mass appears not to be small in comparison with $M_W.$
For this case, one must take into account the $p_\mu p_\nu /M_W^2$ terms in
the $W$ propagator.\cite{IL}

We shall use the above result (and its generalization for heavy top) only to
calculate the {\it imaginary} part of $M_{12},$ which affects CP violation.
The imaginary part is dominated by large momenta in the loop graph, and thus is
an effect of short-distance physics. The {\it real} part, on the other hand, is
dominated by the contribution of the charmed quark and hence of much lower
momenta, and a short-distance calculation is probably unreliable.  The loop
diagrams of Fig.~3 do not contribute {\it too large} a mass difference between
$K_L$ and $K_S.$  This result was one of the bases for the conclusion\cite{GL}
that the charmed quark mass could not be more than a couple of GeV/$c^2.$  A
quantitative calculation of the real part of $M_{12}$ involves contributions
from $\pi^0$, $\eta$, $\eta'$, and other single-particle states; a significant
contribution of the $2 \pi$ intermediate state,\cite{TNT} as well as
contributions from other low-mass states.

The box diagram's contributions to the parameter $\epsilon$, according to the
discussion in Sec.~2, may be evaluated from
\beq
\epsilon \simeq {i ~\I M_{12} \over \mu_S -\mu_L} \simeq
{- \I M_{12} e^{i \pi/4} \over \sqrt 2 \Delta m}~~~.
\eeq
The generalization of Eq.~(\ref{eqn:kmixlo}) to the case of general quark
masses\cite{IL} then leads to
\beq \label{eqn:eps}
|\epsilon| \simeq {G_F^2 m_K f_K^2 B_K M_W^2 \over \sqrt 2 (12 \pi^2) \Delta
m} [ \eta_1 S(x_c) \I \xi_c^2
+ \eta_2 S(x_t) \I \xi_t^2 + 2 \eta_3 S(x_c,x_t) \I \xi_c \xi_t]~~~.
\eeq
The factors $\eta_1 = 0.85,~\eta_2=0.61,~\eta_3=0.36$ are QCD
corrections,\cite{QCDK} while $x_i \equiv m_i^2/M_W^2.$  The functions $S(x)$
and $S(x,y)$ are
\beq \label{eqn:sdef}
S(x) \equiv {x \over 4} \left[ 1 + {3-9x \over (x-1)^2} + {6x^2\ln x \over
(x-1)^3} \right]~~~;
\eeq
\beq
S(x,y) \equiv xy\left\{ \left[ {1 \over 4} + {3 \over 2(1-y)} - {3 \over
4(1-y)^2} \right] {\ln y \over y-x} + (y \leftrightarrow x) - {3 \over
4(1-x)(1-y)} \right\}~~.
\eeq

To evaluate the required imaginary parts of the $\xi_i^2$ or $\xi_i\xi_j,$
we must use a slightly better approximation to the CKM matrix than was
introduced in Sec.~3 C.  The application of the unitarity relation to the
first and second rows of the matrix tells us, in fact, that a more precise
expression for $V_{cd}$ is $V_{cd} = - \lambda - A^2 \lambda^5 (\rho + i
\eta).$  We then find:
$$
\I \xi_c^2 = 2 \R \xi_c \I \xi_c = -2A^2 \lambda^6 \eta~~~;
$$
$$
\I \xi_c \xi_t \simeq \R \xi_c \I \xi_t = A^2 \lambda^6
\eta~~~;
$$
\beq
\I \xi_t^2 = 2 \R \xi_t \I \xi_t = 2A^2 \lambda^6 \eta [
A^2 \lambda^4 (1 - \rho) ]~~~.
\eeq

Eq.~(\ref{eqn:eps}) may then be rewritten as\cite{HRS}
\beq
|\epsilon| = 4.33 A^2 B_K \eta [\eta_3 S(x_c,x_t) - \eta_1 S(x_c) + \eta_2
A^2 \lambda^4 (1-\rho) S(x_t) ]~~~.
\eeq
Using the experimental value\cite{PDG} of $|\epsilon| = (2.26 \pm 0.02) \times
10^{-3}$, the value $B_K = 0.8 \pm 0.2$,\cite{BKlat} and the top quark mass
$m_t = 180 \pm 12$ GeV/$c^2$, we find that CP-violating $K - \bar K$ mixing
leads to the constraint
\beq
\eta(1 + 0.35 - \rho) = 0.48 \pm 0.20~~~,
\eeq
where the first term in parentheses corresponds to the loop diagram with two
top quarks, and the second corresponds to the additional contribution of
charmed quarks.  The major source of error on the right-hand side is the
uncertainty in the parameter $A \equiv V_{cb}/\lambda^2$.
\bigskip

F. ~~ Mixing of $B$ mesons
\medskip

The original evidence for $B^0 - \bar B^0$ mixing came from the presence of
``wrong-sign'' leptons in $B$ meson semileptonic decays.\cite{bbmix}  More
recently, the time-dependence of $B^0 - \bar B^0$ oscillations has been
traced by several groups.\cite{BH}

The splitting $\Delta m_B$ between mass eigenstates is proportional to $f_B^2
m_t^2 |V_{td}|^2$ times a slowly varying function of $m_t$. Here $f_B$ is the
decay constant of the $B$ meson [defined in Eq.~(\ref{eqn:fbdef})].  The
contributions of lighter quarks in the box diagrams, while necessary to cut off
the high-energy behavior of the loop integrals, are numerically insignificant.

The CKM element $|V_{td}|$ is proportional to $|1 - \rho - i \eta|$.  Thus,
exact knowledge of $\Delta m_B,~f_B$ and $m_t$ would specify a circular arc
in the $(\rho,\eta)$ plane with center (1,0).  Errors on all these quantities
spread this arc out into a band.

A general parametrization of the mass matrix for states of two neutral mesons
which can mix with one another can be written by transcribing the results
of Sec.~2 as
\beq
{\cal M} = \left[ \begin{array}{cc}
(\mu_L +\mu_S)/2 & p (\mu_S-\mu_L)/2q \\
q (\mu_S-\mu_L)/2p & (\mu_L +\mu_S)/2
\end{array} \right]~~~.
\eeq
Now, very few intermediate states are common to both $B^0$ and
$\bar B^0$ decays, in contrast to the kaon case for which the $\pi \pi$ state
is dominant.  The $B^0$ (with a $\bar b$ quark) decays mostly to states with a
$\bar c$ (such as $\bar D \pi$), while the $\bar B^0$ (with a $b$ quark) decays
mostly to states with a $c$ (such as $D \pi$). As a result, the elements
$\Gamma_{12}$ and $\Gamma_{21}$ are negligible in comparison with $M_{12}$ and
$M_{21}$, and $\Gamma_S \approx \Gamma_L.$  Consequently, $\mu_S -\mu_L \approx
m_S -m_L,$ a real quantity. Then $\mu_S - \mu_L = 2 \sqrt{{\cal M}_{12} {\cal
M}_{21}}$ is nearly real, so that $p/q \approx (q/p)^*,$ or
\beq
|p/q| \approx 1~~~;~~~|\mu_S - \mu_L| \approx 2|{\cal M}_{12}|~~~.
\eeq

The relative magnitudes of the contributions for different $Q=2/3$ quarks are
easily estimated.  Here, $\xi_c = V_{cd}^*V_{cb} \approx (-\lambda)(A
\lambda^2)$ and $\xi_t = V_{td}^*V_{tb} \approx (A \lambda^3[1-\rho+i
\eta])(1)$ are comparable in magnitude, but $m_t \gg m_c,$ so that the diagram
with $t$ quarks in both internal lines dominates the mass difference.  Here, in
contrast to the situation with kaons, a short-distance calculation of $\Delta
m$ is likely to be reliable.  Furthermore, there are fewer shared intermediate
states between $B^0$ and $\bar B^0$, so that mixing via such
states is unlikely to be very important.  Thus, the mixing
should be given fairly reliably by the contributions shown in Fig.~4.

% This is Figure 4
\begin{figure}
% \vspace{1.3in}
\centerline{\epsfysize = 1.3 in \epsffile {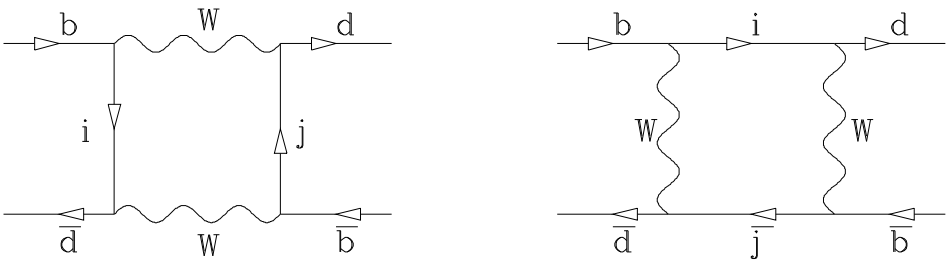}}
\caption{Box diagrams for mixing of $B^0$ and $\bar B^0$.}
\end{figure}

A crucial parameter probed by the graphs of Fig.~4, in addition to the top
quark mass, is the KM element $V_{td}.$  The expression for the ratio of
mass shift to lifetime is
\beq
{\Delta m \over \Gamma} = {G_F^2 \over 6 \pi^2} |V_{td}|^2 M_W^2 m_B f_B^2
B_B \eta_B \tau_B S \left( {m_t^2 \over M_W^2} \right)~~~.
\eeq
Here $f_B$ is defined such that the pion decay constant is 132 MeV.  We shall
take $f_B = 180 \pm 30$ MeV in accord with estimates\cite{FBL,FBQ} to be
described in more detail in Sec.~4 A. $B_B$ is a parameter describing the
degree of vacuum saturation for matrix elements analogous
to (\ref{eqn:bkdef}) for the $B$ mesons.  We shall take $B_B = 1$; the
calculation depends only on the product $f_B B_B^{1/2}$. The parameter $\eta_B$
describes a QCD correction;\cite{QCDB} we take $\eta_B = 0.6 \pm 0.1$. The
$B^0$ lifetime is taken\cite{BH} to be $\tau_B = 1.621 \pm 0.067$ ps. The
function $S$ is the one defined in Eq.~(\ref{eqn:sdef}).

Recent averages \cite{BH} for the $B^0 - \bar B^0$ mixing parameter $\Delta
m_d = 0.462 \pm 0.026$ ps$^{-1}$ and the $B^0$ lifetime $\tau(B^0) = 1.621 \pm
0.067$ ps can be combined to yield $\Delta m/ \Gamma = 0.75 \pm 0.05$.  If
interpreted in terms of the box diagram for $b \bar d \to d \bar b$ (dominated
by the top quark), this value leads to an estimate for $|V_{td}|$ reducing to
$|1 - \rho - i \eta| = 1.03 \pm 0.22$.  The dominant source of uncertainty in
the right-hand side is the error on $f_B$.  Prospects for reducing this
error will be discussed in Sec.~4.
\bigskip

I. ~~ Allowed parameter space
\medskip

Putting the constraints of the above discussion together, we find the
allowed region of the $(\rho,\eta)$ plane illustrated in Fig.~5.  In
each case the $1 \sigma$ outer limits are quoted.

{\it 1.  Charmless $B$ decays} lead to the constraint $(\rho^2 + \eta^2)
^{1/2} = 0.36 \pm 0.09$ mentioned earlier, and hence to the dotted
semicircles with center (0,0).

{\it 2.  The parameter $|\epsilon|$} leads to the constraints denoted by the
solid hyperbolae.

{\it 3.  $B^0 - \bar B^0$ mixing} leads to the constraints shown by the dashed
circular arcs with center (0,1).

% This is Figure 5
\begin{figure}
% \vspace{3.5in}
\centerline{\epsfysize = 3.5 in \epsffile {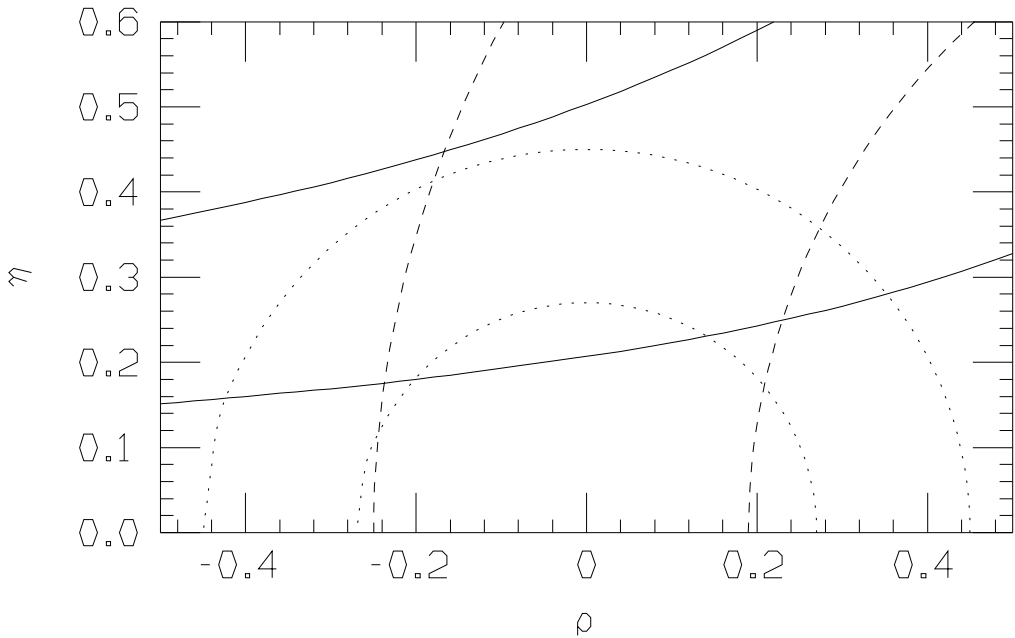}}
\caption{Region in the $(\rho,\eta)$ plane allowed by constraints on
$|V_{ub}/V_{cb}|$ (dotted semicircles), $B^0 - \bar B^0$ mixing (dashed
semicircles), and CP-violating $K - \bar K$ mixing (solid hyperbolae).}
\end{figure}

A large region centered about $\rho \simeq 0$, $\eta \simeq 0.35$ is permitted.
Nonetheless, it could be that the CP violation seen in kaons is due to an
entirely different source, perhaps a superweak mixing of $K^0$ and $\bar K^0$
\cite{sw}. In that case one could probably still accommodate $\eta = 0$, and
hence a real CKM matrix.  In order to confirm the predicted nonzero value of
$\eta$, we turn to other experimental possibilities.
\bigskip

\leftline{4. IMPROVED TESTS OF THE STANDARD PICTURE}
\bigskip

In this section we discuss several types of improved tests of the standard CKM
picture of CP violation.  We leave a discussion of the role of $B$ mesons
to Section 5.
\bigskip

A. ~~ Heavy meson decay constants
\medskip

The decay constant $f_B$ of the $B^0$ meson (the ground state of a $b$ and a
$\bar d$) affects the interpretation of $B^0 - \bar B^0$ mixing in terms of
fundamental parameters of the CKM matrix (particularly $|V_{td}|$). Thus, it is
important to know $f_B$ better.  The decay constants of other pseudoscalar
mesons with one heavy quark and one light antiquark, such as $D_s = c \bar s$,
$D^+ = c \bar d$, and $B_s = b \bar s$) are relevant auxiliary information.
They allow one to check the validity of calculations, and are related to one
another in the heavy-quark-symmetry limit by simple scaling
relations.\cite{HQS}

{\it 1.  The $D_s$ decay constant} has recently been measured by several
groups.  Initial evidence for its value\cite{Dsfact} was obtained by studying
the decays $B \to D_s \bar D$ and assuming that the $D_s$ was produced by the
weak current.  More recently, the decays $D_s \to \mu \nu$ and a few candidates
for $D_s \to \tau \nu$ have been seen.\cite{Dsmeas} One obtains roughly
$f_{D_s} = 300 \pm 50$ MeV.

{\it 2.  The $D^+$ decay constant} is expected on the basis of SU(3)-breaking
estimates\cite{FBL,FBQ} to be 0.8 to 0.9 of $f_{D_s}$.  Quark model
arguments\cite{FBQ} favor the lower value, leading to the prediction
$f_D = 240 \pm 40$ MeV.  This is not far below the value obtained by the
Mark III Collaboration\cite{MkIII} studying the decay $D \to \mu \nu$ at
SPEAR:  $f_D < 290$ MeV (90\% c.l.).

{\it 3.  The $B^0$ decay constant} has been estimated both by means of
lattice gauge theories\cite{FBL} and via spin-dependent isospin splittings in
$D$ and $B$ mesons,\cite{FBQ} which provide a value of the square of the
wave function at zero interquark separation.  Values in the range
$f_B = 190 \pm 40$ MeV are obtained.

{\it 4.  The $B_s$ decay constant} may be related to $f_B$ by the SU(3)
relation mentioned above for charmed mesons:  $f_B/f_{B_s} = 0.8$ to 0.9,
with 0.8 favored by quark models and some recent lattice models.\cite{FBL,FBQ}
We then predict $f_{B_s} = 240 \pm 40$ MeV.

We now give some details of these results.

The factorization hypothesis as applied \cite{Dsfact} to $\bar B$ decays to
$D_s^- D^{(*)}$ is illustrated in Fig.~6.  It is necessary to evaluate the form
factors for emission of the current and to test factorization in other
processes in order to utilize this method.

% This is Figure 6
\begin{figure}
% \vspace{1.6in}
\centerline{\epsfysize = 1.6 in \epsffile {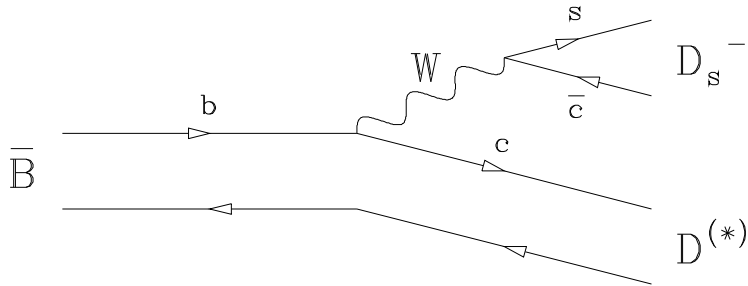}}
\caption{Application of factorization hypothesis to the decay
$\bar B \to D_s^- D^{(*)}$.  The weak current produces the $D_s^-$.}
\end{figure}

The emission of a current with 4-momentum $q$ by a heavy quark $Q_A$ in a
transition to a heavy quark $Q_B$ can be described in the limit of infinitely
heavy quark masses by a universal form factor.\cite{TASI,IW} This quantity
depends only on the invariant square of the difference $w = u - u'$ of the
four-velocities $u \equiv p_A/m_A$ and $u' \equiv p_B/m_B$ of the initial and
final heavy mesons.  In terms of $q^2$ one has
\beq
w^2 = \frac{q^2 - (m_A - m_B)^2}{m_A m_B} = \frac{q^2 - q^2_{\rm max}}{m_A
m_b}~~~.
\eeq
For example, the form factors for $0^- \to 0^-$ and $0^- \to 1^-$ heavy-meson
transitions are related to one another.  (Here the symbol $J^P$ denotes spin
and parity.)  Moreover, quark models or the heavy-quark-symmetry limit can be
used to relate the production of a $D_s$ by the current to production of a
$D_s^*$.  Thus, by measuring such processes as $B \to D^* \ell \nu$ at various
values of momentum transfer, and using heavy-quark symmetry and the
factorization hypothesis, one can predict the rate for all processes $\bar B
\to D_s^{-(*)} D^{(*)}$ in which each final state meson is a pseudoscalar or a
vector meson.

One can test factorization at $q^2 = m_\pi^2$ for pion emission, relying on the
relation\cite{BJfact}
\beq
\frac{B(B \to D^* \pi)}{dB(B \to D^* \ell \nu)/dq^2} {\large |}_{q^2 = m_\pi^2}
= 6 \pi^2 f_\pi^2 |V_{ud}|^2~~~.
\eeq
The result\cite{BS} $f_\pi = 138 \pm 19$ MeV agrees with the actual value of
132 MeV.  At $q^2 = m_{D_s}^2$ factorization may be more problematic, since
final-state interactions are more likely to influence the result.  Applying
it nonetheless, the first estimates were obtained\cite{BS,JRFDS} for the $D_s$
decay constant: $f_{D_s} = 276 \pm 69$ MeV.

The direct observations of $D_s$ leptonic decays include the
following.\cite{Dsmeas} The WA75 collaboration has seen 6 -- 7 $D_s \to \mu
\nu$ events in emulsion and concludes that $f_{D_s} = 232 \pm 69$ MeV.  A
related experiment at Fermilab (the E653 Collaboration) also has events
consistent with leptonic decays of $D_s$, but analysis is still in
progress.\cite{NSpc} The CLEO Collaboration has a much larger statistical
sample than WA75; the main errors arise from background subtraction and overall
normalization (which relies on the $D_s \to \phi \pi$ branching ratio). Using
several methods to estimate this branching ratio, Muheim and Stone estimate
$f_{D_s} = 315 \pm 45$ MeV.  A recent value from the BES Collaboration is
$f_{D_s} = 434 \pm 160$ MeV (based on one candidate for $D_s \to \mu \nu$ and
two for $D_s \to \tau \nu$).  A reanalysis by F. Muheim of the Muheim-Stone
result using the factorization hypothesis yields $f_{D_s} = 310 \pm 37$ MeV.

Quark models can provide estimates of decay constants and their ratios. In a
non-relativistic model,\cite{ES} the decay constant $f_M$ of a heavy meson $M =
Q \bar q$ with mass $M_M$ is related to the square of the $Q \bar q$ wave
function at the origin by $f_M^2 = 12 |\Psi(0)|^2/M_M$.  The ratios of squares
of wave functions can be estimated from strong hyperfine splittings between
vector and pseudoscalar states, $\Delta M_{\rm hfs} \propto |\Psi(0)|^2/m_Q
m_q$.  The equality of the $D_s^* - D_s$ and $D^* - D$ splittings then suggests
that
\beq
f_D/f_{D_s} \simeq (m_d/m_s)^{1/2} \simeq 0.8 \simeq f_B/f_{B_s}~~~,
\eeq
where we have assumed that similar dynamics govern the light quarks bound to
charmed and $b$ quarks.

An estimate of $|\Psi(0)|^2$ can be obtained using electromagnetic
hyperfine splittings,\cite{FBQ} which are probed by comparing isospin
splittings in vector and pseudoscalar mesons.  Before corrections of order
$1/m_Q$, the values $f_D^{(0)} = 290 \pm 15$ MeV, $f_B^{(0)} = 177 \pm 9$
MeV were obtained.  With $f_M = f_M^{(0)} (1 - \Delta/m_M)$ $(M = D,~B)$, we
use our value of $f_D$ to estimate $\Delta/M_D = 0.20 \pm 0.11$, $\Delta/M_B =
0.07 \pm 0.04$, and hence $f_B = f_B^{(0)} (1- \Delta/m_B) = 164 \pm 11$
MeV\null.  Applying a QCD correction\cite{VSPW} of 1.10 to the ratio $f_B/f_D$,
we finally estimate $f_B = 180 \pm 12$ MeV\null. [This is the basis of the
central value taken in Section 3; the error of 30 MeV quoted there reflects our
estimate of systematic uncertainties.] We also obtain $f_{B_s} = 225 \pm 15$
MeV from the ratio based on the quark model.

In a lattice gauge theory calculation, Bernard \ite\cite{FBL} find
$$
f_B = 187 \pm 10 \pm 34 \pm 15~~{\rm MeV}~~~,
$$
$$
f_{B_s} = 207 \pm 9 \pm 34 \pm 22~~{\rm MeV}~~~,
$$
$$
f_D = 208 \pm 9 \pm 35 \pm 12~~{\rm MeV}~~~,
$$
\beq
f_{D_s} = 230 \pm 7 \pm 30 \pm 18~~{\rm MeV}~~~,
\eeq
where the first errors are statistical, the second are associated with fitting
and lattice constant, and the third arise from scaling from the static $(m_Q =
\infty)$ limit.  Duncan \ite\cite{FBL} find
\beq
f_B = 188 \pm 23~({\rm stat}) \pm 15~({\rm syst})^{~+26}_{~-0}~({\rm extrap})
\pm 14~({\rm pert})~{\rm MeV}~~~,
\eeq
in accord with the above result.  They obtain a ratio of strange to nonstrange
$B$ decay constants more in accord with our quark model estimates: $f_{B_s}/f_B
= 1.22 \pm 0.04~({\rm stat}) \pm 0.02~({\rm syst})$.
\bigskip

B. ~~ Rates and ratios
\medskip

Rates and ratios can constrain $|V_{ub}|$ and possibly $|V_{td}|$. The partial
width $\Gamma(B \to \ell \nu)$ is proportional to $(f_B |V_{ub}|)^2$.  The
expected branching ratios are about $(1/2) \times 10^{-4}$ for $\tau \nu$ and
$2 \times 10^{-7}$ for $\mu \nu$, as mentioned in Sec.~3 D 6.  The measurement
of $B^0 - \bar B^0$ mixing can provide the combination $f_B |V_{td}|$. Dividing
$f_B |V_{ub}|$ by $f_B |V_{td}|$, we can eliminate the decay constants,
obtaining
\beq
\frac{|V_{ub}|}{|V_{td}|} \equiv r = \left[ \frac{\rho^2 + \eta^2}
{(1-\rho)^2 + \eta^2} \right]^{1/2}~~~
\eeq
with smaller errors than in $|V_{ub}|$ or $|V_{td}|$.\cite{PHJR}  Contours
of fixed $r$ are circles in the $(\rho,\eta)$ plane with radius $|r/(1-r^2)|$
and center $\rho_0 = -r^2/(1-r^2),~\eta_0 = 0$.

Another interesting ratio\cite{ALI} is $\Gamma(B \to \rho \gamma)/ \Gamma(B
\to K^* \gamma)$, which, aside from phase space corrections, should be
$|V_{td}/V_{ts}|^2 \simeq 1/20$.  Soni, however,\cite{AS} has argued that
there are likely to be long-distance corrections to this relation.
\bigskip

C. ~~ Rare kaon decays
\medskip

A number of rare kaon decays can provide information on details of the CKM
matrix and on fundamental aspects of CP violation.  Other decays provide
valuable auxiliary data. We give a brief sample of the information here,
updating the experimental situation and referring the reader
elsewhere\cite{TASI,QCDK,Dib,RVW} for more complete discussions of formalism.

{\it 1. The decay $K_L \to \ga$} can proceed in a CP-conserving
manner.  It occurs with a branching ratio\cite{PDG} of $(5.73 \pm 0.27) \times
10^{-4}$, or nearly 2/3 the branching ratio of the CP-violating $\pi^0 \pi^0$
mode.\cite{Kgg}  Calculations of the rate involve prominent roles for hadronic
intermediate states such as $\pi^0$, $\eta$, and $\eta'$ (whose contributions
would sum to zero in the SU(3) limit), and thus involve long-distance
strong-interaction physics.

The $K_L \to \ga$ amplitude is important for several reasons:  (a) It
governs the main contribution to the decay $K_L \to \mu^+ \mu^-$; residual
contributions to this process can probe short-distance physics, as discussed
below.  (b) The related processes $K_L \to \gamma \eep$\cite{Keeg} and $K_L
\to \gamma \mu^+ \mu^-$\cite{Kmmg} provide information on the behavior of the
$K_L \to \ga^*$ amplitude for a virtual photon $\gamma^*$, which is
important in estimating certain long-distance contributions to $K_L \to \mu^+
\mu^-$.  (c) A potential background\cite{HG} to the process $K_L \to \pi^0 e^+
e^-$, to be described below, is the process $K_L \to \gamma_1 \eep$, where
the positron or electron radiates a photon $(\gamma_2)$.

{\it 2.  The decay $K_S \to \ga$}, measured\cite{Kgg} to have a branching ratio
of about $2 \times 10^{-6}$, is reliably predicted\cite{KSggt} from the
imaginary part of its amplitude, which arises as a result of the sequence $K_S
\to \pi^+ \pi^- \to \ga$.  This lends some credence to similar attempts to
estimate the rate for $K_L \to \ga \pi^0$ (a process to be described below).

{\it 3.  The decay $K_L \to \mu^+ \mu^-$} is dominated by the two-photon
intermediate state: $K_L \to \ga \to \mu^+ \mu^-$.  The lower limit on the
predicted branching ratio is given by the ``unitarity bound'' which is based on
neglect of all other contributions (including off-shall photons). The most
recent measurement of the rate\cite{Kmm} yields $B(K_L \to \mu^+ \mu^-) = (6.86
\pm 0.37) \times 10^{-9}$, very close to the unitarity bound and limiting other
contributions.  Since loop diagrams involving the top quark are included in
these contributions, an upper limit on $|V_{td}|$ ensues, entailing a lower
bound on $\rho$ of about $-0.7$.\cite{Kmm}  As we see from Fig.~5, a more
stringent lower bound of about $-0.2$ would begin to provide information more
powerful than that supplied by $B^0 - \bar B^0$ mixing.

{\it 4. The decay $K^+ \to \pi^+ \eep$} proceeds primarily through a virtual
photon via $K^+ \to \pi^+ \gamma^* \to \pi^+ \eep$.  The amplitude does not
have a pole at $q^2 = 0$ since the real transition $K^+ \to \pi^+ \gamma$ is
forbidden by angular momentum conservation.  (Both the $K^+$ and the $\pi^+$
are spinless.)  The $K^+ \to \pi^+ \gamma^*$ vertex has both short- and
long-distance contributions.  The experimental value of the branching
ratio\cite{Kpiee} is $B(K^+ \to \pi^+ \eep) = (2.75 \pm 0.23 \pm 0.13) \times
10^{-7}$ (under some assumptions about the form of the interaction).

If the process $K_S \to \pi^0 \gamma^*$ has the same amplitude as $K^+ \to
\pi^+ \gamma^*$ (a possibility if both transitions are governed by the
short-distance process $s \to d \gamma^*$), then one expects $B(K_S \to \pi^0
\eep) = (\tau_S/\tau_{K^+}) B(K^+ \to \pi^+ \eep) \simeq 2 \times
10^{-9}$.  This estimate becomes important in judging the potential for the
decay $K_L \to \pi^0 \eep$ to yield new information on CP violation, as we
shall see below.

{\it 5. The $K^+ \to \pi^+ \nu \bar \nu$ decay rate} rate is governed by loop
diagrams involving the cooperation of charmed and top quark contributions. An
approximate expression\cite{ER} for the branching ratio, summed over neutrino
species, is
\beq \label{eqn:bkpinunu}
B(K^+ \to \pi^+ \nu \bar \nu) \simeq 1.8 \times 10^{-6} | D(x_c) + D(x_t)
\lambda^4 (1 - \rho - i \eta)|^2~~~,
\eeq
where $x_i \equiv m_i^2/M_W^2$ and
\beq
D(x) \equiv \frac{1}{8} \left[ 1 + \frac{3}{(1-x)^2} - \left( \frac{4-x}{1-x}
\right)^2 \right] x \ln x + \frac{x}{4} - \frac{3}{4} \frac{x}{1-x}~~~.
\eeq
The first and second terms in (\ref{eqn:bkpinunu}) are due to the contributions
of the charm and top quarks.

Experimental observation of the $K^+ \to \pi^+ \nu \bar \nu$ decay would lead
to constraints involving circles in the $(\rho, \eta)$ plane with centers at
approximately (1.4,0).\cite{LL}  The favored branching ratio is about $1.2
\times 10^{-10}$, give or take a factor of 2. Variables affecting the result
include not only $\rho$ and $\eta$ (mainly $\rho$), but also the charmed quark
mass and the value of $A$. A low value of the branching ratio within this range
signifies $\rho > 0$, while a high value signifies $\rho < 0$. The present
upper limit\cite{LL} is $B(K^+ \to \pi^+ \nu \bar \nu) < 3 \times 10^{-9}$
(90\% c.l.).

{\it 6. The decay $K_L \to \pi^0 \ga$} is a potential CP-conserving source of
$K_L \to \pi^0 \eep$, a process expected to be mainly CP-violating.  The key
question is whether the two photons are produced in a $J = 2$ state, which can
produce $\eep$ without a suppression factor which operates in the $J = 0$
state.

Two experiments have detected this mode,\cite{pigg} yielding a branching ratio
of $(1.71 \pm 0.28) \times 10^{-6}$ [for $m(\ga)$ within certain
limits].\cite{PDG} Elementary calculations analogous to the rescattering
model\cite{KSggt} which gave $K_S \to \ga$ correctly or, equivalently, based on
chiral perturbation theory, can be applied to the sequence of processes $K_L
\to \pipe \pi^0$, $\pipe \to \ga$, to estimate the contribution of the
charged-pion pair to $K_L \to \pi^0 \ga.$\cite{piggt} The resulting prediction
of the rescattering model is that $B(K_L \to \pi^0 \ga) = 7.5 \times 10^{-7}$.

The contributions of the $\pipe$ intermediate state lead to a dominantly $J =
0~
{}~\ga$ final state. Because of the chirality conservation in electromagnetic
interactions, this final state is very inefficient in producing $\eep$ pairs.
The corresponding $\ga \to \eep$ rescattering prediction for $B(K_L \to
\pi^0 \eep )$ is about $10^{-13}$. This is far lower than the main
contributions
to $K_L \to \pi^0 \eep$, which we shall see are expected to be CP-violating.
On the other hand, if there is any $J = 2$ component to the $\ga$ final state
in $K_L \to \pi^0 \ga$, it can produce $\eep$ much more readily, and the
CP-conserving background to $K_L \to \pi^0 \eep$ becomes significant.

A $J_{\ga} = 2$ contribution to $K_L \to \pi^0 \ga$ is provided by
intermediate states of vector mesons $V$, e.g., via $K_L \to V \gamma \to
\pi^0 \ga$, where the weak Hamiltonian can act at either the first or the
second stage. Opinion seems divided,\cite{Ko,piggy} but typical branching
ratios predicted in such models are $B(K_L \to \pi^0 \ga ) \simeq (1-3) \times
10^{-6}$, in somewhat better accord with the experimental value than the
predictions of the $\pipe$ rescattering model or chiral perturbation theory.
However, the models with large vector meson contributions to the decay $K_L \to
\pi^0 \ga$ tend to involve much larger contributions for $m_{\ga} \leq 2m_\pi$
than the rescattering or chiral pertubation theory models. Although the
acceptance in the CERN experiment\cite{pigg} is highly non-uniform, the
authors are able to place an upper limit of about 12\% on the amount of the
decay $K_L \to \pi^0 \ga$ which comes for $m_{\ga} \leq 2m_\pi$.

The CP-conserving process $K_L \to \pi^0 \ga \to \pi^0 \eep$ therefore is
unlikely to be a major source of $K_L \to \pi^0 \eep$. This remote possibility
could be laid to rest by more detailed studies of the decay $K_L \to \pi^0
\ga$.

{\it 7. The decay $K_L \to \pi^0 \eep$} is expected to be dominated by
CP-violating contributions. Two types of such contributions are expected:
``indirect,'' via the CP-positive $K_1$ component of $K_L = K_2 + \epsilon
K_1$, and ``direct,'' whose presence would be a detailed verification of the
CKM theory of CP violation.

The indirect contribution is expected to lead to a branching
ratio\cite{Dib,RVW} $B^{\rm in} \simeq 6 \times 10^{-12}$, if one uses the
estimate quoted above that $B(K_S \to \pi^0 \eep) \simeq 2 \times 10^{-9}$.
(One calculation\cite{Ko} finds the $K_{S,L} \to \pi^0 \eep$ branching ratios
an order of magnitude larger.)  The phase of the indirect contribution is
expected to be that of $\epsilon$ (i.e., about 45$^{\circ}$) modulo $\pi$. The
direct contribution should be proportional to $i \eta$, and should be
comparable in magnitude to the indirect contribution in most estimates.  One
potential background\cite{HG} is the process $K_L \to \gamma_1 \eep$, where the
positron or electron radiates a photon $(\gamma_2)$. If $m(\gamma_1 \gamma_2)$
is too close to $m_{\pi^0}$, this process can be confused with the signal.

The present 90\% c.l. upper limit to $B(K_L \to \pi^0 \eep)$ is $1.8 \times
10^{-9}$, where results from several experiments have been combined.\cite{DHE}

{\it 8. The decay $K_L \to \pi^0 \mu^+ \mu^-$} should have less background than
$K_L \to \pi^0 \eep$ from photons radiated by the charged leptons, and
should have a comparable rate (aside from phase space differences).  The
present 90\% c.l. upper limit\cite{DHM} is $5.1 \times 10^{-9}$.

{\it 9. The decay $K_L \to \pi^0 \nu \bar \nu$} should have a branching ratio
of about $3 \times 10^{-10} \eta^2$ for $m_t = 180$ GeV.\cite{Dib,RVW}  It
should have only a very small indirect contribution, and would be
incontrovertible evidence for the CKM theory if observed at the predicted
level.  At present, experimental bounds\cite{pinunu} are $B(K_L \to \pi^0 \nu
\bar \nu) < 5.8 \times 10^{-5}$ (90\% c.l., summed over neutrino species).
\bigskip

D. ~~ $\epsilon'/\epsilon$
\medskip

The ratio $\epsilon'/\epsilon$ for kaons has long been viewed as one of the
most promising ways to disprove a ``superweak'' theory of CP violation in
neutral kaon decays\cite{sw,RVW}.  We sketch the way in which $\epsilon'$
arises in the standard electroweak picture.\cite{TASI,eps}

Several types of weak decay amplitudes contribute to $K \to \pi \pi$.

{\it 1.  The penguin graph}, involving the transition $s \to d$ with emission
of at least one gluon, leads only to an $I_{\pi \pi} = 0$ final state, since it
can only change isospin by 1/2 unit.  (Recall that the isospins of the two-pion
$J=0$ final states in kaon decays can be $I=0$ and $I=2.$)  Since it involves
three different generations of quarks as intermediate states, the penguin graph
can give rise to a decay amplitude which is complex at the quark level.  As we
have seen from the discussion in Sec.~3, three generations are the minimum
number for which this can occur.\cite{KM}

{\it 2.  The $W$ exchange diagram}, describing the subprocess $s \bar d
\to u \bar u$, also leads exclusively to an $I_{\pi \pi} = 0$ final state
(since a $q \bar q$ state cannot have $I = 2$), and to a real amplitude $A_0$.

{\it 3.  Free quark transitions} of the form $s \to u \bar u d$, in which
the $\bar d$ in the initial kaon acts as a spectator, can contribute a
real part to both $A_0$ and $A_2$.

% This is Figure 7
\begin{figure}
% \vspace{1.5in}
\centerline{\epsfysize = 1.5 in \epsffile {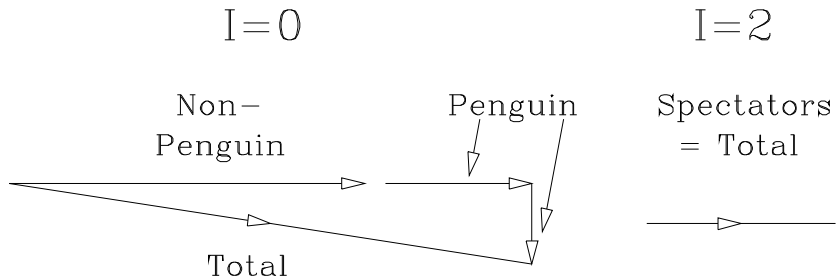}}
\caption{Summary of contributions to $I = 0$ and $I = 2$ final states in
$K \to \pi \pi$ decays.}
\end{figure}

Simple diagrams in the complex plane (Fig.~7) show the difference between the
total $I=0$ and $I=2$ amplitudes.  As a result of the complex penguin diagram
contribution, the two isospin amplitudes have a {\it nonzero relative phase}.
Adopting the convention\cite{WY} in which $A_0$ is real, we may rotate the
figures simultaneously by a small redefinition of the $K$ and $\bar K$ fields,
so that now it is $A_2$ which has a small imaginary part.  As mentioned in
Sec.~2, the parameter $\epsilon'$ describing direct CP violation (that not
due to mixing) in neutral kaon decays is then given by
\beq
\epsilon' = {i \over \sqrt 2} {\I A_2 \over A_0} e^{i(\delta_2 - \delta_0)}
{}~~~.
\eeq
The problem of calculating $\epsilon'$ then reduces to that of evaluating
the relative contribution of penguin and non-penguin amplitudes to the {\it
real} part of the $I=0$ amplitude.\cite{pen} It is more straightforward to
evaluate the phase of the penguin amplitude.  The magnitude of the $I=2$
amplitude could, in principle, be evaluated from first principles (though QCD
corrections would be important), but in practice it may be simply taken from
experiment.

Aadditional contributions and uncertainties are likely to decrease the overall
magnitude of $\epsilon'/\epsilon,$ and can even change its sign for top quarks
heavier than about 200 GeV.  The result should be corrected for
isospin-breaking effects and for terms involving the replacement of the gluon
emitted in the $s \to d$ transition by a photon or $Z,$ whose coupling to the
remaining quarks in the system is not isoscalar.  Such diagrams can contribute
directly to the $I=2$ amplitude and constitute an important effect on
$\epsilon'$.\cite{epscorr}

The latest estimates\cite{BEPS} are equivalent (for a top mass of about 180
GeV/$c^2$) to $[\epsilon'/\epsilon]|_{\rm kaons} = (6 \pm 3) \times 10^{-4}
\eta$, with an additional factor of 2 uncertainty associated with hadronic
matrix elements. The Fermilab E731 Collaboration's\cite{E731} value
$\epsilon'/\epsilon = (7.4 \pm 6) \times 10^{-4}$ is consistent with $\eta$ in
the range (0.2 to 0.45) we have already specified. The CERN NA31
Collaboration's\cite{NA31} value $\epsilon'/\epsilon = (23.0 \pm 6.5) \times
10^{-4}$ is higher than theoretical expectations. Both groups are preparing new
experiments, which should begin taking data around 1996.
\bigskip

\leftline{5. IMPORTANCE OF $B$ HADRONS}
\bigskip

A. ~~ Mixing of strange $B$ mesons
\medskip

$B_s - \bar B_s$ mixing probes the ratio $(\Delta m)|_{B_s}/(\Delta
m)|_{B_d} = (f_{B_s}/f_{B_d})^2 (B_{B_s}/B_{B_d})$ $|V_{ts}/V_{td}|^2$, which
should be a very large number (of order 20 or more). Thus, strange $B$'s should
undergo many particle-antiparticle oscillations before decaying.

The main uncertainty in an estimate of $x_s \equiv (\Delta m/ \Gamma)_{B_s}$ is
associated with $f_{B_s}$.  The CKM elements $V_{ts} \simeq -0.04$ and $V_{tb}
\simeq 1$ which govern the dominant (top quark) contribution to the mixing are
known reasonably well. We show in Table 3 the dependence of $x_s$ on $f_{B_s}$
and $m_t$. To measure $x_s$, one must study the time-dependence of decays to
specific final states and their charge-conjugates with resolution equal to a
small fraction of $\tau(B_s) = 1.55 \pm 0.13$ ps.\cite{BH}

% This is Table 3
\begin{table}
\caption{Dependence of mixing parameter $x_s$ on top quark mass and
$B_s$ decay constant.}
\begin{center}
\renewcommand{\arraystretch}{1.3}
\begin{tabular}{c c c c} \hline
\null \qquad $m_t$ (GeV/$c^2$)&  168  &  180  &  192  \\ \hline
$f_{B_s}$ (MeV)               &       &       &       \\
150                           &   9   &   10  &   11  \\
200                           &  15   &   17  &   19  \\
250                           &  24   &   27  &   30  \\ \hline
\end{tabular}
\end{center}
\end{table}

The estimate $f_B/f_{B_s} = 0.8 - 0.9$ and an experimental value for $x_s$
would allow us to tell whether the unitarity triangle had non-zero area by
specifying $|1 - \rho - i \eta|$\cite{DPF}. Present bounds are not yet strong
enough for this purpose. (See, e.g., Ref.~\cite{ALBs}, for which the largest
claimed lower bound is $x_s > 9$.) Assuming that $|V_{ub}/V_{cb}| > 0.06$, one
must show $0.73 < |1 - \rho - i \eta| < 1.27$. Taking the $B_s$ and $B_d$
lifetimes to be equal, and assuming $0.7 < x_d < 0.8$, this will be so if $16 <
x_s < 34$.  An ``ideal'' measurement would thus be $x_s = 25 \pm 3$.

The decay of a $\bar B_s = b \bar s$ meson via the quark subprocess $b (\bar s)
\to c \bar c s (\bar s)$ gives rise to neutral final states which turn out to
be predominantly CP-even.\cite{CPeven}  The mixing of $\bar B_s$ and $B_s$
leads to eigenstates $B_s^{\pm}$ of even and odd CP; the predominance of
CP-even final states formed of $c \bar c s \bar s$ means that the CP-even
eigenstate will have a shorter lifetime.  With $\Delta \Gamma(B_s) \equiv
\Gamma(B_s^+) - \Gamma(B_s^-)$ and $\bar \Gamma(B_s) \equiv [\Gamma(B_s^+) +
\Gamma(B_s^-)]/2$ , Bigi \ite~\cite{Blifes} estimate
\beq \label{eqn:widthdiff}
\frac{\Delta \Gamma}{\bar \Gamma(B_s)} \simeq 0.18 \frac{f_{B_s}^2}
{(200~{\rm MeV})^2}~~~,
\eeq
possibly the largest lifetime difference in hadrons containing $b$ quarks.

One could measure $\bar \Gamma(B_s)$ using semileptonic decays, while the
decays to CP eigenstates could be measured by studying the correlations between
the polarization states of the vector mesons in $B_s^{\pm} \to J/\psi K_S$.
[For a similar method applied to decays of other pseudoscalar mesons see, e.g.,
Ref.~\cite{Nelson}.]

The ratio of the mass splitting to the width difference between CP eigenstates
of strange $B$'s is predicted to be large and independent of CKM matrix
elements\cite{IsiBs,BP} (to lowest order, neglecting QCD corrections which may
be appreciable):
\beq
\frac{\Delta m}{\Delta \Gamma} \simeq - \frac{2}{3 \pi} \frac{m_t^2 h(m_t^2
/M_W^2)}{m_b^2} \left( 1 - \frac{8}{3} \frac{m_c^2}{m_b^2} \right)^{-1}
\simeq 200~!
\eeq
Here $h(x)$ decreases monotonically from 1 at $x=0$ to $1/4$ as $x \to \infty$;
it is about 0.53 for the present value of $m_t$. If the mass difference $\Delta
m / \bar \Gamma \approx 20$ turns out to be too large to measure, the width
difference $\Delta \Gamma / \bar \Gamma \approx 1/10$ may be large enough to
detect.
\bigskip

B. ~~ General time-dependent formalism for $B$ mesons
\medskip

{\it 1.  Time dependences.\cite{BCP}}
We may adapt the formalism introduced in Sec.~2 for eigenstates of the mass
matrix to describe the time evolution of states which are initially $B$ or
$\bar B$ in terms of the evolution of mass eigenstates $B_L$ and $B_H$.
Here $L$ and $H$ stand for ``light'' and ``heavy''. The corresponding
eigenvalues of the mass matrix are $\mu_{L,H} \equiv m_{L,H} - i
\Gamma_{L,H}/2$. In terms of basis states $| B \rangle $ and $| \bar B
\rangle $, we write
\beq
| B_L \rangle  = p | B \rangle  +q | \bar B \rangle~~~,~~~
| B_H \rangle  = p | B \rangle  -q | \bar B \rangle~~~,
\eeq
where we have assumed CPT invariance as in Sec.~2. It is convenient to define
$\Delta \mu \equiv \mu_H - \mu_L = \Delta m - i \Delta \Gamma /2$; $\bar \mu
\equiv (\mu_H + \mu_L )/2$; $ m \equiv (m_H + m_L )/2$; and $\Gamma \equiv
(\Gamma_H + \Gamma_L )/2$. States $B_L$ and $B_H$ evolve as $|B_{L,H} \rangle
\to | B_{L,H} \rangle  e^{-i \mu_{L,H}t}$, so that
\beq \label{eqn:tevol}
| B \rangle  \to f_+ (t) | B \rangle  + {q \over p}f_- (t) | \bar B
\rangle~~;~~~|\bar B \rangle  \to {p \over q} f_- (t) | B \rangle  + f_+ (t) |
\bar B \rangle~~~,
\eeq
where $f_{\pm}(t) \equiv (e^{-i \mu_L t} \pm e^{-i \mu_H t})/2$, or
\beq
f_+ (t) = e^{-imt} e^{- \Gamma t/2} \cos (\Delta \mu t/2)~~~;~~~
f_- (t) = e^{-imt} e^{- \Gamma t/2}i \sin  (\Delta \mu t/2)~~~.
\eeq
The time-integrated probabilities for detection of $B$ or $\bar B$, given
an initial $\bar B,$ are
\beq
\int_0^\infty dt | f_{\pm} (t) |^2 =
\frac{(\Delta m/\Gamma )^2 + \left \{
\begin{array}{c}
2 \\ 0
\end{array}
\right \}}{2 + 2 (\Delta m /\Gamma )^2 }~~~,
\eeq
so (for $|p/q| =1$, which we shall see is approximately true for
$B$'s) the ratio of ``wrong-sign'' to ``right-sign'' decays is equal to
\beq \label{eqn:rat}
\frac{\int_0^\infty |f_- (t) |^2 dt}{\int_0^\infty |f_+ (t) |^2 dt} =
\frac{(\Delta m /\Gamma )^2}{2 + (\Delta m / \Gamma )^2 }~~~.
\eeq
For $\Delta m /\Gamma \simeq 0.75$, which is approximately the case for $B^0
- \bar B^0$ mixing, this ratio is about $0.22$, while for $\Delta m /
\Gamma \simeq 20$ (a typical value expected for $B_s - \bar{B}_s$
mixing, as noted in Sec.~5 A) the ratio (\ref{eqn:rat}) is $0.995$.

{\it 2. CP-violating asymmetries for $B$'s.}
We shall be interested in comparing rates for production of a final state $| f
\rangle $ with those for the CP-conjugate final states $| \bar f \rangle \equiv
CP | f \rangle $. Denote a state which is initially $B^0$ as $B_{\rm phys}^0
(t)$, and correspondingly for $\bar{B}^0$. These states evolve in time as
described by Eq.~(\ref{eqn:tevol}). Then
$$
\langle f | B_{\rm phys}^0 (t) \rangle = f_+ (t) \langle f | B \rangle +
{q \over p} f_- (t) \langle f | \bar{B} \rangle~~~,
$$
\beq
\langle \bar{f} | \bar{B}_{\rm phys}^0 (t) \rangle =
{p \over q} f_- (t) \langle \bar{f} | B \rangle
+  f_+ (t) \langle \bar{f} | \bar{B} \rangle~~~.
\eeq
Now define
\beq
x \equiv \frac{\langle f | \bar{B} \rangle }{\langle f | B \rangle}
{}~~~;~~ \lambda_0 \equiv {q \over p } x~~~;~~~
x \equiv \frac{\langle \bar{f} | \bar{B} \rangle }{\langle \bar{f} |
\bar{B} \rangle}~~~;~~ \bar{\lambda}_0 \equiv {p \over q } x~~~.
\eeq
Then
$$
\langle f | B^0_{\rm phys} (t) \rangle =
\langle f | B \rangle  \{ f_+ (t) + \lambda_0 f_- (t) \}~~~,
$$
\beq
\langle \bar{f} | \bar{B}^0_{\rm phys} (t) \rangle =
\langle \bar{f} |\bar{B} \rangle \{f_+(t)+\bar{\lambda}_0 f_-(t)\}~~~.
\eeq

Several simplifications often are possible. If each factor in $x$ and $\bar x$
is dominated by a strong eigenchannel, final state phases cancel, and one can
use the discussion of Sec.~2 E to show that $\bar x = x^*$. For $B$ mesons,
where the mixing is dominated by the box diagrams containing internal top
quarks, one has
\beq
{p \over q} \simeq \left (
\frac{M_{12}}{M_{21}} \right )^{1/2} \simeq
\frac{V_{td}^* V_{tb}}{V_{td} V_{tb}^*}
\eeq
so that $|p/q | \simeq 1$. Combining this result with that for $x$, we find
$\bar{\lambda}_0 = \lambda_0^*$.

The time-integrated asymmetry $C_f$ for production of a final state $f$ is
defined as
\beq
C_f \equiv
\frac{\Gamma (B_{\rm phys}^0 \to f) - \Gamma (\bar{B}_{\rm phys}^0
\to \bar f )}{\Gamma (B_{\rm phys}^0 \to f) + \Gamma (\bar{B}_{\rm
phys}^0 \to \bar f )}~~~.
\eeq
With $\bar \lambda_0 = \lambda_0^*$, and neglecting $\Delta \Gamma /\Gamma$
with respect to $\Delta m / \Gamma$, we find
\beq \label{eqn:cfeq}
C_f \equiv
\frac{- 2z ~{\rm Im} ~ \lambda_0}{ 2 + z^2 (1 + | x |^2 )}~~~,
\eeq
where $z \equiv \Delta m /\Gamma$. Thus, there is an optimum value of $z$ for
observing a given time-integrated asymmetry, depending on the value of $| x |$.
For $|x| = 1$ (as happens when $f$ is a CP eigenstate, for instance), this
optimum value is $z = 1$. The case of $B^0 - \bar{B}^0$ mixing $(z \simeq
0.75$ is not far from this value.

{\it 3. Time-dependent asymmetry.}
The decays of neutral $B$ mesons are expressed as a function of proper time as
$$
d \Gamma (B_{\rm phys}^0 \to f )dt \sim | f_+ (t) + \lambda_0 f_- (t) |^2
{}~~~;
$$
\beq
d \Gamma (\bar{B}_{\rm phys}^0 \to f )dt \sim | f_+ (t) +
\bar{\lambda}_0 f_- (t) |^2~~~.
\eeq
A great simplification occurs when $f$ is a CP eigenstate: $ \bar f = \pm f$.
In this case $x^* = x^{-1}$, so $| x | = 1$, hence $ | \bar \lambda_0 | = |
\lambda_0 | = 1$ and $\bar \lambda_0 = \lambda_0^*$.
We have
$$
| f_+ + \lambda f_- |^2 =  e^{- \Gamma t} \left |
\cos \frac{\Delta m t}{2} + i \lambda_0 \sin
\frac{\Delta m t}{2} \right |^2
$$
\beq
=  e^{- \Gamma t} \left [ 1 + 2 {\rm Re} (i \lambda_0 ) \sin
\frac{\Delta m t}{2} \cos
\frac{\Delta m t}{2} \right ]
= e^{-\Gamma t} \left [ 1 - {\rm Im} \lambda_0 \sin (\Delta m t) \right ]
\eeq
or
$$
d \Gamma (B_{\rm phys}^0 (t) \to f )/dt  \sim e^{- \Gamma t}
[ 1 - {\rm Im} \lambda_0 \sin (\Delta m t ) ]
$$
\beq \label{eqn:dgdt}
d \Gamma (\bar{B}_{\rm phys}^0 (t) \to f )/dt  \sim e^{- \Gamma t}
[ 1 + {\rm Im} \lambda_0 \sin (\Delta m t ) ]~~~,
\eeq
where we have again neglected $\Delta \Gamma / \Gamma$ in comparison with
$\Delta m / \Gamma$. This step is justified for $B$'s but not for
$K$'s. In contrast to the situation for $K$'s, the final states to which both
$B$ and $\bar{B}$ can decay are only a small fraction of those to which
$B$ or $\bar B$ normally decay, and so (in accord with Eq.~(\ref{eqn:g12}),
which give rise to $\Delta \Gamma$) one should expect quite similar lifetimes
$\Gamma_H$ and $\Gamma_L$ for the two mass eigenstates.

Integration of (\ref{eqn:dgdt}) gives
\beq
C_f = \frac{-(\Delta m /\Gamma){\rm Im}\lambda_0}{1 + (\Delta m / \Gamma)^2}
\eeq
for the total asymmetry (in accord with Eq.~(\ref{eqn:cfeq}) for $|x| = 1$).
This asymmetry is suppressed both when $\Delta m / \Gamma$ is very small (e.g.,
for $D^0$) and when it is very large (e.g., as is expected for $B_s$). For
$B_s$, in order to see an asymmetry, one must not integrate with respect to
time. Many experiments are planned with detection of $B_s$ as their focus; they
will require precise vertex detection to measure mixing as a function of proper
time.

When more than one eigenchannel contributes to a decay, there can appear terms
of the form $\cos (\Delta m t)$ as well as $\sin (\Delta m t)$ in results
analogous to Eqs.~(\ref{eqn:dgdt}).\cite{PP} These complicate the analysis
somewhat, but information can be obtained from them\cite{pipi} on the relative
contributions of various channels to decays.
\bigskip

C. ~~ CP-violating $B$ meson decays
\medskip

Two main avenues for detecting CP violating in systems involving $b$ quarks
involve 1) decays to CP eigenstates,\cite{BCP} and 2) decays to CP
non-eigenstates. In both cases, partial rates for particle and antiparticle
decays are compared, but experimental aspects and interpretations differ.

{\it 1. In decays to CP eigenstates}, one compares the partial rate for a decay
of an initial $B^0$ with that for an initial $\bar B^0$.  The decays are
described by time-dependent functions whose difference when integrated over all
time is responsible for the rate asymmetry.  We have formulated this
time-dependence above, and will illustrate the calculation of the asymmetry in
some specific examples.

The interference of direct decays (such as $B^0 \to J/\psi K_S$) and those
involving mixing (such as $B^0 \to \bar B^0 \to J/\psi K_S$) gives rise to rate
asymmetries which can be easily interpreted in terms of the angles $\alpha,~
\beta,~\gamma$.  Thus, if we define
\beq
A(f) \equiv \frac{\Gamma(B_{t=0} \to f) - \Gamma(\bar B_{t=0} \to f)}
{\Gamma(B_{t=0} \to f) + \Gamma(\bar B_{t=0} \to f)}~~~,
\eeq
we have, in the limit of a single direct contribution to decay amplitudes,
\beq \label{eqn:asymms}
A(J/\psi K_S, \pi^+ \pi^-) = - \frac{x_d}{1+x_d^2} \sin(2\beta,2\alpha)~~~,
\eeq
where $x \equiv \Delta m/\Gamma$.  This limit is expected to be very good for
$J/\psi K_S$, but some correction for penguin contributions is probably needed
for $\pi^+ \pi^-$.  The value $x_d = 0.75 \pm 0.05$ is nearly optimum to
maximize the coefficient of $\sin(2\beta,2\alpha)$.

We now give some details of the calculation of the asymmetries in
Eq.~(\ref{eqn:asymms}).\cite{TASI} The decay amplitudes for $B^0 \to \jape$ are
governed by the subprocesses
\beq
\bar B^0 : ~~ a (b \to c \bar c s ) \sim V_{cb} V_{cs}^*~~~;~~~
B^0 : ~~ a (\bar b \to \bar c  c \bar s ) \sim V_{cb}^* V_{cs}~~~.
\eeq
We are interested in the parameter $ x \equiv \langle \jape K_S | \bar B^0
\rangle / \langle \jape K_S | B^0 \rangle$. We express $K_S$ in terms of $K$
and $\bar K: ~~ | K_S \rangle = p_K | K \rangle  + q_K | \bar K \rangle $, so
$\langle K_S | \bar K \rangle  = q_K ^*$; $\langle K_S | K \rangle = p_K^*$;
and
\beq
x = \left( {q \over p} \right)_K^*
{}~ \frac{\langle \jape \bar K| \bar B \rangle}{\langle \jape K | B
\rangle}~~~.
\eeq
Let us, for the moment, imagine $ (q/p)_K$ to be dominated by the charmed quark
loop. The value of $(q/p)_K^*$ is then $V_{cd}^*V_{cs} / V_{cd} V_{cs}^*$. The
ratio of matrix elements in (10.23) is $V_{cb} V_{cs}^* / V_{cb}^* V_{cs}$.
Finally, the CP-violating asymmetry depends on $\lambda_0 \equiv
x (q/p)_B$, so
\beq \label{eqn:phases}
\lambda_0 \simeq \frac{V_{cd}^* V_{cs}}{V_{cd} V_{cs}^* } ~
 \frac{V_{cb} V_{cs}^*}{V_{cb}^* V_{cs}} ~
 \frac{V_{td} V_{tb}^*}{V_{td}^* V_{tb}} ~~~,
\eeq
where we have used the dominance of the $t$ quark loop in  $ B - \bar B$ mixing
when calculating $(q/p)_B$. Cancelling some terms, we find that $\lambda_0$ is
manifestly invariant under redefinition of quark phases. A similar conclusion
would be drawn if $\epsilon_K$ were dominated by the top quark loop. In fact,
the phase of $(q/p)_K^*$ may be calculated in either manner if one neglects
$m_u$ in comparison with $m_c$ and $m_t$. In practical calculations, $(q/p)_K =
(1 - \epsilon)/(1 + \epsilon)$ is very close to $1$, so the phase of the first
term in (\ref{eqn:phases}) is negligible in comparison with that of the other
two.

The result is that $\lambda_0 \simeq V_{td} / V_{td}^* = e^{-2 i \beta}$, where
$\beta$ is the angle of the unitarity triangle (Fig.~2) opposite the side
corresponding to $V_{ub}^*$. The CP-violating asymmetry, whether time-dependent
or time-integrated, measures ${\rm Im} ~ \lambda_0 =- \sin (2 \beta)$, i.e.,
the ``best-known'' angle of the triangle.

Let us assume that the decay $\bar B^0 ( \equiv b \bar d) \to \pipe$ is
dominated by a process like that shown in Fig.~6, where the current creates a
$\pi^-$, and the $\pi^+$ is composed of the $u$ quark in the subprocess $b \to
u d \bar u$ and the $\bar d $ quark from the initial $\bar B^0$. We shall
neglect the effects of penguin graphs, though they could be
significant.\cite{PP} (Methods have been proposed for detecting such
contributions, based on searches for a cos $(\Delta m t)$ term characteristic
of the presence of more than one eigenchannel contributing to a decay to a CP
eigenstate like $\pipe$.)

Proceeding as for the $\jape K_S$ case, we compare $A(\bar B^0 \to
\pipe) \sim V_{ub} V_{ud}^*$ with $A(B^0 \to \pipe) \sim V_{ub}^*
V_{ud}$, so $x = V_{ub} V_{ud}^* / V_{ub}^* V_{ud}$ and
\beq
\lambda_0 =  {q \over p} x =
\frac{V_{td} V_{tb}^* V_{ub} V_{ud}^*}{V_{td}^* V_{tb} V_{ub}^* V_{ud}}
\simeq  \frac{V_{ub} V_{td}}{V_{ub}^* V_{td}^*} =
e^{-2i (\gamma + \beta)} = e^{-2i (\pi - \alpha)} = e^{2 i \alpha} ~~~.
\eeq
Then ${\rm Im} \lambda_0 = \sin (2 \alpha)$, which can have either sign,
depending on the shape of the unitarity triangle.

For the range of parameters noted in Fig.~5, we expect $-0.4 \le A(J/\psi K_S)
\le -0.1$, i.e., an asymmetry of a definite sign, and $-0.48 \le A(\pi^+ \pi^-)
\le 0.3$, i.e., nearly any asymmetry within the possible limits imposed by the
factor $x_d/(1+x_d^2)$.  Thus, the $J/\psi K_S$ asymmetry is likely to provide
a consistency check, while the $\pi^+ \pi^-$ asymmetry should be more useful in
specifying the parameters $\rho,\eta$ (unless it lies outside the expected
limits).

In order to employ this method it is necessary to know the flavor of the
produced $B$ meson.  We shall remark presently on possible ``tagging''
methods.

{\it 2.  Decays to CP non-eigenstates} can exhibit rate asymmetries only if
there are two different weak decay amplitudes {\it and} two different strong
phase shifts associated with them.  Comparing
\beq
\Gamma \equiv \Gamma(B^0 \to f)
= |a_1 e^{i(\phi_1 + \delta_1)} +  a_2 e^{i(\phi_2 + \delta_2)}|^2
\eeq
with
\beq
\bar \Gamma \equiv \Gamma(\bar B^0 \to \bar f)
= |a_1 e^{i(-\phi_1 + \delta_1)} +  a_2 e^{i(-\phi_2 + \delta_2)}|^2
\eeq
we notice that only the weak phases $\phi_i$ change sign under CP inversion,
not the strong phases $\delta_i$.  Defining $\phi \equiv \phi_1 - \phi_2$,
$\delta \equiv \delta_1 - \delta_2$, we find
\beq
\Gamma - \bar \Gamma \propto 2 a_1 a_2 \sin \phi \sin \delta~~~,
\eeq
so both $\phi$ and $\delta$ must be not equal to 0 or $\pi$ in order to see a
rate difference.

A pair of channels in which both weak and strong phase shift differences
might be nonzero, for instance, could be the $I = 1/2$ and $I = 3/2$ amplitudes
for $B \to \pi K$. Here one is able to compare decays of charged $B$ mesons
with those of their antiparticles, so the identification of the flavor of the
decaying meson does not pose a problem.  On the other hand, $\delta$ is
generally expected to be small and quite uncertain for the energies
characteristic of $B$ decays.  We shall outline below some recent progress in
using decays of charged $B$ mesons to provide information on CKM phases {\em
without} necessarily having to observe a CP-violating decay rate asymmetry.
\bigskip

D. ~~ $B$ Flavor tagging in $C = \pm 1~B \bar B$ states
\medskip

In the decays of neutral $B$ mesons to CP eigenstates, it is necessary to know
the flavor of the meson at time of production.  A conventional means for
``tagging'' the flavor of a $B$ is to identify the flavor of the meson produced
in association with it.  At a hadron collider or in high energy $\eep$
collisions as at LEP, this method suffers only from the possible dilution of
the ``tagging'' signal by $B^0 - \bar B^0$ mixing, and from the difficulty of
finding the ``tagging'' hadron.

In $\eep$ annihilations, one frequently can produce a $B \bar B$ state with a
definite eigenvalue of the charge-conjugation operator $C$. For example, at the
$\Upsilon (4 S),~C(B \bar{B}) = - 1$, since the pair is produced from a virtual
photon, while the production of a $B^* \bar{B}$ pair by a virtual photon
followed by $B^* \to \gamma + B$ leads to a $ B \bar B$ pair with $C = + 1$.

We present an abbreviated discussion, of which a more complete version can be
found elsewhere.\cite{BKUS} Consider the process in which an $\eep$ collision
produces a $B^0 \bar{B}^0$ pair, one of which decays to $\jape K_S$ and the
other of which decays semileptonically. We can tell what the flavor is of the
meson decaying semileptonically, since $\bar b$ (in a $B^0$) gives $\bar c$ (in
$D^{*-} ) \ell^+ \bar \nu_\ell$, while $b$ (in a $\bar{B}^0$) gives $c$ (in
$D^{*+}) \ell^- \nu_\ell$. However, because of the possibility of $B^0
\leftrightarrow \bar{B}^0$ mixing, the flavor of the meson decaying to $\jape ~
K_S$ is not uniquely labelled.

We can write the time-dependent partial decay rates as
$$
d^2 \Gamma [ \jape K_S (t) ~D^{*+} \ell^- \bar \nu_\ell (\bar t ) ]_{C =
\mp 1} / dt d \bar t
\sim e^{-\Gamma ( t + \bar t )} [ 1 \mp \sin \Delta m (t
\mp \bar t ) {\rm Im} \lambda ] ~~~;
$$
\beq
d^2 \Gamma [ \jape K_S (t) ~D^{*-} \ell^+  \nu_\ell (\bar t ) ]_{C =
\mp 1} / dt d \bar t
\sim e^{-\Gamma ( t + \bar t )} [ 1 \pm \sin \Delta m (t
\mp \bar t ) {\rm Im} \lambda ] ~~~.
\eeq
The term contributing to the time-dependent asymmetry is even or odd in $ t -
\bar t$ depending on whether $C ( B \bar{B})$ is even or odd. Thus, for a $C =
-$ eigenstate [such as $\Upsilon (4S)$], the time-integrated asymmetry
vanishes. This curtails the utility of the $\Upsilon (4S)$ in producing $ B
\bar{B}$ pairs for the study of CP-violating asymmetries, unless time-dependent
effects can be studied. If we can distinguish which of the times $t$ and $\bar
t$ is earlier, we can study an asymmetry proportional to
\beq
\Gamma^2 \int_0^\infty dt d \bar t e^{-\Gamma ( t + \bar t )} \sin \Delta m
( | t - \bar t | ) = \frac{z}{1 + z^2}
\simeq 0.48 ~~ {\rm for} ~~ z \equiv \Delta m / \Gamma \simeq 0.75~~.
\eeq
For a $C = +$ eigenstate [as might be reachable just above the $\Upsilon (4S)$
via $\eep \to B \bar{B}^* \to B \bar{B} \gamma ]$, the
time-integrated asymmetry is proportional to
\beq
\Gamma^2 \int_0^\infty dt d \bar t e^{-\Gamma ( t + \bar t )} \sin \Delta m
( | t + \bar t | ) = \frac{2z}{(1 + z^2)^2}
\simeq 0.61 ~~ {\rm for} ~~ z \equiv \Delta m / \Gamma \simeq 0.75~~.
\eeq

The need to distinguish which of the times $t$ and $\bar t$ is earlier has
been addressed by the construction of asymmetric ``$B$ factories'' at
SLAC (Stanford) and KEK (Tsukuba), in which two beams of unequal energies
collide, leading to a Lorentz boost of the center of mass.  It may also
be possible to employ ingenious schemes\cite{KB} for enhancing vertex
resolution in symmetric machines such as CESR (Cornell).
\bigskip

E. ~~ Flavor tagging by means of correlations with ``nearby'' hadrons
\medskip

In this section I would like to discuss recent progress in tagging neutral
$B$ mesons by means of the hadrons produced nearby in phase space.  This
method, also proposed\cite{AB} for tagging strange $B$'s via
associated kaons, has been the subject of recent papers devoted
to correlations of nonstrange $B$'s with charged pions.\cite{Correls}

{\it 1.  Fragmentation vs. resonances.}
The existence of correlations between $B$ mesons and pions can be visualized
either in terms of a fragmentation picture or in terms of explicit resonances.

In a fragmentation picture, if a $b$ quark picks up a $\bar d$ quark from the
vacuum to become a $\bar B^0$ meson, and a charged pion containing the
corresponding $d$ quark is generated, that pion will be a $\pi^-$.  It is
likely to lie ``near'' the $\bar B^0$ in phase space, in the sense that its
transverse momentum with respect to the $B$ is low, its rapidity is correlated
with that of the $B$, or the effective mass of the $\pi B$ system is low.
Similarly, if a $\bar b$ quark picks up a $d$ quark to become a $B^0$, the
charged pion containing the corresponding $\bar d$ will be a $\pi^+$.

If a $b$ quark picks up a strange antiquark to become a $\bar B_s$, the next
hadron down the fragmentation chain will contain an $s$ quark.  If this hadron
is a charged kaon, it will be a $K^-$.\cite{AB}  A $B_s = \bar b s$ will
correspondingly be associated with a $K^+$.  By similar arguments\cite{HLB} a
$\bar B^0 = b \bar d$ can be associated with a proton $p = uud$, while a $B^0 =
\bar b d$ can be associated with an antiproton $\bar p = \bar u \bar u \bar d$.

The signs of the pions in the above correlations are those which would have
resulted from the decays $B^{**+} \to B^{(*)0} \pi^+$ or $B^{**-} \to \bar
B^{(*)0} \pi^-$. We utilize the double-asterisk superscript to distinguish
$B^{**}$'s from the hyperfine partners of the $B$'s, the $B^*$'s, which are
only about 46 MeV heavier than the $B$'s and cannot decay to them via pions.

The importance of explicit {\em narrow} $B^{**}$ resonances is that they
permit reduction of combinatorial backgrounds.  Thus, we turn to what is
expected (and, more recently, observed) about such resonances.

{\it 2. Spectroscopic predictions.}
We shall briefly recapitulate material which has been presented in more detail
elsewhere.\cite{JRCharm,Fest} In a hadron containing a single heavy quark, that
quark ($Q = c$ or $b$) plays the role of an atomic nucleus, with the light
degrees of freedom (quarks, antiquarks, gluons) analogous to the electron
cloud.  The properties of hadrons containing $b$ quarks then can calculated
from the corresponding properties of charmed particles by taking account
\cite{HQS} of a few simple ``isotope effects.''  If $q$ denotes a light
antiquark, the mass of a $Q \bar q$ meson can be expressed as
\beq
M(Q \bar q) = m_Q + {\rm const.}[n,L] + \frac{\langle p^2 \rangle}{2 m_Q} +
a \frac{\langle {\bf \sigma_q \cdot \sigma_Q} \rangle}{m_q m_Q} + {\cal O}
(m_Q^{-2})~~~.
\eeq
Here the constant depends only on the radial and orbital quantum numbers $n$
and $L$.  The $\langle p^2 \rangle /2m_Q$ term expresses the dependence of
the heavy quark's kinetic energy on $m_Q$, while the last term is a hyperfine
interaction.  The expectation value of $\langle {\bf \sigma_q \cdot \sigma_Q}
\rangle$ is $(+1,~-3)$ for $J^P = (1^-,~0^-)$ mesons. If we define
$\bar{M} \equiv [3 M(1^-) + M(0^-)]/4$, we find
\beq
m_b - m_c + \frac{\langle p^2 \rangle}{2 m_b} - \frac{\langle p^2 \rangle}
{2 m_c} = \bar{M}(B \bar q) - \bar{M}(c \bar q) \simeq 3.34~{\rm
GeV}~~.
\eeq
so $m_b - m_c > 3.34$ GeV, since $\langle p^2 \rangle > 0$.  Details of
interest include (1) the effects of replacing nonstrange quarks with strange
ones, (2) the energies associated with orbital excitations, (3) the size of the
$\langle p^2 \rangle$ term, and (4) the magnitude of hyperfine effects.  In all
cases there exist ways of using information about charmed hadrons to predict
the properties of the corresponding $B$ hadrons.  A recent comparison of the
charmed and beauty spectra may be found in Ref.~\cite{Fest}.  For S-wave states
the predictions of the heavy-quark symmetry approach work rather well.

The $B^* - B$ hyperfine splitting scales as the inverse of the heavy-quark
mass: $B^* - B = (m_c/m_b)(D^* - D)$.  Consequently, while $D^{*+} \to D^0
\pi^+$ and $D^{*+} \to D^+ \pi^0$ are both allowed, leading to a useful
method\cite{SN} for identifying charmed mesons via the soft pions often
accompanying them, the only allowed decay of a $B^*$ is to $B \gamma$.  No soft
pions are expected to accompany $B$ mesons.  One must look to the next-higher
set of levels, the $B^{**}$ resonances consisting of a $\bar b$ quark and a
light quark in a P-wave, or the fragmentation process mentioned above, for the
source of pions correlated with the flavor of $B$ mesons.

One can use heavy-quark symmetry or explicit quark models to extrapolate from
the properties of known $D^{**}$ resonances to those of $B^{**}$ states. Two
classes of such resonances are expected,\cite{DGG} depending on whether the
total angular momentum $j = s_q + L$ of the light quark system is 1/2 or 3/2.
Here $s_q$ is the light quark's spin and $L=1$ is its orbital angular momentum
with respect to the heavy antiquark. The light quark's $j = 1/2,~3/2$ can
couple with the heavy antiquark's spin $S_{\bar Q} = 1/2$ to form states with
total angular momentum and parity $J^P_j = 0^+_{1/2},~1^+_{1/2},~1^+_{3/2},
{}~2^+_{3/2}$.

The $0^+_{1/2}$ and $1^+_{1/2}$ states are expected to decay to a ground-state
heavy meson with $J^P = 0^-$ or $1^-$ and a pion via an S-wave, and hence to be
quite broad. No evidence for these states exists in the $\bar c q$ or the $\bar
b q$ system.  By contrast, the $1^+_{3/2}$ and $2^+_{3/2}$ states are expected
to decay to a ground-state heavy meson and a pion mainly via a D-wave, and
hence to be narrow.  Candidates for all the nonstrange and strange $D^{**}$
states of this variety have been identified.\cite{Dstars} The known nonstrange
$D^{**}$ resonances (identified in both charged and neutral states) are a $2^+$
state around 2.46 GeV/$c^2$, decaying to $D \pi$ and $D^* \pi$, and a $1^+$
state around 2.42 GeV/$c^2$, decaying to $D^* \pi$. In addition, strange
$D_s^{**+}$ resonances have been seen, at 2.535 GeV/$c^2$ (a candidate for
$1^+_{3/2}$) and 2.573 GeV/$c^2$ (a candidate for $2^+_{3/2}$). Thus, adding a
strange quark adds about 0.1 GeV/$c^2$ to the mass.

Once the masses of $D^{**}$ resonances are known, one can estimate those of the
corresponding $B^{**}$ states by adding about 3.32 GeV (the quark mass
difference minus a small binding correction).  The results of this exercise are
shown in Table 4.  The reader should consult Ref.~\cite{EHQ} for more detailed
predictions based on potential models and for relations between decay widths of
$D^{**}$ states and those of the $B^{**}$'s.  Thus, the study of excited
charmed states can play a crucial role in determining the feasibility of
methods for identifying the flavor of neutral $B$ mesons.

% This is Table 4
\begin{table}
\caption{P-wave resonances of a heavy antiquark and a light quark $q=u,~d$.
In final states $P,~V$ denote a heavy $0^-,~1^-$ meson.  For strange states,
add about 0.1 GeV$/c^2$ to the masses.}
\begin{center}
\renewcommand{\arraystretch}{1.3}
\begin{tabular}{c c c c} \hline
$J^P_j$     & $ \bar c q$ mass & $\bar b q$ Mass  &  Allowed final \\
            & (GeV/$c^2$)      & (GeV/$c^2$)      &     state(s)   \\ \hline
$2^+_{3/2}$ &    $2.46^{a)}$   & $\sim 5.77^{b)}$ &  $P \pi, V \pi$ \\
$1^+_{3/2}$ &    $2.42^{a)}$   & $\sim 5.77^{b)}$ &     $V \pi$     \\
$1^+_{1/2}$ &   $<2.42^{c)}$   &  $< 5.77^{b)}$   &     $V \pi$     \\
$0^+_{1/2}$ &   $<2.42^{c)}$   &  $< 5.77^{b)}$   &     $P \pi$     \\ \hline
\end{tabular}
\end{center}
\leftline{$^{a)}$ Observed value.\protect\cite{Dstars}.}
\leftline{$^{b)}$ Predicted by extrapolation from corresponding $D^{**}$ using
heavy-quark symmetry.}
\leftline{$^{c)}$ Predicted in most quark models.}
\end{table}

{\it 3.  Spectroscopic observations}

The OPAL,\cite{OPB} DELPHI,\cite{DELB} and ALEPH\cite{ALB} Collaborations
have now observed $B \pi$ correlations which can be interpreted in terms
of the predicted $J^P_j = 1^+_{3/2},~2^+_{3/2}$ states.  The OPAL
data are shown in Fig.~8, while the DELPHI data are shown in Fig.~9.  (The
ALEPH data were presented since the Summer School.)  Note that OPAL also sees a
$B K$ correlation.

% This is Figure 8
\begin{figure}
% \vspace{3.0in}
\centerline{\epsfysize = 3.0 in \epsffile {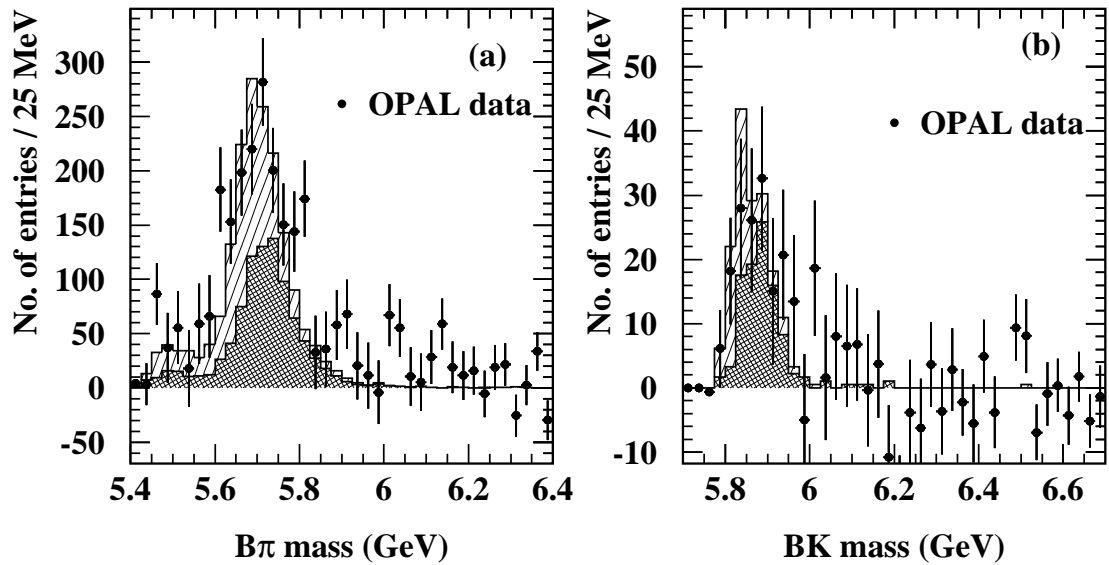}}
\caption{Evidence for B$^{**}$ resonances obtained by the OPAL Collaboration
\protect\cite{OPB}. (a) $B \pi$ background-subtracted mass distribution; (b)
$B K$ background-subtracted mass distribution.  The solid and cross-hatched
histograms correspond to the contributions of $2^+$ and $1^+$ resonances in
two-resonance fits based on the mass splittings and branching ratios
predicted in Ref.~\protect\cite{EHQ}.}
\end{figure}

% This is Figure 9
\begin{figure}
% \vspace{6.5in}
\centerline{\epsfysize = 6.5 in \epsffile {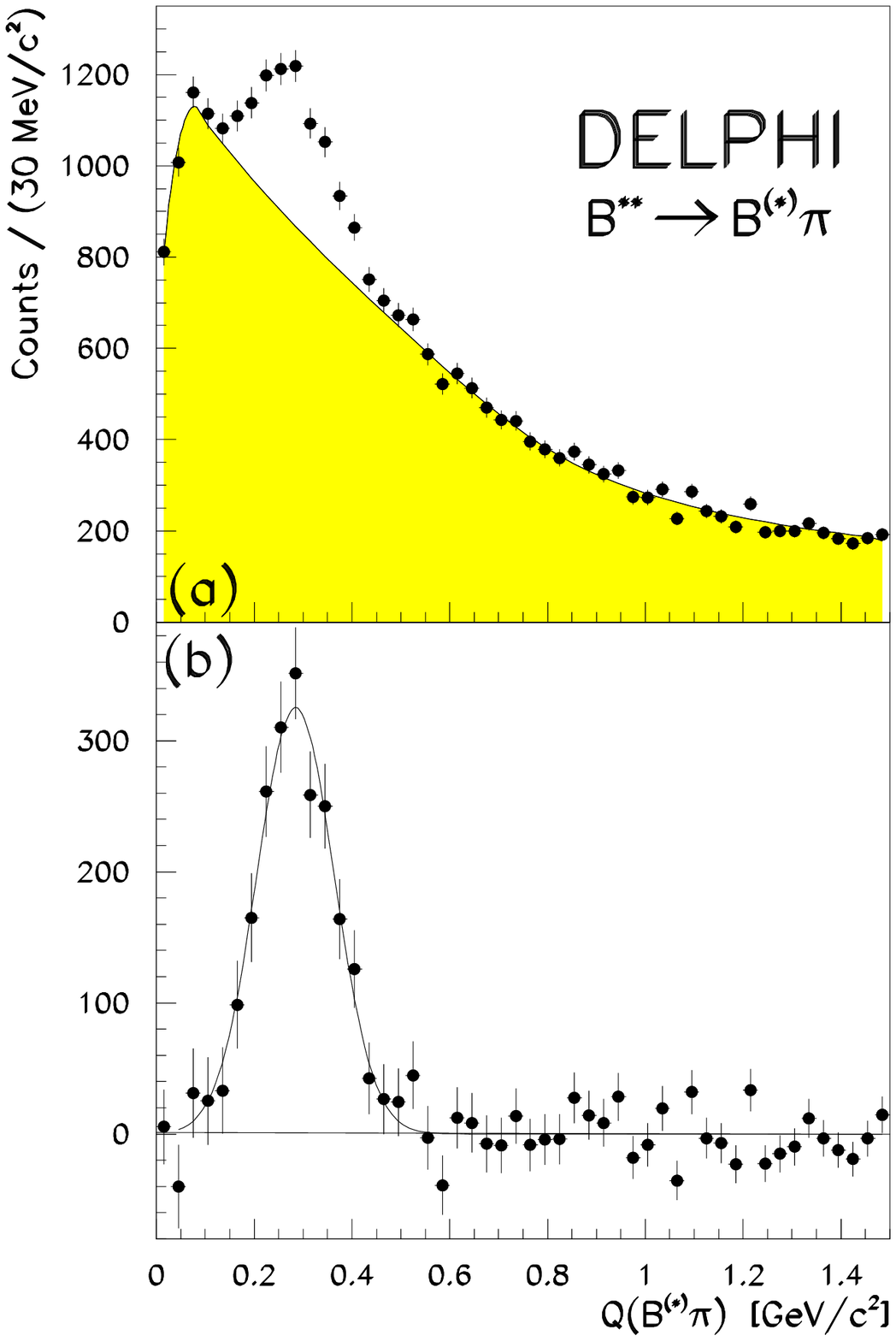}}
\caption{Evidence for B$^{**}$ resonances obtained by the DELPHI Collaboration
\protect\cite{DELB}.  (a) Unsubtracted distribution in $Q(B^{(*)} \pi) \equiv
M(B \pi) - M_B - m_\pi$.  (b) Background-subtracted distribution.}
\end{figure}

In all experiments one is able to measure only the effective mass of a $B \pi$
system.  If a $B^{**}$ decays to $B^* \pi$, the photon in the $B^* \to \gamma
B$ decay is lost, leading to an underestimate of the $B^{**}$ mass by about 46
MeV/$c^2$ but negligible energy smearing.\cite{Correls}  Thus, the
contributions to the $B \pi$ mass distribution of $1^+$ and $2^+$ resonances
with spacing $\delta \equiv M(2^+) - M(1^+)$ can appear as three peaks, one at
$M(2^+)$ due to $2^+ \to B \pi$, one at $M(2^+) - 46$ MeV/$c^2$ due to $2^+ \to
B^* \pi$, and one at $M(2^+) - \delta - 46$ MeV/$c^2$ due to $1^+ \to B^* \pi$.

The OPAL Collaboration fits their $B \pi$ mass distribution either with one
peak with $M = 5681 \pm 11$ MeV/$c^2$ and width $\Gamma = 116 \pm 24$ MeV, or
two resonances, a $1^+$ candidiate at 5725 MeV/$c^2$ with width $\Gamma = 20$
MeV and a $2^+$ candidiate at 5737 MeV/$c^2$ with width $\Gamma = 25$ MeV.  The
widths, mass splittings, and branching ratios to $B \pi$ and $B^* \pi$ in this
last fit are taken from Ref.~\cite{EHQ}, and only the overall mass and
production cross sections are left as free parameters.  The OPAL $B K$ mass
distribution is fit either with a single resonance at $M = 5853 \pm 15$
MeV/$c^2$ with width $\Gamma = 47 \pm 22$ MeV, or two narrow resonances, a
$1^+$ candidate at 5874 MeV/$c^2$ and a $2^+$ candidate at 5886 MeV/$c^2$.

The fitted masses of the nonstrange and strange resonances are respectively
30 MeV/$c^2$ lower and 40 MeV/$c^2$ higher than the predictions of
Ref.~\cite{EHQ}. The difference could well be due to additional contributions
to the nonstrange channel from the lower-lying $0^+_{1/2}$ and $1^+_{1/2}$
states or from nonresonant fragmentation. The corresponding strange states
might lie below $BK$ threshold.

The OPAL results imply that $(18 \pm 4)\%$ of the observed $B^+$ mesons are
accompanied by a ``tagging $\pi^-$'' arising from $B^{**0}$ decay.  By isospin
reflection, one should then expect $(18 \pm 4)\%$ of $B^0$ to be accompanied by
a ``tagging $\pi^+$'' arising from $B^{**+}$ decay.  This is good news for the
possibility of ``same-side tagging'' of neutral $B$ mesons.  [Another $(2.6 \pm
0.8)\%$ of the observed $B^+$ mesons are accompanied by a ``tagging $K^-$.''
The isospin-reflected kaon is neutral, and unsuitable for tagging.]

The DELPHI data can be fit with a single peak having a mass of $M = 5732
\pm 5 \pm 20$ MeV/$c^2$ and width $\Gamma = 145 \pm 28$ MeV.  The number of
$B^{**}_{u,d}$ per $b$ jet is quoted as $0.27 \pm 0.02 \pm 0.06$. The ALEPH
peak occurs at $M(B \pi) - M(B) = 424 \pm 4 \pm 10$ MeV/$c^2$, with Gaussian
width $\sigma = 53 \pm 3 \pm 9$ MeV/$c^2$.  The ALEPH $B^{**}$ signal is
characterized by a production rate
\beq
\frac{B(Z \to b \to B^{**}_{u,d})}{B(Z \to b \to B_{u,d})} = (27.9 \pm
1.6 \pm 5.9 \pm 3.8)\%~~~,
\eeq
where the first error is statistical, the second is systematic, and the
third is associated with uncertainty in ascribing the peak to the contribution
of various resonances.  Multiplying the DELPHI and ALEPH results by the isospin
factor of 2/3 to compare with the OPAL result, we find complete agreement among
the three.

{\it 4. Isospin and Correlations}

In principle it should be possible to calibrate the correlations between
charged pions and neutral $B$'s by comparing them with the isospin-reflected
correlations betwen charged pions and {\it charged} $B$'s (whose flavor may be
easier to identify).  Thus, the enhancement of the non-exotic $\pi^+ B^0$
channel with respect to the exotic $\pi^- B^0$ channel should be the same as
that of the non-exotic $\pi^- B^+$ channel with respect to the exotic $\pi^+
B^+$ channel.  What can spoil this relation? I. Dunietz and I\cite{Ispin}
have explored several instances in which care is warranted in making this
comparison.  Some of the differences could be real, but there are many
sources of potential instrumental error against which one has concrete
remedies.

{\it a) Interaction with the producing system}
can lead to final states which need not be invariant under isospin
reflection.  For example, although a pair of gluons would produce a $b \bar b$
pair with isospin $I = 0$, the fragmentation process could involve picking up
quarks from the producing system (e.g., proton or antiproton fragments) in a
manner not invariant with respect to $u \leftrightarrow d$ substitution.
Similarly, the production of a $B$ meson through diffractive dissociation of
a proton (which has more valence $u$ quarks than $d$ quarks) need not be
invariant under isospin reflection.

{\it b) Misidentification of associated charged kaons as pions}
can lead one to overestimate the charged $B$ -- charged pion correlations. One
expects $B^+ K^-$ correlations, as seen by OPAL, but not $B^0 K^+$
correlations. As mentioned, the isospin reflection of a charged kaon is
neutral, and would not contribute to a correlation between charged particles
and neutral $B$'s.

{\it c) Pions in the decay of the associated $B$} will not be produced
in an isospin-reflection-symmetric manner. One must be careful not to
confuse them with primary pions.

{\it d) Different time-dependent selection criteria} for charged and
neutral $B$'s can lead one to mis-estimate the mixing of neutral $B$'s with
their antiparticles.  (I thank P. Derwent for pointing this out.)  It is
possible to make an unfortunate cut on $B^0$ lifetime which enhances the
mixing considerably with respect to the value obtained by integrating over
all times.

{\it e) Overestimates of particle identification efficiencies} can lead
to confusion in identification of the flavor of a neutral $B$ through the
decay $B^0 \to J/\psi K^{*0} \to J/\psi K^+ \pi^-$.  It is possible,
especially for $K^+$ and $\pi^-$ with equal laboratory momenta, to confuse
them with $\pi^+$ and $K^-$, while still keeping them in a $K^*$ peak.

The CDF Collaboration at Fermilab has been studying charged pion -- $B$
correlations ever since the reports of Ref.~\cite{Correls} appeared, but
no public announcement of these results has yet appeared.  Such correlations
certainly should be present in hadron collider data.
\bigskip

F. ~~ $B$ decays to pair of light mesons
\medskip

In this subsection we turn to decays of $B$ mesons to CP non-eigenstates. We
have suggested\cite{BPP} that relations between decays of charged $B$'s to
pairs of light pseudoscalar mesons based on flavor SU(3)\cite{OldSU} could
provide information on weak phases by means of {\em rate measurements alone}.
The latest chapter in this story has been written since the Summer School.

{\it 1.  $\pi \pi$ and $\pi K$ final states.}
Two years ago the CLEO Collaboration\cite{Battle} presented evidence
for a combination of $B^0 \to K^+ \pi^-$ and $\pi^+ \pi^-$ decays, generically
known as $B^0 \to h^+ \pi^-$. On the basis of 2.4 fb$^{-1}$ of data, the most
recent result\cite{Wurt} is $B(B^0 \to h^+ \pi^-) = (1.81^{~+0.6~+0.2}
_{~-0.5~-0.3} \pm 0.2) \times 10^{-5}$.  Although one still cannot conclude
that either decay mode is nonzero at the $3 \sigma$ level, the most likely
solution is roughly equal branching ratios (i.e., about $10^{-5}$) for each
mode.

Other results\cite{CLEOGlas} of the CLEO Collaboration on related modes
include the upper bounds
$B(B^0 \to \pi^0 \pi^0) < 1.0 \times 10^{-5}$,
$B(B^+ \to \pi^+ \pi^0) < 2.3 \times 10^{-5}$, and
$B(B^+ \to K^+ \pi^0) < 3.2 \times 10^{-5}$.
Interesting levels for the last two modes\cite{BPP} are probably around
$(1/2) \times 10^{-5}$, and probably $10^{-6}$ or less for $\pi^0 \pi^0$.
With good particle identification and a factor of several times more data,
it appears that CLEO will be able to make a systematic study of decay modes
with two light pseudoscalars.  What can it teach us?

{\it 2. SU(3) relations and the phase $\gamma$.}
We mentioned earlier that rate asymmetries in the decays $B^0 \to \pi^+ \pi^-$
and $\bar B^0 \to \pi^+ \pi^-$ could provide information on the weak angle
$\alpha$, as long as a single quark subprocess dominated the decay.  Additional
contributions from penguin diagrams\cite{PP} can be taken into account by
means of an isospin triangle construction\cite{pipi} involving the relation
$A(B^0 \to \pi^+ \pi^-) = \sqrt{2} A(B^0 \to \pi^0 \pi^0) + \sqrt{2} A(B^+
\to \pi^+ \pi^0)$, and the corresponding relation for the charge-conjugate
processes.  Here we define amplitudes such that a partial width is always
proportional to the square of an amplitude with the same coefficient.

An amplitude quadrangle applies to the decays $B \to \pi K$:  $A(B^+
\to \pi^+ K^0) + \sqrt{2} A(B^+ \to \pi^0 K^+) = A(B^0 \to \pi^- K^+) +
\sqrt{2} A(B^0 \to \pi^0 K^0)$.  When combined with the corresponding relation
for $\bar B^0$ and $B^-$ decays, and used in conjunction with the
time-dependence of the decays $(B^0~{\rm or}~\bar B^0) \to \pi^0 K_S$, these
quadrangles are useful in extracting the weak phase $\alpha$.\cite{pik}

In examining SU(3) relations among $B \to PP$ amplitudes, where $P$ is a
pseudoscalar meson, we found that one of the diagonals of the amplitude
quadrangle for $B \to \pi K$ (corresponding to an amplitude with isospin $I =
3/2$) could be related to the purely $I = 2$ amplitude for $B^+ \to \pi^+
\pi^0$.  We obtained the relation
\beq  \label{eqn:tri}
A(B^+ \to \pi^+ K^0) + \sqrt{2} A(B^+ \to \pi^0 K^+) = \tilde r_u \sqrt{2}
A(B^+ \to \pi^+ \pi^0)~~~,
\eeq
where $\tilde r_u \equiv (f_K/f_\pi) | V_{us}/V_{ud}|$.  The $B^+ \to \pi^+
K^0$ amplitude is expected to be dominated by a (gluonic) penguin contribution,
involving mainly the combination of CKM elements $V_{tb}^* V_{ts}$, whose
electroweak phase is $\pi$.  The electroweak phase of the $B^+ \to \pi^+ \pi^0$
amplitude is just $\gamma = {\rm Arg}(V_{ub}^*)$.  Thus, in the absence of
strong-interaction phase shift differences, the shape of the amplitude triangle
would give $\gamma$.  One could account for strong-interaction phases by
comparing the amplitude triangle for $B^+$ decays with that for $B^-$ decays
\cite{BPP}.  This construction is illustrated in Fig.~10.  The
angle between the sides corresponding to $B^+ \to \pi^+ \pi^0$ and $B^- \to
\pi^- \pi^0$ is $2 \gamma$. The most likely of two possible orientations for
the triangles is shown; the other orientation corresponds to reflection of
the $B^-$ triangle about the common base.

% This is Figure 10
\begin{figure}
% \vspace{2.35in}
\centerline{\epsfysize = 2.35 in \epsffile {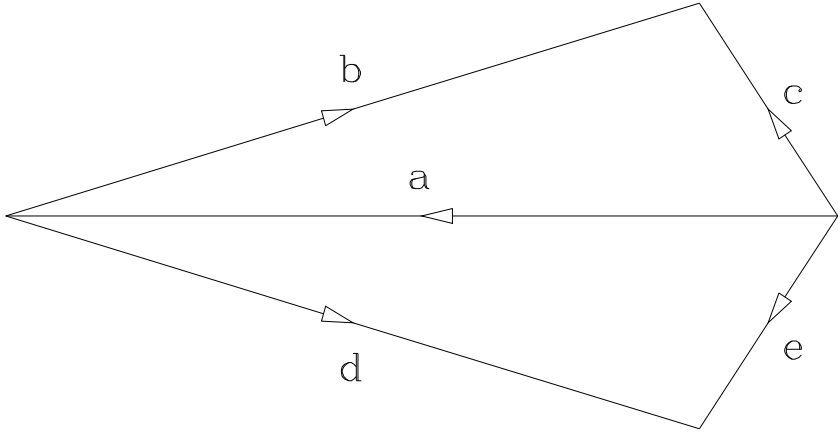}}
\caption{Amplitude triangles illustrating the extraction of the weak
phase $\gamma$ from charged $B$ decays to $\pi \pi$ and $\pi K$.
(a) $A(B^+ \to \pi^+ K^0) = A(B^- \to \pi^- \bar K^0)$;
(b) $\protect \sqrt{2} A(B^+ \to \pi^0 K^+)$;
(c) $\protect \sqrt{2} \tilde{r}_u A(B^+ \to \pi^+ \pi^0)$;
(d) $\protect \sqrt{2} A(B^- \to \pi^0 K^-)$;
(e) $\protect \sqrt{2} \tilde{r}_u A(B^- \to \pi^- \pi^0)$.
The exterior angle between sides c and e is $2 \gamma$.}
\end{figure}

The amplitude $A(\pi^0 K^+)$ which is the sum of ``tree'' and penguin graph
contributions can be expressed\cite{LWC} as
\beq
\sqrt{2}A(\pi^0 K^+) = A e^{i \gamma} e^{i \delta_3} + B e^{i \delta_P}
= e^{i \delta_P} (A e^{i \gamma} e^{i \delta} + B)~~,
\eeq
where $\delta_3$ is the strong $I = 3/2$ phase, $\delta_P$ is the strong
penguin phase, and $\delta \equiv \delta_3 - \delta_P$.  In the corresponding
expression for $A(\pi^0 K^-)$, only the sign of $\gamma$ is changed.  The
$\pi^0 K^{\pm}$ decay rates are then
\beq
\Gamma(\pi^0 K^{\pm}) = (1/2)[A^2 + B^2 + 2 A B \cos(\delta \pm \gamma)]~~~.
\eeq
The sum and difference of these rates are
\beq
S \equiv \Gamma(\pi^0 K^+) + \Gamma(\pi^0 K^-) = A^2 + B^2 + 2AB \cos \delta
\cos \gamma~~~,
\eeq
\beq
D \equiv \Gamma(\pi^0 K^+) - \Gamma(\pi^0 K^-) = 2 A B \sin \delta \sin
\gamma~~~.
\eeq
The CP-violating rate difference $D$ is probably very small as a result of the
likely smallness of the phase difference $\delta$.

For $|A/B| = 1/3,~\delta = 0$, and $\cos \gamma \simeq 0$ (realistic values),
and expressing $\cos \gamma = (S - A^2 - B^2)/2AB$, we found that in order to
measure $\gamma$ to $10^{\circ}$ one needs a sample consisting of about 100
events in the channel $\pi^0 K^{\pm}$ corresponding to $S$.\cite{PASCOS}

{\it 3. Electroweak penguins.}
The analyses of Ref.~\cite{BPP} assumed that the only penguin contributions to
$B$ decays were gluonic in nature.  Consequently, one could treat the
flavor-dependence in terms of an effective $\bar b \to \bar d$ or $\bar b \to
\bar s$ transition since the gluon couples to light quarks in a
flavor-symmetric manner.  Thus, the $I = 3/2$ amplitude in $B \to \pi K$
(the diagonal of the $\pi K$ amplitude quadrangle mentioned above) was due
entirely to the Cabibbo-suppressed ``tree-diagram'' process $\bar b \to \bar u
u \bar s$, whose weak phase was well-specified.

It was pointed out\cite{RF,DH} that in certain penguin-dominated $B$ decays
such as $B \to \pi K^*~{\rm and}~\pi K$, electroweak penguin amplitudes were
large enough to compete favorably with the tree amplitude in the $I = 3/2$
channel.  In contrast to gluonic penguins, the virtual photon or $Z$ emitted in
an electroweak penguin diagram does not couple to light quarks in a
flavor-symmetric manner, and possesses an $I = 1$ component. Specifically, if
one decomposes amplitudes into isospin channels,
$$
A(B^+ \to \pi^+ K^0) = (1/3)^{1/2}A_{3/2} - (2/3)^{1/2}A_{1/2}~~,
$$
\beq
A(B^+ \to \pi^0 K^+) = (2/3)^{1/2}A_{3/2} + (1/3)^{1/2}A_{1/2}~~,
\eeq
Deshpande and He\cite{DH} find, in a specific calculation, that
$$
A_{1/2} \sim -0.75 e^{i \gamma} e^{i \delta_{T,1/2}} + 7.3 e^{i \delta_{P,1/2}}
{}~~~,
$$
\beq
A_{3/2} \sim -1.06 e^{i \gamma} e^{i \delta_{T,3/2}}
+ 0.84 e^{i \delta_{P,3/2}}~~~,
\eeq
where the first term in each equation is the ``tree'' contribution (of lowest
order in electroweak interactions), while the second term is the penguin
contribution.  Only the electroweak penguin contributes to the $I = 3/2$
amplitude, but with magnitude comparable to the tree contribution.  The
electroweak penguin spoils the relation of Eq.~(\ref{eqn:tri}).

Recently several of us re-examined the effects of SU(3) breaking\cite{SUbr}
and electroweak penguins,\cite{EWP} to see if one could
extract electroweak penguin effects {\em directly from the data}.
Since our previous SU(3) decomposition gave a complete set of
reduced amplitudes, electroweak penguins only changed the interpretation of
these amplitudes, so that a separation of electroweak penguin effects was
not possible merely on the basis of SU(3).

As pointed out by Deshpande and He,\cite{DHP} certain amplitudes [notably
those for $B_s \to (\pi^0~{\rm or}~\rho^0) + (\eta~{\rm or}~\phi)$] are
expected to be {\rm dominated} by electroweak penguins. We noted that the $\pi
K$ amplitude quadrangle could be written in such a manner that one if its
diagonals was equal to $\sqrt{3}A(B_s \to \pi^0 \eta_8)$, where $\eta_8$
denotes an unmixed octet member. The shape of the quadrangle, shown in Fig.~11,
is uniquely determined, up to possible discrete ambiguities. The case of
octet-singlet mixtures in the $\eta$ requires us to replace the $\sqrt{3}$ by
the appropriate coefficient; one can show that the SU(3) singlet contribution
of the $\eta$ is unimportant in this case.

% This is Figure 11
\begin{figure}
% \vspace{2.2in}
\centerline{\epsfysize = 2.2 in \epsffile {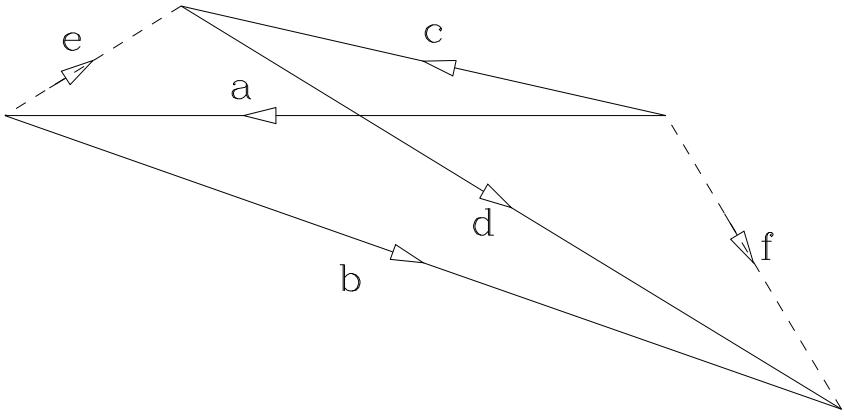}}
\caption{Amplitude quadrangle for $B \to \pi K$ decays. (a) $A(B^+ \to
\pi^+ K^0)$; (b) $\protect \sqrt 2 A(B^+ \to \pi^0 K^+)$; (c) $\protect
\sqrt 2 A(B^0 \to \pi^0 K^0)$; (d) $A(B^0 \to \pi^- K^+)$; (e) the diagonal
$D_2 = \protect\sqrt 3 A(B_s \to \pi^0 \eta_8)$; (f) the diagonal $D_1 =
A_{3/2}$ corresponding to the $I = 3/2$ amplitude.}
\label{piKquad}
\end{figure}

The quadrangle has been written in such a way as to illustrate the
fact\cite{BPP} that the $B^+ \to \pi^+ K^0$ amplitude receives only penguin
contributions in the absence of ${\cal O}(f_B/m_B)$ corrections. The weak
phases of $\bar b \to \bar s$ penguins, which are dominated by a top quark in
the loop, are expected to be $\pi$. We have oriented the quadrangle to subtract
out the corresponding strong phase, and define corresponding strong phase shift
differences $\tilde \delta$ with respect to the strong phase of the $B^+ \to
\pi^+ K^0$ amplitude.

The $I = 3/2~\pi K$ amplitude is composed of two parts, as noted above.  We
can rewrite it slightly as
\beq
A_{3/2} = |A_T| e^{i \gamma} e^{i \tilde \delta_{T,3/2}} - |A_P|~~~.
\eeq
The corresponding charge-conjugate quadrangle has one diagonal equal to
\beq
\bar A_{3/2} = |A_T| e^{- i \gamma} e^{i \tilde \delta_{T,3/2}} - |A_P|~~~,
\eeq
so that one can take the difference to eliminate the electroweak penguin
contribution:
\beq  \label{eqn:diff}
A_{3/2} - \bar A_{3/2} = 2 i |A_T| \sin \gamma~e^{i \tilde \delta_{T,3/2}}~~~.
\eeq
The quantity $|A_T|$ can be related to the $I = 2$ $\pi \pi$ amplitude
in order to obtain $\sin \gamma$. Specifically, if we neglect electroweak
penguin effects in $B^+ \to \pi^+ \pi^0$ (a good approximation), we find that
\beq  \label{eqn:pipi}
|A_T| = \sqrt{2} \tilde{r}_u |A(B^+ \to \pi^+ \pi^0)|~~~.
\eeq
Thus, we can extract not only $\sin \gamma$, but also a strong phase shift
difference $\tilde \delta_{T,3/2}$, by comparing Eqs.~(\ref{eqn:diff}) and
(\ref{eqn:pipi}). If such a strong phase shift difference exists, the $B$ and
$\bar B$ quadrangles will have different shapes, and CP violation in the $B$
system will already have been demonstrated.

The challenge in utilizing the amplitude quadrangle in Fig.~11 is to measure
$B(B_s \to \pi^0 \eta)$, which has been estimated\cite{DHP} to be only $2
\times 10^{-7}$!  Very recently (since the Summer School) Deshpande and He
\cite{DHeta} have pointed out that the amplitude triangle
\beq
2 A(B^+ \to \pi^+ K^0) + \sqrt{2} A(B^+ \to \pi^0 K^+) = \sqrt{6} A(B^+ \to
\eta_8 K^+)~~~,
\eeq
implied by the SU(3) relations of Refs.~\cite{BPP,EWP}, where $\eta_8$ is
the octet component of the $\eta$, permits one to specify the
quantity $A_{3/2} - \bar A_{3/2}$ in Eq.~(\ref{eqn:diff}) and extract $\gamma$
as above.  Here there is some delicacy associated with the SU(3) singlet
component of the physical $\eta$.

{\it 4. Other final states.}

{\it a) $PV$ final states} ($V = \rho,~\omega,~K^*,~\phi$) are characterized
by more graphs (and hence more reduced SU(3) amplitudes), since one no longer
has the benefit of Bose statistics as in $B \to P P$ decays.  There still
exist quadrangle relations in $\rho K$ and $K^* \pi$ decays, however.
Remarkably, if $\Delta S = 0$ gluonic penguin diagrams (small in any case) are
approximately equal for the cases in which the spectator quark ends up in a
vector meson and in a pseudoscalar [see Ref.~\cite{EWP} for details], the
previous quadrangle construction still holds if we replace $\sqrt{3} A(B_s \to
\pi^0 \eta_8$ with $\sqrt{2} A(B_s \to \pi^0 \phi)$, $A(B \to \pi K)$ with $A(B
\to \pi K^*)$, and $A(B^+ \to \pi^+ \pi^0)$ with $A(B^+ \to \pi^0 \rho^+)$.

Deshpande and He\cite{DHP} predict $B(B_s \to \pi^0 \phi) \approx 2 \times
10^{-8}$, which probably means that the quadrangle reduces to two nearly
overlapping triangles (whose shapes will consequently be difficult to specify).
On the other hand, in $PV$ decays, the effects of electroweak penguins then
may not be so important if the dominant processes are characterized by
branching ratios of order $10^{-5}$ as in $B \to PP$ decays.

Hints of some signals have been seen in some $PV$ channels in the latest
CLEO data,\cite{Wurt} but only upper limits are being quoted.  These are
fairly close to theoretical expectations in the case of some $\pi \rho$
channels.

{\it b) $VV$ final states} satisfy Bose statistics. Since the total angular
momentum of the decaying particle is zero, the (space) $\times$ (spin) part of
the $VV$ wave function will be symmetric, as in $PP$ final states.\cite{HJLpc}
Thus, there should exist amplitude relations for each relative orbital angular
angular momentum $\ell$.  If one $\ell$ value dominates the decays, such
relations might be tested using triangles constructed of square roots of decay
rates, as in the $PP$ case.
\bigskip

G. ~~ Progress in studies of CP violation in $B$ physics
\medskip

Since the earliest observations of $B$ mesons, the goal of utilizing them for
CP-violation studies\cite{BCP,Brevs} has steadily marched toward realization.
For perspective, the situation as of January 1987\cite{SLCrev} is compared with
that of February 1995 in Table 5.  Moreover, in recent years there have arisen
further possibilities based on symmetric $e^+ e^-$ colliders,\cite{KB} on the
``tagging'' of neutral $B$ mesons using hadrons produced nearby in phase space,
and on the measurement of CKM phases without explicit detection of CP-violating
asymmetries.  This area is rich indeed!

% This is Table 5
\begin{table}
\caption{Progress in the study of CP violation in $B$ meson systems.}
\begin{center}
\begin{tabular}{||l | c | c||} \hline
{}~~~Step           &     1/87         & 2/95        \\ \hline
1.~~See $B$'s    &  $B \to \jape + \ldots $  & Good signals of $B \to
\jape$ \\
                  & a good tag       & $+ K$ in hadron colliders \\ \hline
2.~~Measure      & Expected $D \pi$,& $B(D \pi) \sim 0.3\%$ \\
\qquad branching ratios  & $\jape ~ K, \ldots$ & $B (\jape ~K) \simeq 0.1\%$ \\
                  & level of $10^{-3}$  &      \\ \hline
3.~~$B_u : B_d : B_s$  & Expected      & Still not known.  CP viol.\\
\qquad needed            & $\sim 2:2:1$     & hard to see for $B_s$. \\ \hline
4.~~See $B_d - \bar B_d$ and & Some UA1 evidence  & $(\Delta m /\Gamma )_B
= 0.75 \pm 0.15$ \\
\qquad $B_s - \bar B_s$ mixing & for $B_s - \bar B_s$ mixing &
$(\times \stackrel{>}{\sim} 15$ for $B_s $) \\ \hline
5.~~See $b \to u$ & Neither seen   & $|V_{ub}/V_{cb}| = 0.08 \pm 0.02 $ \\
\qquad and $t$ quark   & yet   & $m_t = 180 \pm 12$ GeV/$c^2$ \\ \hline
6.~~See CP violation   & Optimistic estimate: & Realism:  need
$\stackrel{>}{\sim}$ \\
\qquad in $B$ system   & need $\stackrel{>}{\sim} 10^7 ~B$'s & (few $\times
10^8$) $B$'s in \\
                &                        & any practical expt. \\ \hline
7.~~Measure time-dep.  & Importance stressed of & Asymmetric $B$ factories \\
\qquad CP violation    & cuts on decay times & under construction \\ \hline
\end{tabular}
\end{center}
\end{table}
\newpage

\leftline{6. NON-STANDARD AND SPECULATIVE ASPECTS}
\bigskip

A. ~~ Superweak theory
\medskip

It is possible to explain the nonzero value of $\epsilon$ in the neutral kaon
system by means of an {\it ad hoc} $\Delta S = 2$ interaction leading directly
to CP-violating $K^0 - \bar K^0$ mixing.\cite{sw}  The phase of $\epsilon$ will
then automatically be the superweak phase mentioned in Section 2, and one will
see no difference between $\eta_{+-}$ and $\eta_{00}$. The only evidence
against this possibility so far is the $> 3 \sigma$ observation of nonzero
$\epsilon'/\epsilon$ by the CERN NA31 experiment,\cite{NA31} a result not
confirmed by Fermilab E731.\cite{E731}

A superweak interaction (of considerably greater strength) could in principle
lead to observable CP-violating $B^0 - \bar B^0$ mixing.  If this were so,
one would expect\cite{Win} $A(\pi^+ \pi^-) = - A(J/\psi K_S)$ as a result of
the opposite CP eigenvalues of the two final states.  In order for this
relation to hold in the standard model, one would need $\eta = (1-\rho)
[\rho/(2-\rho)]^{1/2}$.  Taking account of possible errors in checking that
the asymmetries are actually equal and opposite, one concludes that a portion
of the allowed region of parameters shown in Fig.~5 could not be distinguished
from a superweak theory.  The ratio $A(\pi^+ \pi^-)/A(J/\psi K_S)$ is
informative in a more general context:  for example, if it exceeds
1, then $\rho$ must be negative.\cite{PHJR}

If $\epsilon$ arises entirely from a superweak interaction, there is no need
for CKM phases, and one will see no ``direct'' effects in kaon or $B$ decays.
There will also be no neutron or electron electric dipole moments, though such
effects also will be well below experimental capabilities in the standard CKM
picture.
\bigskip

B. ~~ Right-handed $W$'s
\medskip

The standard electroweak theory involves coupling of $W$ bosons only to
left-handed fermions, through terms in the Lagrangian such as $\bar d
\gamma^{\mu} (1 - \gamma_5) W^{(+)}_{\mu} u$.  The left-handed quarks and
leptons are members of a doublet of left-handed isospin SU(2)$_L$, while
right-handed quarks and leptons are singlets. The corresponding formula for the
charge of a fermion then is $Q = I_{3L} + (Y/2)$, where $I_{3L}$ is the third
component of the SU(2)$_L$ group, while $Y$ is the weak hypercharge [a U(1)
quantum number].  The neutral boson $W^0$ coupling to the SU(2) group and a
neutral boson $B^0$ coupling to the U(1) charge mix with one another to form
the photon and the $Z^0$. There is no need in such a theory for a right-handed
neutrino; nothing couples to it.  There is a single set of charged $W$ bosons,
which we may call $W_L^{\pm}$.  Their leptonic decays are of the form $W \to
\ell \nu_\ell$.

The asymmetry of the standard theory has led to the proposal\cite{RHW} that
there also exist {\it right-handed $W$'s}, coupling to right-handed fermions
through terms in the Lagrangian such as $\bar d \gamma^{\mu} (1 + \gamma_5)
W_{R \mu}^{(+)} u$.  The right-handed quarks and leptons are members of a
doublet of right-handed isospin SU(2)$_R$, while left-handed quarks and leptons
are singlets of this group.  The fermion charge is now given by the much more
symmetric (and easy-to-remember!) formula $Q = I_{3L} + I_{3R} + (B-L)/2$,
where $B$ and $L$ are baryon and lepton number: $B = 1/3$ for a quark; $L = 1$
for a lepton.

There are now two neutral bosons $W_L^0$ and $W_R^0$ coupling to the two SU(2)
groups, and a neutral boson coupling to the U(1)$_{B - L}$ quantum number.
Consequently, there are not one but two $Z$'s in such a theory; one of them
must have a mass of least a few hundred GeV in order not to have already been
seen. A right-handed neutrino $N_{R \ell}$ (for each lepton family) is needed
in the theory for cancellation of triangle anomalies.  There are two sets of
charged $W$ bosons, $W_L^{\pm}$ and $W_R^{\pm}$.  The leptonic decay of the
$W_R$ must involve the $N_R$.  The $N_R$ and the ordinary left-handed $\nu_L$
need not have the same masses if they are {\it Majorana particles} (equal to
their antiparticles), as is permitted for neutral particles. If the $N_R$ turns
out to be too heavy to permit $W_R \to \ell N_{R \ell}$, one should look
for the $W_R$ instead through the decay $W_R \to t \bar b$.\cite{RT}

Limits on right-handed $W$'s stem, for example, from the detailed study of the
electron spectrum in $\mu \to e \nu \bar \nu$.\cite{WRlims}  In this manner one
concludes that a $W_R$ would have to be heavier than nearly 1/2 TeV {\it if
right-handed neutrinos are light}. The limits also depend on the degree to
which $W_L$ and $W_R$ bosons mix with one another.  An even more stringent
lower limit of a couple of TeV follows from the remarkable efficiency of box
diagrams (cf. Fig.~3) with one $W_L$ and one $W_R$ in inducing $K - \bar K$
mixing.\cite{Mixbox}  This limit is degraded considerably, however,\cite{LS} if
the couplings of $W_R$ do not follow the same CKM pattern as those of $W_L$.

If right-handed $W$'s couple to right-handed fermions, one can obtain new
CP-violating interactions (for example, via box diagrams involving $W_R$ and
the usual left-handed $W_L$).  The right-handed $W$ mass scale must be tens of
TeV or less in order for a large enough contribution to $\epsilon$ to be
generated.  In contrast to the situation described in Section 3, one can
generate CP violation using only two quark families, since redefinitions of
quark phases are constrained by the right-handed couplings.

An amusing feature of right-handed $W$ couplings is that their participation
(or even dominance) in $b$ quark decays is surprisingly hard to rule
out.\cite{GW}  Some suggestions have been made to test the usual picture of
left-handed $b \to c$ decays using the polarization of $\Lambda_b$ baryons
produced in $b$ quark fragmentation.\cite{ARWW}
\bigskip

C. ~~ Multi-Higgs and supersymmetric models
\medskip

If there is more than one Higgs doublet, complex vacuum expectation values of
Higgs fields can lead to CP-violating effects.\cite{LW}  It appears that in
order to explain $\epsilon \ne 0$ in neutral kaon decays by this mechanism, one
expects too large a neutron electric dipole moment.\cite{Sanda}  The
possibility of such effects remains open, however, and the best test for them
remains the study of dipole moments.  Current models\cite{Hay,YLWu} tend to be
constrained by the present limits\cite{limits} of
\beq
|d_n| < 1.1 \times 10^{-25}~e{\cdot \rm cm}~(95\%~{\rm c.l.})~~,~~~
|d_e| < 2 \times 10^{-26}~e{\cdot \rm cm}~(95\%~{\rm c.l.})~~.
\eeq
Other CP-violating effects in multi-Higgs-boson models include transverse
lepton polarization in $K_{\mu 3}$ decays\cite{FKG} and various asymmetries
in charm decays,\cite{YLWu} which we discuss in Sec.~6 E below.

One needs complex vacuum expectation values in some multi-Higgs models in order
to generate CP violation.  Supersymmetric models, in which for every particle
there is a ``superpartner'' differing by 1/2 unit of spin, have two Higgs
doublets whose ratio of vacuum expectation values is real and thus cannot
generate CP violation.\cite{Gunion} However, such models can generate dipole
moments at one loop\cite{SMB} order in perturbation theory.  Thus, the
possibility of dipole moments showing up just below present bounds is
considerably enchanced in such models relative to the standard CKM predictions,
which are less than about $10^{-32}~e{\cdot \rm cm}$ for the neutron and
$10^{-38}~e{\cdot \rm cm}$ for the electron and rely on contributions at the
three-or-greater-loop level.

A useful constraint on multi-Higgs models arises from the
$b \to s \gamma$ transition.\cite{AKG}  The observed transition rate\cite{bsg}
corresponding to a branching ratio $B(b \to s \gamma) = (2.32 \pm 0.51 \pm 0.29
\pm 0.32) \times 10^{-4}$, is consistent with the standard-model prediction
based on the penguin graph,\cite{Bubsg} limiting the contributions from other
sources such as charged Higgs bosons. Corrections to the standard model's
predictions for the $Z b \bar b$ vertex (in which loops involving top quarks
play a role) also set limits on masses and couplings in multi-Higgs models,
while supersymmetric models can improve small discrepancies between experiment
and theory for the $Z b \bar b$ vertex.\cite{AKG}

In Fig.~12 we show the constraints on the mass $m_+$ of the charged Higgs boson
and the ratio of vacuum expectation values of the two Higgs doublets, $\tan
\beta = v_2/v_1$, from various sources.\cite{AKG}  For large $m_+$ one has
$\tan \beta > 3/4$, while for large $\tan \beta$ the charged Higgs boson is
restricted to lie above about 100 GeV.
\bigskip

% This is Figure 12
\begin{figure}
% \vspace{5.8in}
\centerline{\epsfysize = 5.8 in \epsffile {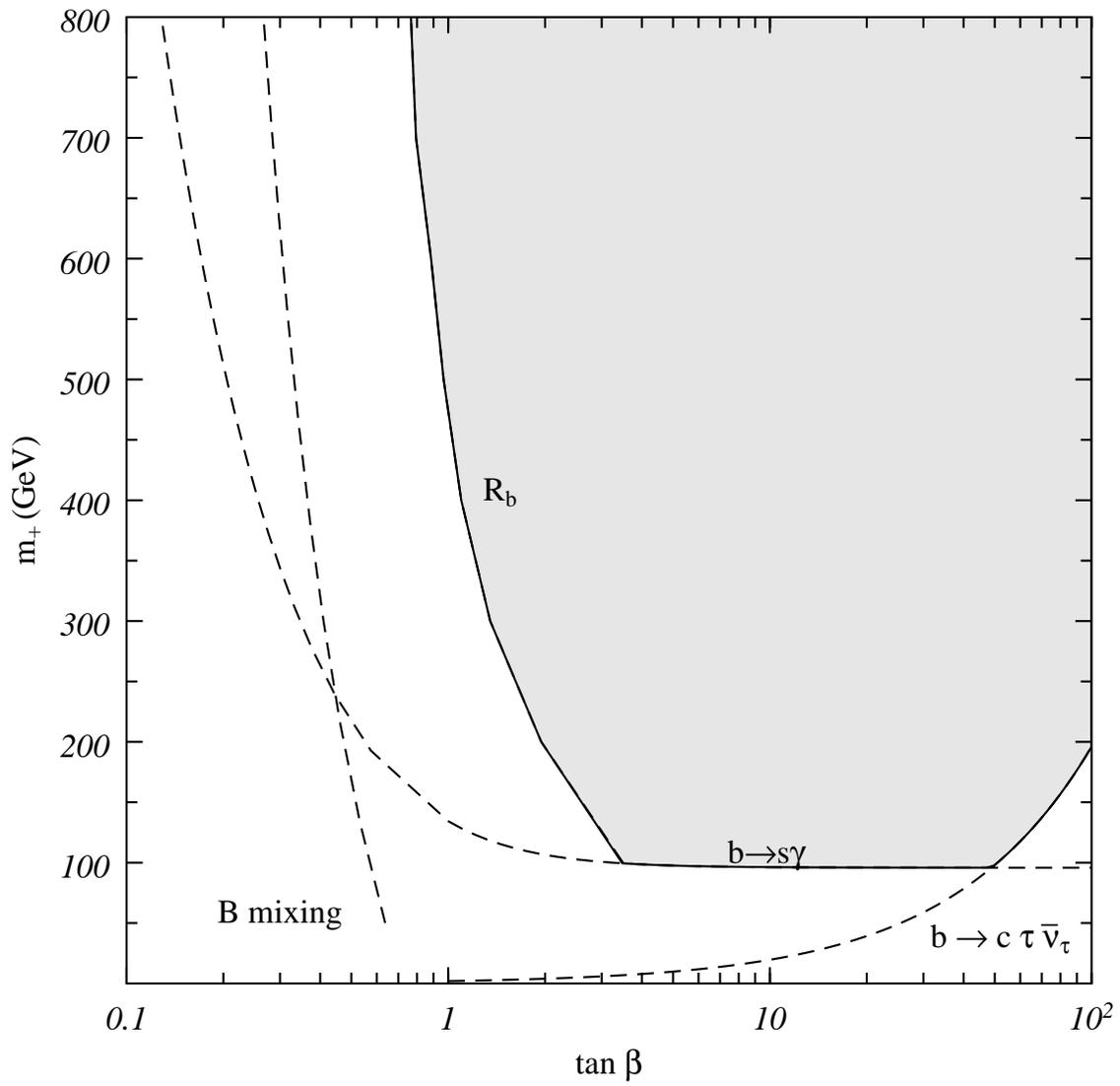}}
\caption{Constraints on two-Higgs models from various experimental
sources.\protect\cite{AKG}  The allowed region is shaded.  $R_b$:  constraint
from the $Z b \bar b$ vertex.}
\end{figure}

D. ~~ Dipole moments
\medskip

The Hamiltonian for the interaction of a particle with spin $\bf S$ with
homogeneous electric and magnetic fields can be written
\beq
h = -(d~{\bf E} + \mu~{\bf B}) \cdot {\bf S}/S~~~,
\eeq
where $d$ is the electric dipole moment and $\mu$ is the magnetic dipole
moment.  Under time reversal, $t \to -t$, the fields and spin transform as
\beq
{\bf E} \to {\bf E}~~~,~~~{\bf B} \to -{\bf B}~~~,~~~{\bf S} \to - {\bf S}~~~,
\eeq
so that a nonzero electric dipole moment would violate time reversal
invariance.  (In fact, a nonzero expectation value $\langle {\bf E} \cdot
{\bf S} \rangle \ne 0$ would violates $P$ as well as $T$.)

We have mentioned present limits on the neutron and electron dipole moments.
The neutron limits are obtained in experiments in which ultra-cold neutrons
are confined in a magnetic ``bottle'' by means of their magnetic moments.
The electron limits are based on experiments in atoms, in which fortunate
overlaps of levels often lead to substantial amplification of the electron's
moment.  In the case of the limit cited above, based on atomic thallium,
the amplification factor is $d_{\rm Atom}/d_e \simeq -600$.

An example of a recent stringent limit, based on comparison of Larmor
frequencies for parallel and antiparallel electric and magnetic
fields,\cite{Hg} is $d(^{199}{\rm Hg}) < 1.3 \times 10^{-27} e{\cdot \rm cm}$
(95\% c.l.).
\bigskip

E. ~~ Charm decays
\medskip

The standard model predicts very small CP-violating effects in charmed particle
decays.  $D^0 - \bar D^0$ mixing is expected to be small and uncertain,
dominated by long-distance effects.\cite{Dmix}  The short-distance contribution
to CP-violating mixing should be of order $(m_b/m_t)^2$ times that in neutral
kaons, while the lifetime of a neutral $D$ meson is about 0.4 ps in comparison
with 52 ns for a $K_L$.  The tree-level decays $c \to s u \bar d$, $c \to d u
\bar d$, $c \to d u \bar s$ have zero or negligible weak phases in the standard
convention.

For precisely these reasons, CP-violating charmed particle decays are an
excellent place to look for new physics.\cite{YLWu,BigC}  New effects tend to
be accompanied with flavor-changing neutral currents, which may or may not be
an advantage in specific cases.  Experiments are easy to perform and
undersubscribed in comparison with the many proposed studies of $B$ physics.
Information on rate asymmetries $A(f) \equiv [\Gamma(i \to f) - \Gamma(\bar{i}
\to \bar f)]/[\Gamma(i \to f) + \Gamma(\bar{i} \to \bar f)]$ is just now
beginning to appear, as illustrated in Table 6.\cite{Asymms}

% This is Table 6
\begin{table}
\caption{Rate asymmetries in charmed meson decays.}
\begin{center}
\renewcommand{\arraystretch}{1.3}
\begin{tabular}{c c c} \hline
Charmed &      Final       &          Asymmetry \\
meson   &      state       &                    \\ \hline
$D^+$   & $K^-K^+\pi^+$    & $-0.031 \pm 0.068$ \\
$D^+$   & $\bar K^{*0}K^+$ & $-0.12 \pm 0.13$   \\
$D^+$   & $\phi \pi^+$     & $0.066 \pm 0.086$  \\
$D^0$   &    $K^+ K^-$     & $0.024 \pm 0.084$  \\ \hline
\end{tabular}
\end{center}
\end{table}
\bigskip

F. ~~ Baryogenesis, neutrino masses and grand unification
\medskip

The ratio of baryons to photons in our Universe is a few parts in $10^9$. If
baryons and antibaryons had been produced in equal numbers, mutual
annihilations should have reduced this quantity to a much smaller
number,\cite{JP} of order a part in $10^{18}$. In 1967 Sakharov\cite{Sak}
proposed three ingredients of any theory which sought to explain the
preponderance of baryons over antibaryons in our Universe:  (1) violation of C
and CP; (2) violation of baryon number, and (3) a period in which the Universe
was out of thermal equilibrium.  Thus our very existence may owe itself to CP
violation.  However, no consensus exists on a specific implementation of
Sakharov's suggestion.

A toy model illustrating Sakharov's idea can be constructed within a model
wherein color SU(3) and electroweak SU(2) $\times$ U(1) are embedded in a
``grand unified'' SU(5) gauge group.\cite{GG} This group contains ``$X$''
bosons which can decay both to $uu$ and to $e^+ \bar d$.  By CPT, the total
decay rates of $X$ and $\bar X$ must be equal, but CP-violating rate
differences $\Gamma(X \to uu) \ne \Gamma(\bar X \to \bar u \bar u)$ and
$\Gamma(X \to e^+ \bar d) \ne \Gamma(\bar X \to e^- d)$ are permitted.  This
example conserves $B - L$, where $B$ is baryon number (1/3 for quarks) and $L$
is lepton number (1 for electrons).

It was pointed out by 't Hooft\cite{tH} that the electroweak theory contains an
anomaly as a result of nonperturbative effects.  This anomaly conserves $B - L$
but violates $B + L$.  If a theory leads to $B - L = 0$ but $B + L \ne 0$ at
some primordial temperature $T$, the anomaly can wipe out any $B+L$ as $T$
sinks below the electroweak scale.\cite{KRS}  Thus, the toy model mentioned
above and many others are unsuitable in practice.  Proposed solutions include
(1) the generation of baryon number directly at the electroweak scale rather
than at a higher temperature,\cite{FS} and (2) the generation of nonzero $B -
L$ at a high temperature, e.g., through the generation of nonzero lepton number
$L$ which is then reprocessed into nonzero baryon number by the `t Hooft
anomaly mechanism.\cite{Yana} The first scenario, based on standard model
CP-violating interactions (as manifested in the CKM matrix), is widely regarded
as inadequate to generate the observed baryon asymmetry at the electroweak
scale.\cite{HS}  We illustrate in Fig.~13 some aspects of the second scenario.
(Missing ingredients are denoted by question marks.) The existence of a baryon
asymmetry, when combined with information on neutrinos, could provide a window
to a new scale of particle physics.

If neutrinos have masses at all, they are much lighter than their charged
counterparts or the corresponding leptons.  (See Fig.~1.) One possibility for
the suppression of neutrino masses\cite{seesaw} is the so-called ``seesaw''
mechanism, by which light neutrinos acquire Majorana masses of order $m_M =
m_D^2/M_M$, where $m_D$ is a typical Dirac mass and $M_M$ is a large Majorana
mass acquired by right-handed neutrinos.  Such Majorana masses change lepton
number by two units and therefore are ideal for generating a lepton asymmetry
if Sakharov's other two conditions are met.

% This is Figure 13
\begin{figure}
% \vspace{6in}
\centerline{\epsfysize = 6 in \epsffile {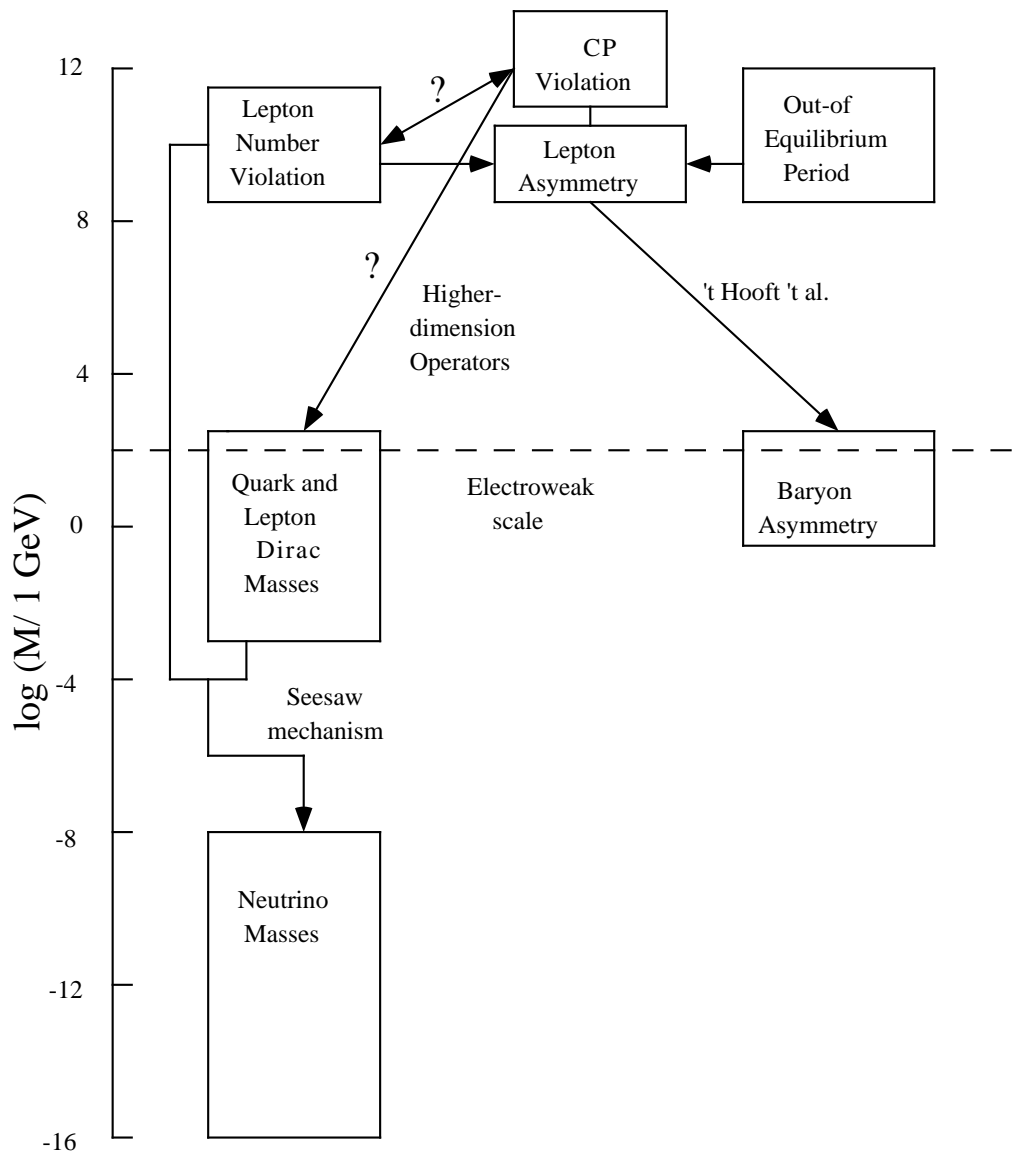}}
\caption{Mass scales associated with one scenario for baryogenesis.}
\end{figure}

The question of baryogenesis is thus shifted onto the leptons:  Do neutrinos
indeed have masses?  If so, what is their ``CKM matrix''?  Do the properties of
heavy Majorana right-handed neutrinos allow any new and interesting natural
mechanisms for violating CP at the same scale where lepton number is violated?
Majorana masses for right-handed neutrinos naturally violate left-right
symmetry and could be closely connected with the violation of $P$ and $C$ in
the weak interactions.\cite{BKCP}

An open question in this scenario, besides the precise form of CP violation at
the lepton-number-violating scale, is how this CP violation gets communicated
to the lower mass scale at which we see CKM phases.  Presumably this occurs
through higher-dimension operators which imitate the effect of Higgs boson
couplings to quarks and leptons.

The presence of a suitable mass scale for lepton number violation is suggested
by certain patterns of electroweak-strong unification.  If the strong and
electroweak coupling constants are evolved to high mass scales in accord with
the predictions of the renormalization group,\cite{GQW} as shown in
Fig.~14(a), they approach one another in the simplest SU(5) model,\cite{GG}
but do not really cross at the same point.  This ``astigmatism'' can be cured
by invoking supersymmetry,\cite{Amaldi} as illustrated in Fig.~14(b).  Here
the cure is effected not just by the contributions of superpartners, but by the
richer Higgs structure in supersymmetric theories.  The theory predicts many
superpartners below the TeV mass scale, some of which ought to be observable in
the next few years.

% This is Figure 14
\begin{figure}
% \vspace{5in}
\centerline{\epsfysize = 5 in \epsffile {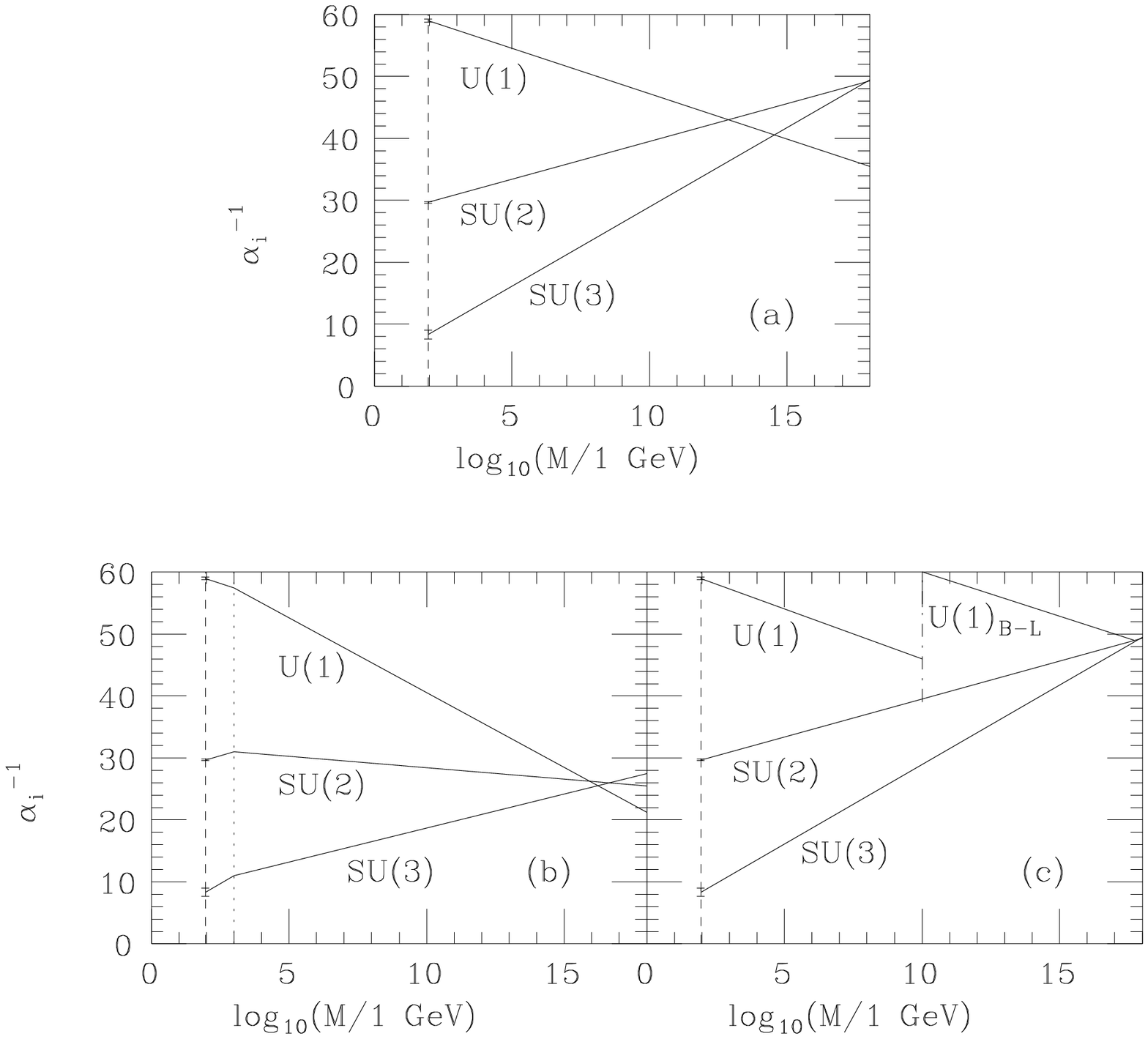}}
\caption{Behavior of coupling constants predicted by the renormalization group
in various grand unified theories. Error bars in plotted points denote
uncertainties in coupling constants measured at $M = M_Z$ (dashed vertical
line).  (a)  SU(5); (b) supersymmetric SU(5) with superpartners above 1 TeV
(dotted line) (c) example of an SO(10) model with an intermediate mass scale
(dot-dashed vertical line).}
\end{figure}

Alternatively, one can embed SU(5) in an SO(10) model,\cite{SOten} in which
each family of quarks and leptons (together with a right-handed neutrino for
each family) fits into a 16-dimensional spinor representation.  Fig.~14(c)
illustrates one scenario for breaking of SO(10) at two different scales, the
lower of which is a comfortable scale for the breaking of left-right symmetry
and the generation of right-handed neutrino Majorana masses.
\bigskip

G. ~~ The strong CP problem
\medskip

We will be very brief about this subject.  There are fine reviews
elsewhere.\cite{PecR,MD} As a result of nonperturbative effects, the QCD
Lagrangian acquires an added CP-violating term $g_s^2 \bar \theta F_{\mu \nu}^a
\tilde{F}^{\mu \nu a}/32 \pi^2$, where $\bar \theta = \theta + {\rm
Arg~det}~M$, $\theta$ is a term describing properties of the QCD vacuum, and
$M$ is the quark mass matrix. The limit on the observed neutron electric dipole
moment, together with the estimate\cite{PecR} $d_n \simeq 10^{-16} \bar
\theta~e$ cm, implies that $\theta \le 10^{-9}$, which looks very much like
zero.  How can one undertand this?  Several proposals exist.

{\it 1. Vanishing $m_u$.}
If one quark mass vanishes (the most likely candidate being $m_u$), one can
rotate away any effects of $\bar \theta$.\cite{PecR}  However, it is
generally though not universally felt that this bends the constraints of
chiral symmetry beyond plausible limits.\cite{Leut}  My own guess is that
light-quark masses are in the ratios $u:d:s = 3:5:100$.

{\it 2.  Axion.}
One can introduce a continuous U(1) global symmetry\cite{PQ} such that
$\bar \theta$ becomes a dynamical variable which relaxes to zero as a result of
new interactions.  The spontaneous breaking of this symmetry then leads to
a pseudo-Nambu-Goldstone boson, the {\it axion},\cite{AX} for which searches
may be performed in many ways.  My favorite is via the Primakoff
effect,\cite{PS} in which axions in the halo of our galaxy interact with a
static man-made strong magnetic field to produce photons with frequency
equal to the axion mass (up to Doppler shifts).  These photons can be detected
in resonant cavities.  Present searches would have to be improved by about
a factor of 100 to detect axions playing a significant role in the mass of the
galaxy.\cite{MD}

{\it 3. Boundary conditions.}
It has been proposed\cite{RGS} that one consider not the $\theta$-vacuum, but
an incoherent mixture consisting of half $\theta$ and half $-\theta$, in the
manner of a sum over initial spins of an unpolarized particle.  The
experimental consequences of this proposal are still being worked out.
\bigskip

H. ~~ Possible origin of the CKM Matrix
\medskip

Numerous attempts have been made to understand the structure of the CKM matrix
by imposing some discrete (``family'') symmetries on the quarks and leptons.
One thereby obtains strategically placed zeroes in the mass matrices, which can
lead to relations between quark masses and CKM matrix elements.  We have
reviewed such approaches previously.\cite{JRCKM}  Here we would like to
mention a favorite class of models which is considerably less popular, but
worthy of consideration nonetheless.

We imagine quarks as being composite objects.  States of definite mass and
charge (such as the quarks $d,~s$, and $b$) are imagined to involve mixing of
configurations of the constituents.  Thus we envision families as corresponding
to orthogonal mixed configurations.  There should be a gap between the
light-fermion families and additional states at the compositeness
scale.\cite{tHc}

Let us imagine that a weak charge-changing transition changes the identity of
one of the constituents, thereby altering its interaction with its neighbors.
As a consequence, a rotation will be induced in the basis of states, so that
$u,~c$, and $t$ will involve different mixed configurations than $d,~s$, and
$b$.  The simplest example in the quark model is provided by the lowest-lying
states of the charmed baryon $\Xi_c^+ = csu$ or $\Xi_c^0 = csd$, which
correspond approximately (but not exactly) to basis states in which the two
light quarks are coupled to spin 0 or 1.  If $m_u \ne m_d$, the $\Xi_c^+$ and
$\Xi_c^0$ will be slightly different mixtures of the basis states.\cite{PTP}
This two-family system thus provides an illustration of the origin of the
Cabibbo angle.

We have constructed a toy model of the CKM matrix\cite{RW} with the above
mechanism in mind.  One gets satisfactory relations between CKM elements and
quark masses, but at the price of discrete choices of interactions between
the subunits.  There is no obvious source of the CKM phase in such a model;
it has to be put in by hand. The interested reader is referred to the original
article for more details.
\bigskip

I. ~~ Overview of nonstandard and speculative aspects
\medskip

Each of the suggested alternatives to the standard picture of CP violation may
be tested against particular experimental touchstones.  A key experiment to
disprove the superweak model of CP violation in the kaon system will be the
observation of a nonzero value of $\epsilon'$.  The scale at which right-handed
$W$'s make their appearance, if at all, is inversely correlated with neutrino
masses in ``seesaw'' types of models for those masses. Multi-Higgs models for
CP violation are constrained by existing limits on electric dipole moments, so
that such effects could show up at any time as experimental sensitivities are
improved.

The standard model of CP violation predicts very small effects in charmed
particle decays.  Since charmed particles are easier to produce than $B$'s
and the standard expectations for CP asymmetries in their decays are small,
they are a very good place to look for non-standard effects.

The scenario we have presented for baryogenesis, based on its correlation with
lepton number violation, again hinges on the values of neutrino masses.  We
have proposed that fundamental CP and lepton number violation arise together
at a high mass scale, for which neutrino masses would provide an estimate.
The strong CP problem may be addressed experimentally by means of axion
searches, of which those based on RF cavities are particularly appealing.
Even the possibility that quarks and leptons are composite can be addressed
experimentally, though the problems in such models are well-known and
serious.\cite{tHc,RS}
\bigskip

\leftline{7. ~~CONCLUSIONS}
\bigskip

The observed CP violation in the neutral kaon system has been successfully
parametrized in terms of the Cabibbo-Kobayashi-Maskawa (CKM) matrix. The
problem has been shifted to one of understanding the magnitudes and phases of
CKM elements.  Even before this more ambitious question is addressed, however,
one seeks independent tests of the CKM picture of CP violation.  Rare $K$
decays and $B$ decays will provide many such tests.

Alternative (non-CKM) theories of CP violation are much more encouraging for
some CP-violating quantities like the neutron electric dipole moment or effects
in charmed particle decays. However, most of these alternative theories do not
predict observable direct CP-violating effects in $K$ or $B$ decays.

No real understanding exists yet of baryogenesis or of the strong CP problem.
Fortunately, there exist many possibilities for experiments bearing on these
questions, including searches for neutrino masses and for axions.

The CKM picture suggests that we may understand CP violation better when the
pattern of fermion masses itself is understood.  Why is the top quark heavier
than all the other quarks and leptons (or why are the others so much lighter
than the top?)  The observation of the top quark\cite{CDFt,D0t} has forced us
to confront this problem, and perhaps we will figure out the answer some day.
\bigskip

\leftline{\bf Acknowledgments}
\bigskip

I would like to thank Jim Amundson, Isi Dunietz, Aaron Grant, Michael Gronau,
Oscar Hern\'andez,  Nahmin Horowitz, Mike Kelly, David London, Alex Nippe,
Sheldon Stone, Tatsu Takeuchi, and Mihir Worah for enjoyable collaborations on
some of the topics mentioned in these lectures. In addition I am grateful to
Harry Lipkin, Bob Sachs, Bruce Winstein, and Lincoln Wolfenstein for fruitful
discussions. This work was supported in part by the United States Department of
Energy under Grant No. DE FG02 90ER40560.

\bigskip

%%%%%%%%%%%%%%%%%%%%%%%%%%%
% Journal and other miscellaneous abbreviations for references
% World Scientific format for journals
\def \ap#1#2#3{{\it Ann. Phys. (N.Y.)} {\bf#1} (#3) #2}
\def \apny#1#2#3{{\it Ann.~Phys.~(N.Y.)} {\bf#1} (#3) #2}
\def \app#1#2#3{{\it Acta Physica Polonica} {\bf#1} (#3) #2}
\def \arnps#1#2#3{{\it Ann. Rev. Nucl. Part. Sci.} {\bf#1} (#3) #2}
\def \arns#1#2#3{{\it Ann. Rev. Nucl. Sci.} {\bf#1} (#3) #2}
\def \ba88{{\it Particles and Fields 3} (Proceedings of the 1988 Banff Summer
Institute on Particles and Fields), edited by A. N. Kamal and F. C. Khanna
(World Scientific, Singapore, 1989)}
\def \baps#1#2#3{{\it Bull. Am. Phys. Soc.} {\bf#1} (#3) #2}
\def \be87{{\it Proceedings of the Workshop on High Sensitivity Beauty
Physics at Fermilab,} Fermilab, Nov. 11--14, 1987, edited by A. J. Slaughter,
N. Lockyer, and M. Schmidt (Fermilab, Batavia, IL, 1988)}
\def \cn{Collaboration}
\def \corn{{\it Lepton and Photon Interactions:  XVI International Symposium,
Ithaca, NY 1993,} edited by P. Drell and D. Rubin (AIP, New York, 1994)}
\def \cp89{{\it CP Violation,} edited by C. Jarlskog (World Scientific,
Singapore, 1989)}
\def \dpf91{{\it The Vancouver Meeting - Particles and Fields '91}
(Division of Particles and Fields Meeting, American Physical Society,
Vancouver, Canada, Aug.~18--22, 1991), ed. by D. Axen, D. Bryman, and M. Comyn
(World Scientific, Singapore, 1992)}
\def \dpfa{{\it The Albuquerque Meeting:  DPF 94} (Division of Particles and
Fields Meeting, American Physical Society, Albuquerque, NM, August 2--6,
1994), ed. by S. Seidel (World Scientific, River Edge, NJ, 1995)}
\def \dpff{{\it The Fermilab Meeting -- DPF 92} (Division of Particles and
Fields Meeting, American Physical Society, Fermilab, 10--14 November, 1992),
ed. by C. H. Albright \ite~(World Scientific, Singapore, 1993)}
\def \efi{Enrico Fermi Institute Report No.~}
\def \hb87{{\it Proceeding of the 1987 International Symposium on Lepton and
Photon Interactions at High Energies,} Hamburg, 1987, ed. by W. Bartel
and R. R\"uckl (Nucl.~Phys.~B, Proc. Suppl., vol. 3) (North-Holland,
Amsterdam, 1988)}
\def \ib{{\it ibid.}~}
\def \ibj#1#2#3{{\it ibid.} {\bf#1} (#3) #2}
\def \ijmpa#1#2#3{{\it Int.~J. Mod.~Phys.}~A {\bf#1} (#3) #2}
\def \jpg#1#2#3{{\it J. Phys.} G {\bf#1} (#3) #2}
\def \kdvs#1#2#3{{\it Kong.~Danske Vid.~Selsk., Matt-fys.~Medd.} {\bf #1}
(#3) No #2}
\def \ky85{{\it Proceedings of the International Symposium on Lepton and
Photon Interactions at High Energy,} Kyoto, Aug.~19-24, 1985, edited by M.
Konuma and K. Takahashi (Kyoto Univ., Kyoto, 1985)}
\def \lat90{{\it Results and Perspectives in Particle Physics} (Proceedings of
Les Rencontres de Physique de la Vallee d'Aoste [4th], La Thuile, Italy, Mar.
18-24, 1990), edited by M. Greco (Editions Fronti\`eres, Gif-Sur-Yvette,
France,
1991)}
\def \lg91{International Symposium on Lepton and Photon Interactions, Geneva,
Switzerland, July, 1991}
\def \lkl87{{\it Selected Topics in Electroweak Interactions} (Proceedings of
the Second Lake Louise Institute on New Frontiers in Particle Physics, 15--21
February, 1987), edited by J. M. Cameron \ite~(World Scientific, Singapore,
1987)}
\def \mpla #1#2#3{{\it Mod. Phys. Lett.} A {\bf#1} (#3) #2}
\def \nc#1#2#3{{\it Nuovo Cim.} {\bf#1} (#3) #2}
\def \np#1#2#3{{\it Nucl. Phys.} {\bf#1} (#3) #2}
\def \oxf65{{\it Proceedings of the Oxford International Conference on
Elementary Particles} 19/25 Sept.~1965, ed.~by T. R. Walsh (Chilton, Rutherford
High Energy Laboratory, 1966)}
\def \pascos{{\it PASCOS 94} (Proceedings of the Fourth International
Symposium on Particles, Strings, and Cosmology, Syracuse University, 19--24
May 1994), ed.~by K. C. Wali (World Scientific, Singapore, 1995)}
\def \pisma#1#2#3#4{{\it Pis'ma Zh. Eksp. Teor. Fiz.} {\bf#1} (#3) #2 [{\it
JETP Lett.} {\bf#1} (#3) #4]}
\def \pl#1#2#3{{\it Phys. Lett.} {\bf#1} (#3) #2}
\def \plb#1#2#3{{\it Phys. Lett.} B {\bf#1} (#3) #2}
\def \ppnp#1#2#3{{\it Prog. Part. Nucl. Phys.} {\bf#1} (#3) #2}
\def \pr#1#2#3{{\it Phys. Rev.} {\bf#1} (#3) #2}
\def \prd#1#2#3{{\it Phys. Rev.} D {\bf#1} (#3) #2}
\def \prl#1#2#3{{\it Phys. Rev. Lett.} {\bf#1} (#3) #2}
\def \prp#1#2#3{{\it Phys. Rep.} {\bf#1} (#3) #2}
\def \ptp#1#2#3{{\it Prog. Theor. Phys.} {\bf#1} (#3) #2}
\def \rmp#1#2#3{{\it Rev. Mod. Phys.} {\bf#1} (#3) #2}
\def \rp#1{~~~~~\ldots\ldots{\rm rp~}{#1}~~~~~}
\def \si90{25th International Conference on High Energy Physics, Singapore,
Aug. 2-8, 1990, Proceedings edited by K. K. Phua and Y. Yamaguchi (World
Scientific, Teaneck, N. J., 1991)}
\def \slac75{{\it Proceedings of the 1975 International Symposium on
Lepton and Photon Interactions at High Energies,} Stanford University, Aug.
21-27, 1975, edited by W. T. Kirk (SLAC, Stanford, CA, 1975)}
\def \slc87{{\it Proceedings of the Salt Lake City Meeting} (Division of
Particles and Fields, American Physical Society, Salt Lake City, Utah, 1987),
ed. by C. DeTar and J. S. Ball (World Scientific, Singapore, 1987)}
\def \smass82{{\it Proceedings of the 1982 DPF Summer Study on Elementary
Particle Physics and Future Facilities}, Snowmass, Colorado, edited by R.
Donaldson, R. Gustafson, and F. Paige (World Scientific, Singapore, 1982)}
\def \smass90{{\it Research Directions for the Decade} (Proceedings of the
1990 DPF Snowmass Workshop), edited by E. L. Berger (World Scientific,
Singapore, 1991)}
\def \smassb{{\it Proceedings of the Workshop on $B$ Physics at Hadron
Accelerators}, Snowmass, Colorado, 21 June--2 July 1994, ed.~by P. McBride
and C. S. Mishra, Fermilab report FERMILAB-CONF-93/267 (Fermilab, Batavia, IL,
1993)}
\def \stone{{\it B Decays}, edited by S. Stone (World Scientific, Singapore,
1994)}
\def \tasi90{{\it Testing the Standard Model} (Proceedings of the 1990
Theoretical Advanced Study Institute in Elementary Particle Physics),
edited by M. Cveti\v{c} and P. Langacker (World Scientific, Singapore, 1991)}
\def \yaf#1#2#3#4{{\it Yad. Fiz.} {\bf#1} (#3) #2 [Sov. J. Nucl. Phys. {\bf #1}
 (#3) #4]}
\def \zhetf#1#2#3#4#5#6{{\it Zh. Eksp. Teor. Fiz.} {\bf #1} (#3) #2 [Sov.
Phys. - JETP {\bf #4} (#6) #5]}
\def \zhetfl#1#2#3#4{{\it Pis'ma Zh. Eksp. Teor. Fiz.} {\bf #1} (#3) #2 [JETP
Letters {\bf #1} (#3) #4]}
\def \zp#1#2#3{{\it Zeit. Phys.} {\bf#1} (#3) #2}
\def \zpc#1#2#3{{\it Zeit. Phys.} C {\bf#1} (#3) #2}


\begin{thebibliography}{249}

\bibitem{CPT} J. Schwinger, \pr{91}{713}{1953} (see esp.~p.~720 ff);
\pr{94}{1362}{1954} (see esp.~Eq (54) on p.~1366 and p.~1376 ff); G. L\"uders,
\kdvs{28}{5}{1954}; \apny{2}{1}{1957}; W. Pauli, in W. Pauli, ed. {\it Niels
Bohr and the Development of Physics} (New York, Pergamon, 1955), p.~30.

\bibitem{LY} T. D. Lee and C. N. Yang, \pr{104}{254}{1956}.

\bibitem{Pviol} C. S. Wu, E. Ambler, R. W. Hayward, D. D. Hoppes, and R. P.
Hudson, \pr{105}{1413}{1957}; R. L. Garwin, L. M. Lederman, and M. Weinrich,
\pr{105}{1415}{957}; J. I. Friedman and V. L. Telegdi, \pr{105}{1681}{1957};
\pr{106}{1290}{1957}.

\bibitem{VA} R. P. Feynman and M. Gell-Mann, \pr{109}{193}{1958}; E. C. G.
Sudarshan and R. E. Marshak, \pr{109}{1860}{1958}.

\bibitem{CCFT} J. H. Christenson, J. W. Cronin, V. L. Fitch, and R. Turlay,
\prl{13}{138}{1964}.

\bibitem{CL} T. P. Cheng and L. F. Li, {\it Gauge Theory of Elementary
Particles} (Oxford University Press, 1984).

\bibitem{CJ} \cp89.

\bibitem{GP} M. Gell-Mann and A. Pais, \pr{97}{1387}{1955}.

\bibitem{KCP} T. D. Lee, R. Oehme and C. N. Yang, \pr{106}{340}{1957};
B. L. Ioffe, L. B. Okun' and A. P. Rudik, \zhetf{32}{396}{1957}{5}{328}{1957}.

\bibitem{Revs} T. T. Wu and C. N. Yang, \prl{13}{380}{1964};
J. S. Bell and J. Steinberger in \oxf65, p.~193;
T. D. Lee and C. S. Wu, \arns{16}{511}{1966};
P. K. Kabir, {\it The CP Puzzle: Strange Decays of the Neutral Kaon} (Academic
Press, New York, 1968);
V. L. Fitch, \rmp{53}{367}{1981}; J. W. Cronin, \ibj{53}{373}{1981};
J. W. Cronin, \app{B15}{419, 721}{1984};
V. V. Barmin \ite, \np {B247}{293}{1984};
L. Wolfenstein, \arnps{36}{137}{1986};
R. G. Sachs, {\it The Physics of Time Reversal} (University of Chicago Press,
Chicago, 1987);
K. Kleinknecht, in \cp89, p.~41; T. Nakada in \corn, p.~425.

\bibitem{Cab} N. Cabibbo, \prl{10}{531}{1963}.

\bibitem{Charm} B. J. Bjorken and S. L. Glashow, \pl{11}{255}{1964}; Y. Hara,
\pr{134}{B701}{1964}; Z. Maki and Y. Ohnuki, \ptp{32}{144}{1964}; S. L.
Glashow, J. Iliopoulos, and L. Maiani, \prd{2}{1285}{1970}.

\bibitem{KM} M. Kobayashi and T. Maskawa, \ptp{49}{652}{1973}.

\bibitem{GL} M. K. Gaillard and B. W. Lee, \prd{10}{897}{1974}.

\bibitem{WP} L. Wolfenstein, \prl{51}{1945}{1983}.

\bibitem{CKMans} G. Harris and J. L. Rosner, \prd{45}{946}{1992}; M. Lusignoli,
L. Maiani, G. Martinelli, and L. Reina, \np{B369}{139}{1992}; A. Buras, M. E.
Lautenbacher, and G. Ostermaier, \prd{50}{3433}{1994}; J. L. Rosner, in \stone,
p.~470; A. Ali and D. London, \zpc{65}{431}{1995}.

\bibitem{BCP} J. Ellis, M. K. Gaillard, D. V. Nanopoulos, and S. Rudaz,
\np{B131}{285}{1977}; \ibj{B132}{541(E)}{1978}; A. B. Carter and A. I. Sanda,
\prl{45}{952}{1980}; \prd{23}{1567} {1981}; I. I. Bigi and A. I. Sanda, \np
{B193}{85}{1981}.

\bibitem{Brevs} I. Dunietz and J. L. Rosner, \prd{34}{1404}{1986}; I. I. Bigi
and A. I. Sanda, \np{B281}{41}{1987}; I. Dunietz, \ap{184}{350}{1988}; M. B.
Wise, in \ba88, p.~124; J. L. Rosner, in \tasi90, p.~91; Y. Nir and H. R.
Quinn, \arnps{42}{211}{1992}; I. I. Bigi, V. A. Khoze, N. G. Uraltsev, and A.
I. Sanda, in \cp89, p.~175; B. Winstein and L. Wolfenstein,
\rmp{65}{1113}{1993}.

\bibitem{PecR} R. D. Peccei, in \cp89, p.~503.

\bibitem{PQ} R. D. Peccei and H. R. Quinn, \prl{38}{1440}{1977};
\prd{16}{1791}{1977}.

\bibitem{AX} S. Weinberg, \prl{40}{223}{1978}; F. Wilczek,
\prl{40}{279}{1978}.

\bibitem{Sak} A. D. Sakharov, \pisma{5}{32}{1967}{24}.

\bibitem{TASI} J. L. Rosner, in \tasi90, p.~91.

\bibitem{JRCKM} J. L. Rosner, in \stone, p.~470.

\bibitem{PASCOS} J. L. Rosner, in \pascos, p.~37.

\bibitem{JRCharm} J. L. Rosner, in {\it The Future of High-Sensitivity Charm
Experiments}, Proceedings of the CHARM2000 Workshop, Fermilab, June 7--9 1994,
ed.~by D. M. Kaplan and S. Kwan (Fermilab, Batavia, IL, 1994, report
no.~FERMILAB-Conf-94/190), p.~297.

\bibitem{DPF} J. L. Rosner, in \dpfa.

\bibitem{Fest} J. L. Rosner, \efi 95-02, Jan.~1995, to be published in Comments
on Nucl.~and Part.~Phys., 1995.

\bibitem{GN} M. Gell-Mann, \pr{92}{833}{1953}; ``On the Classification of
Particles'' (1953, unpublished); M. Gell-Mann and A. Pais in {\it Proceedings
of the 1954 Glasgow Conference on Nuclear and Meson Physics}, ed.~by E. H.
Bellamy and R. G. Moorhouse (Pergamon, London and New York, 1955); M.
Gell-Mann, \nc{4}{Suppl.~848}{1956}; T. Nakano and K. Nishijima,
\ptp{10}{581}{1953}; K. Nishijima, \ptp{12}{107}{1954}; \ibj{13}{285}{1955}.

\bibitem{KL} K. Lande, E. T. Booth, J. Impeduglia, and L. M. Lederman,
\pr{103}{1901}{1956}.

\bibitem{PDG} Particle Data Group, L. Montanet \ite, \prd{50}{1174}{1994}.

\bibitem{pen} J. Ellis, M. K. Gaillard, and D. V. Nanopoulos, \np{B100}{313}
{1975}; \ibj{B109}{213}{1976}; A. I. Va\u{\i}nshte\u{\i}n, V. I. Zakharov, and
M. A. Shifman, \pisma{22}{123}{1975}{55};
\zhetf{72}{1275}{1977}{45}{670}{1977}; J. Ellis, M. K. Gaillard, D. V.
Nanopoulos, and S. Rudaz, \np {B131}{285}{1977}; \ibj{B132}{541(E)}{1978}.

\bibitem{CG} S. Coleman and S. L. Glashow, \prl{6}{423}{1961}; \pr{134}{B671}
{1964}.

\bibitem{BSS} M. Bander, D. Silverman, and A. Soni, \prl{44}{7}{1980}.

\bibitem{PQCD} M. K. Gaillard and B. W. Lee, \prl{33}{108}{1974}; G. Altarelli
and L. Maiani, \pl{52B}{351}{1974}.

\bibitem{DIH} For reviews and further references see A. J. Buras, in \cp89,
p.~575; G. Martinelli, \ib, p.~706.

\bibitem{Sachs} See, e.g., R. G. Sachs, {\it The Physics of Time Reversal
Invariance} (University of Chicago Press, Chicago, 1988).

\bibitem{Kabir} See, e.g., P. K. Kabir, {\it The CP Puzzle} (Academic Press,
New York, 1968).

\bibitem{WolfCPT} L. Wolfenstein, in {\it Theory and Phenomenology in Particle
Physics,} edited by A. Zichichi (Academic Press, New York, 1969).  For a more
recent discussion, see L. Wolfenstein, \prd{43}{151}{1991}.

\bibitem{WY} T. T. Wu and C. N. Yang, \prl{13}{380}{1964}.

\bibitem{NonWY} For more extensive discussions, see L. L. Chau, \prp{95}{1}
{1983}; E. A. Paschos and U. T\"urke, \prp{178}{145}{1989}.

\bibitem{phs} W. Ochs, $\pi N$ Newsletter {\bf 3} (1991) 25.

\bibitem{Ke4} L. Rosselet \ite, \prd{15}{574}{1977}.

\bibitem{OPE} P. Estabrooks and A. D. Martin, \np{B79}{301}{1974};
W. Ochs, thesis, Ludwig-Maximilians Universit\"at, Munich, 1973;
G. Grayer \ite, \np{B75}{189}{1974};
B. Hyams \ite, \ib~{\bf B100} (1977) 126.

\bibitem{E731} Fermilab E731 \cn, L. K. Gibbons \ite, \prl{70}{1203}{1993}.

\bibitem{NA31} CERN NA31 \cn, G. D. Barr \ite, \plb{317}{233}{1993}.

\bibitem{sw} L. Wolfenstein, \prl{13}{562}{1964}.

\bibitem{Schwing} Fermilab E773 \cn, B. Schwingenheuer \ite,
\prl{74}{4376}{1995}.

\bibitem{dsdq} R. G. Sachs, {\it The Physics of Time Reversal Invariance}
(University of Chicago Press, Chicago, 1988), p.~190.

\bibitem{KSbd} T. Nakada and L. Wolfenstein, in Particle Data Group, L.
Montanet \ite, \prd{50}{1536}{1994}.

\bibitem{CPLEAR} CPLEAR \cn, presented by P. Pavlopoulos at Europhysics
Conference on High Energy Physics, Marseille, 1993, as reported by T. Nakada
in \corn, p.~425; K. Jon-And \ite, \np{A558}{437c}{1993}.

\bibitem{E621} Fermilab E621 \cn, Y. Zou \ite, \plb{339}{519}{1994}.

\bibitem{phifact} DA$\Phi$NE Project, Laboratori Nazionali di Frascati, 1990.
Plans for other $\phi$ factories are discussed by M. Fukawa
\ite, KEK report KEK-90-12, August, 1990; W. A. Barletta \ite, UCLA report,
May, 1989 (unpublished); D. B. Cline, \np{24A}{1991}{150 (Proc.
Suppl.)}; E. P. Solodov, \yaf{55}{}{1992}{899}.

\bibitem{GWS} S. L. Glashow, \np{22}{579}{1961}; S. Weinberg,
\prl{19}{1264}{1967}; A. Salam, in {\it Proceedings of the Eighth Nobel
Symposium}, edited by N. Svartholm (Almqvist and Wiksell, Stockholm; Wiley, New
York, 1978), p. 367.

\bibitem{QP} L.-L. Chau and W.-Y. Keung, \prl{53}{1802}{1984}; H. Harari and M.
Leurer, \plb{181}{123}{1986}; J. D. Bjorken and I. Dunietz,
\prd{36}{2109}{1987}.

\bibitem{UT} L.-L. Chau and W.-Y. Keung, \prl{53}{1802}{1984}; M. Gronau and J.
Schechter, \ibj{54}{385}{1985}; M. Gronau, R. Johnson, and J. Schechter,
\prd{32}{3062}{1985}; C. Jarlskog, in {\it Physics at LEAR with Low Energy
Antiprotons,} proceedings of the workshop, Villars-sur-Ollon, Switzerland,
1987, edited by C. Amsler \ite ~(Harwood, Chur, Switzerland, 1988), p. 571; J.
D. Bjorken and I. Dunietz, \prd{36}{2109}{1987}.

\bibitem{NQ} Y. Nir and H. Quinn, \arnps{42}{211}{1992}.

\bibitem{GZ} S. S. Gershtein and Ya. B. Zel'dovich, \zhetf{29}{698}{1955}
{2}{576}{1956}.

\bibitem{WMVUD} W. J. Marciano, \arnps{41}{469}{1991}.

\bibitem{pib} W. K. McFarlane \ite, \prd{32}{547}{1985}; W. J. Marciano and Z.
Parsa, \arnps{36}{171}{1986}.

\bibitem{AG} M. Ademollo and R. Gatto, \pl{13}{264}{1964}.

\bibitem{VUSfit} M. Bourquin \ite, \zpc{21}{27}{1983}; H. Leutwyler and M.
Roos, \zpc{25}{91}{1984}; J. F. Donoghue, B. R. Holstein, and S. W. Klimt,
\prd{35}{934}{1987}.

\bibitem{VCDdis} CDHS \cn, H. Abramowicz \ite, \zpc{15}{19}{1982};
CCFR \cn, S. A. Rabinowitz \ite, \prl{70}{134}{1993}.

\bibitem{VCDslc} Mark III \cn, J. Adler \ite, \prl{62}{1821}{1989};
CLEO \cn, M. S. Alam \ite, \prl{71}{1311}{1993}.

\bibitem{VCS} M. Witherell, in \corn, p.~198.

\bibitem{kin} J. L. Cortes, X. Y. Pham, and A. Tounsi, \prd{25}{188}{1982}.

\bibitem{BH} T. Browder and K. Honscheid, University of Hawaii
report UH-511-816-95, March, 1995, to appear in {\it Progress in
Nuclear and Particle Physics}, v.~35.

\bibitem{HQS} B. Grinstein, \arnps{42}{101}{1992}, and references therein.

\bibitem{PBall} P. Ball, M. Beneke, and V. M. Braun, CERN report CERN-TH-95-65,
March, 1995 (unpublished).

\bibitem{spect} W. Kwong, L. H. Orr, and J. L. Rosner, \prd{40}{1453}{1989}.

\bibitem{corrs} See, e.g., M. Wise, in \corn, p.~253, and references therein.

\bibitem{ARGVub} ARGUS \cn, H. Albrecht \ite, \plb{234}{409}{1990};
\ib~{\bf 255} (1991) 297.

\bibitem{CLVub} CLEO \cn, as reported by D. Besson in \corn, p.~221.

\bibitem{Vubmods} G. Altarelli \ite, \np{B208}{365}{1982};
B. Grinstein, N. Isgur, and M. Wise, \prl{56}{298}{1986};
N. Isgur, D. Scora, B. Grinstein, and M. Wise, \prd{39}{799}{1989};
M. Wirbel, B. Stech, and M. Bauer, \zpc{29}{637}{1985};
J. G. K\"orner and G. A. Schuler, \zpc{38}{511}{1988};
D. Scora and N. Isgur, CEBAF report CEBAF-TH-94-14, March, 1995
(unpublished).

\bibitem{pilnu} CLEO \cn, presented by L. K. Gibbons at XXX Rencontres
de Moriond, Les Arcs, France, March, 1995.

\bibitem{FBL} See, e.g., UKQCD \cn, R. M. Baxter \ite, \prd{49}{1594}{1994};
P. B. Mackenzie, plenary talk, in \dpfa;
C. W. Bernard, J. N.  Labrenz, and A. Soni, \prd{49}{2536}{1994};
A. Duncan \ite, \prd{51}{5101}{1995}.

\bibitem{FBQ} J. F. Amundson \ite, \prd{47}{3059}{1993}; J. L. Rosner, \dpff,
p.~658.

\bibitem{CDFt} CDF \cn, F. Abe \ite, \prl{74}{2626}{1995}.

\bibitem{D0t} D0 \cn, S. Abachi \ite, \prl{74}{2632}{1995}.

\bibitem{IL} T. Inami and C. S. Lim, \ptp{65}{297}{1981};
\ibj{65}{1772(E)}{1981}.

\bibitem{TNT} T. N. Truong, \prl{17}{1102}{1966}; \ibj{18}{300(E)}{1967}.

\bibitem{QCDK} See G. Harris and J. L. Rosner, \prd{45}{946}{1992}, and
references therein.
% QCD Corrections to K box

\bibitem{HRS} C. Hamzaoui, J. L. Rosner, and A. I. Sanda, in \be87, p. 97; J.
L. Rosner, A. I. Sanda, and M. P. Schmidt, \ib, p. 165.

\bibitem{BKlat} See, e.g., R. Gupta, D. Daniel, G. Kilcup, A. Patel, and S. R.
Sharpe, \prd{47}{5113}{1993}; P. B. Mackenzie, in \corn, p.~634. Our quoted
error reflect the spread in values obtained by different means.

\bibitem{bbmix} ARGUS \cn, H. Albrecht \ite, \plb{192}{245}{1987}.

\bibitem{QCDB} A. Buras, M. Jamin, and P. H. Weisz, \np{B347}{491}{1990}.

\bibitem{Dsfact} ARGUS \cn, H. Albrecht \ite, \plb{219}{121}{1989}; D.
Bortoletto, Ph.~D. Thesis, Syracuse University, 1989; D. Bortoletto and S. L.
Stone, \prl{65}{2951}{1990}; CLEO \cn, D. Bortoletto \ite,
\prd{45}{2212}{1992}; J. L. Rosner, \prd{42}{3732}{1990}; in \smass90, p.~255;
M. Paulini \ite, in {\it Proceedings of the Joint International Symposium and
Europhysics Conference on High Energy Physics,} ed.~by S. Hegarty, K. Potter,
and E. Quercigh (World Scientific, Singapore, 1992), p.~592;  CLEO \cn, A. Bean
\ite, \prl{70}{2681}{1993}.

\bibitem{Dsmeas} CERN WA75 \cn, S. Aoki \ite, \ptp{89}{131}{1993}; CLEO \cn, D.
Acosta \ite, \prd{49}{5690}{1994}; F. Muheim and S. Stone,
\prd{49}{3767}{1994}; BES \cn, J. Z. Bai \ite, \prl{74}{4599}{1995}; F. Muheim,
parallel session on $B$ physics, in \dpfa.

\bibitem{MkIII} Mark III \cn, J. Adler \ite, \prl{60}{1375}{1988};
\ibj{63}{1658(E)}{1989}.

\bibitem{IW} N. Isgur and M. Wise, \plb{237}{527}{1990}; H. Georgi,
\plb{240}{447}{1990}; A. F. Falk, H. Georgi, B. Grinstein, and M. B. Wise,
\np{343}{1}{1990}; H. Georgi and M. B. Wise, \plb{243}{279}{1990}; J. D.
Bjorken, SLAC Report No. SLAC-PUB-5278, June, 1990, invited talk presented at
Les Rencontres de Physique de la Vallee d'Aoste, La Thuile, Aosta Valley,
Italy, Mar. 18--24, 1990; SLAC Report No. SLAC-PUB-5362, presented at \si90;
SLAC Report No. SLAC-PUB-5389, presented at 18th SLAC Summer Institute, July
16--27, 1990.  See also L. Durand III, P. C. De Celles, and R. B. Marr,
\pr{126}{1882}{1962}; S. Nussinov and W. Wetzel, \prd{36}{130}{1987}.

\bibitem{BJfact} J. D. Bjorken, Nucl.~Phys.~Proc.~Suppl.~{\bf 11} (1989) 325.
% Bjorken factorization test

\bibitem{BS} D. Bortoletto and S. L. Stone, \prl{65}{2951}{1990}.

\bibitem{JRFDS} J. L. Rosner, \prd{42}{3732}{1990}; in \smass90, p.~255.

\bibitem{NSpc} N. Stanton, private communication.

\bibitem{ES} E. V. Shuryak, \np{B198}{83}{1982}.

\bibitem{VSPW} M. B. Voloshin and M. A. Shifman, \yaf{45}{463}{1987}{292};
H. D. Politzer and M. B. Wise, \plb{206}{681}{1988}; \ibj{208}{504}{1988}

\bibitem{PHJR} P. F. Harrison and J. L. Rosner, \jpg{18}{1673}{1992}.

\bibitem{ALI} A. Ali, \jpg{18}{1065}{1992}.

\bibitem{AS} A. Soni, paper no. 282, in \dpfa.

\bibitem{Dib} C. O. Dib, Ph.D. Thesis, Stanford University, 1990, SLAC Report
SLAC-364, April, 1990 (unpublished).

\bibitem{RVW} B. Winstein and L. Wolfenstein, \rmp{65}{1113}{1993}.

\bibitem{Kgg} CERN NA31 \cn, H. Burkhardt \ite, \plb{199}{139}{1987}.

\bibitem{Keeg} CERN NA31 \cn, G. D. Barr \ite, \plb{240}{283}{1990};
Brookhaven E845 \cn, K. D. Ohl \ite, \prl{65}{1407}{1990}.

\bibitem{Kmmg} Fermilab E799 \cn, M. B. Spencer \ite, \prl{74}{3323}{1995}.

\bibitem{HG} H. B. Greenlee, \prd{42}{3724}{1990}.

\bibitem{KSggt} R. L. Goble, \prd{7}{931}{1973}; R. L. Goble, R. Rosenfeld,
and J. L. Rosner, \prd{39}{3264}{1989}.

\bibitem{Kmm} BNL E791 \cn, A. P. Heinson \ite, \prd{51}{985}{1995}.

\bibitem{Kpiee} C. Alliegro \ite, \prl{68}{278}{1992}.

\bibitem{LL} L. Littenberg, plenary talk, in \dpfa.

\bibitem{ER} J. Ellis and S. Rudaz, \np{B304}{205}{1988}.  See also
J. S. Haggerty, in \smass90, p.~275; I. I. Bigi and F. Gabbiani,
\np{B367}{3}{1991}; G. Buchalla and A. J. Buras, \np{B412}{106}{1994}.

\bibitem{pigg} CERN NA31 \cn, G. D. Barr \ite, \plb{242}{523}{1990};
Fermilab E731 \cn, V. Papadimitriou \ite, \prd{44}{R573}{1991}.

\bibitem{piggt} L. M. Sehgal, \prd{6}{367}{1972};
P. Ko and J. L. Rosner, \prd{40} {3775}{1989};
G. Ecker, A. Pich, and E. De Rafael, \plb{189}{363}{1987};
\np{B291}{692}{1987}; \ibj{B303}{665}{1988}.

\bibitem{Ko} P. Ko, \prd{44}{139}{1991}.

\bibitem{piggy} L. M. Sehgal, \prd{38}{808}{1988}; J. Flynn and L. Randall,
\plb{216}{221}{1989}; G. Ecker, A. Pich, and E. De Rafael, \plb{237}{481}
{1990}.

\bibitem{DHE} Fermilab E799 \cn, D. Harris \ite, \prl{71}{3918}{1993}.

\bibitem{DHM} Fermilab E799 \cn, D. Harris \ite, \prl{71}{3914}{1993}.

\bibitem{pinunu} Fermilab E799 \cn, M. Weaver \ite, \prl{72}{3758}{1994}.

\bibitem{eps} M. B. Wise, in \ba88, p.~124.

\bibitem{epscorr} J. M. Flynn and L. Randall, \plb{224}{221}{1989};
G. Buchalla, A. J. Buras, and M. Harlander, \np{B337}{313}{1990}.

\bibitem{BEPS} A. J. Buras, M. Jamin, and M. E. Lautenbacher,
\np{B408}{209}{1993}.  See also M. Ciuchini, E. Franco, G. Martinelli, and
L. Reina, \plb{301}{263}{1993}.

\bibitem{ALBs} ALEPH \cn, Y. B. Pan, paper no. 476, in \dpfa.

\bibitem{CPeven} R. Aleksan, A. Le Yaouanc, L. Oliver, O. P\`ene, and J. C.
Raynal, \plb{316}{567}{1993}.

\bibitem{Blifes} I. I. Bigi \ite, in \stone, p.~132.

\bibitem{Nelson} C. A. Nelson, Jr., \prd{30}{107, 1937}{1984};
\ibj{32}{1848(E)} {1985}.

\bibitem{IsiBs} I. Dunietz, Fermilab report FERMILAB-PUB-94/36-T, Jan.~1995
(unpublished).

\bibitem{BP} T. Browder and S. Pakvasa, University of Hawaii report
UH-511-814-95, Jan.~1995 (unpublished).

\bibitem{PP} D. London and R. Peccei, \plb{223}{257}{1989}; M. Gronau,
\prl{63}{1451}{1989}; B. Grinstein, \plb{229}{280}{1989}; M. Gronau,
\plb{300}{163}{1993}.

\bibitem{pipi} M. Gronau and D. London, \prl{65}{3381}{1990}.

\bibitem{BKUS} I. I. Bigi, V. A. Khoze, N. G. Uraltsev, and A. I. Sanda, in
\cp89, p.~175.

\bibitem{KB} K. Berkelman, \mpla{10}{165}{1995}.

\bibitem{AB} A. Ali and F. Barreiro, \zpc{30}{635}{1986}.

\bibitem{Correls} M. Gronau, A. Nippe, and J. L. Rosner, \prd{47}{1988}{1992};
M. Gronau and J. L. Rosner, in \smassb, p.~701; \prl{72}{195}{1994};
\prd{49}{254}{1994}.

\bibitem{HLB} H. J. Lipkin, private communication.

\bibitem{SN} S. Nussinov, \prl{35}{1672}{1975}.

\bibitem{DGG} A. De R\'ujula, H. Georgi, and S. Glashow, \prl{37}{785}{1976}.

\bibitem{Dstars} CLEO \cn, J. Bartelt, parallel session on charm, in
\dpfa.

\bibitem{EHQ} C. T. Hill, in \smassb, p.~127; C. Quigg, {\it ibid.}, p.~443; E.
Eichten, C. T. Hill, and C. Quigg, \prl{71}{4116}{1994}, and in {\it The Future
of High-Sensitivity Charm Experiments}, Proceedings of the CHARM2000 Workshop,
Fermilab, June 7--9 1994, ed.~by D. M. Kaplan and S. Kwan (Fermilab, Batavia,
IL, 1994, report no.~FERMILAB-Conf-94/190), pp.~345, 355.

\bibitem{OPB} OPAL \cn, R. Akers \ite, \zpc{66}{19}{1995}.

\bibitem{DELB} DELPHI \cn, P. Abreu \ite, \plb{345}{598}{1995}.

\bibitem{ALB} ALEPH \cn, presented by S. Schael at Moriond Workshop, March,
1995.

\bibitem{Ispin} I. Dunietz and J. L. Rosner, \prd{51}{2471}{1995}.

\bibitem{BPP} M. Gronau, J. L. Rosner, and D. London, \prl{73}{21}{1994}; O. F.
Hernandez, D. London, M. Gronau, and J. L. Rosner, \plb{333}{500}{1994}; M.
Gronau, O. F. Hern\'andez, D. London, and J. L. Rosner, \prd{50}{4529} {1994}.

\bibitem{OldSU} D. Zeppenfeld, \zpc{8}{77}{1981};
M. Savage and M. Wise, \prd{39}{3346}{1989}; \ibj{40}{3127(E)}{1989};
J. Silva and L. Wolfenstein, \prd{49}{R1151}{1994}.

\bibitem{Battle} CLEO \cn, M. Battle \ite, \prl{71}{3922}{1993}.

\bibitem{Wurt} CLEO \cn, S. Playfer, \baps{40}{929}{1995} (Joint April Meeting
of the APS and AAPT, Washington, DC); F. W\"urthwein, \ib~{\bf 40}, 923 (1995).

\bibitem{CLEOGlas} CLEO \cn,  P. Gaidarev, \baps{40}{923}{1995}.

\bibitem{pik} Y. Nir and H. R. Quinn, \prl{67}{541}{1991};
M. Gronau, \plb{265}{389}{1991};
H. J. Lipkin, Y. Nir, H. R. Quinn and A. E. Snyder, \prd{44}{1454}{1991};
L. Lavoura, \mpla{7}{1553}{1992}.

\bibitem{LWC} L. Wolfenstein, \prd{52}{to be published}{1995}.

\bibitem{RF} R. Fleischer, \zpc{62}{81}{1994}; \plb{321}{259}{1994};
\ibj{332}{419}{1994}.

\bibitem{DH} N. G. Deshpande and X.-G. He, \prl{74}{26, 4099(E)}{1995}.

\bibitem{SUbr} M. Gronau, O. F. Hern\'andez, D. London, and J. L. Rosner,
\efi~95-09, submitted to Phys.~Rev.~D.

\bibitem{EWP} M. Gronau, O. F. Hern\'andez, D. London, and J. L. Rosner,
\efi~95-11, submitted to Phys.~Rev.~D.

\bibitem{DHP} N. G. Deshpande, X.-G. He, and J. Trampetic, \plb{345}{547}
{1995}.

\bibitem{DHeta} N. G. Deshpande and X.-G. He, University of Oregon report
OITS-576, May, 1995, submitted to Phys.~Rev.~Letters.

\bibitem{HJLpc} We thank H. J. Lipkin for a discussion on this point.

\bibitem{SLCrev} J. L. Rosner, in {\it Proceedings of the Salt Lake City
Meeting} (Division of Particles and Fields, American Physical Society, Salt
Lake City, Utah, Jan.~14-17, 1987), edited by C. De Tar and J. S. Ball (World
Scientific, Singapore, 1987), p. 59.

\bibitem{Win} B. Winstein, \prl{68}{1271}{1992}.

\bibitem{RHW} J. C. Pati and A. Salam, \prd{8}{1240}{1973}; R. E. Marshak and
R. N. Mohapatra, \plb {91B}{222}{1980}.

\bibitem{RT} J. L. Rosner and E. Takasugi, \prd{42}{241}{1990}.

\bibitem{WRlims} A. Jodidio \ite, \prd{34}{1967}{1986}; \ibj{37}{237(E)}{1988};
M. Aoki \ite, \prd{50}{69}{1994}.

\bibitem{Mixbox} G. Beall, M. Bander, and A. Soni, \prl{48}{848}{1982}.

\bibitem{LS} P. Langacker and S. Uma Sankar, \prd{40}{1569}{1989}.

\bibitem{GW} M. Gronau and S. Wakaizumi, \prl{68}{1814}{1992};
\plb{280}{79}{1992}; \prd{47}{1262}{1993}.

\bibitem{ARWW} J. Amundson, J. L. Rosner, M. Worah, and M. Wise, \prd{47}
{1260}{1993}.

\bibitem{LW} T. D. Lee, \prd{8}{1226}{1973}; \prp{9C}{143}{1974};
S. Weinberg, \prl{37}{657}{1976}.

\bibitem{Sanda} I. I. Bigi and A. I. Sanda, in \cp89, p.~362.

\bibitem{Hay}T. Hayashi, Y. Koide, M. Matsude, and M. Tanimoto,
\ptp{91}{915}{1994}.

\bibitem{YLWu} Y. L. Wu, Carnegie-Mellon University report CMU-HEP-94-01, 1994.

\bibitem{FKG} S. Weinberg, \prl{37}{657}{1976}; R. Garisto and G. L. Kane,
\prd{44}{2038}{1991}; G. B\'elanger and C. Q. Geng, \prd{44}{2789}{1991}.

\bibitem{limits} I. S. Altarev \ite, \plb{276}{242}{1992}; K. Abdullah \ite,
\prl{65}{2347}{1990}.
% dipole moment limits

\bibitem{Gunion} J. Gunion, lectures at this Summer School.

\bibitem{SMB} S. M. Barr, \ijmpa{8}{209}{1993}.

\bibitem{AKG} See, e.g., A. K. Grant, \prd{51}{207}{1995}.

\bibitem{bsg} CLEO \cn, M. S. Alam \ite, \prl{74}{2885}{1995}.

\bibitem{Bubsg} A. J. Buras, M. Misiak, M. Munz, and S. Pokorski,
\np{B424}{374}{1994}.

\bibitem{Hg} J. P. Jacobs \ite, \prl{71}{3782}{1993}.

\bibitem{Dmix} J. F. Donoghue, E. Golowich, B. Holstein, and J. Trampeti\'c,
\prd{33}{179}{1986}; L. Wolfenstein, \plb{164}{170}{1985}; H. Georgi,
\plb{297}{353}{1992}.

\bibitem{BigC} I. I. Bigi, in {\it Proceedings of the Tau-Charm Factory
Workshop}, Stanford, CA, May 23 -- 27, 1989, ed.~by Lydia V. Beers
(SLAC, Stanford, CA, 1989), p.~169.

\bibitem{Asymms} Fermilab E687 \cn, P. L. Frabetti \ite, \prd{50}{R2953}{1994}.

\bibitem{JP} J. Preskill, \prl{43}{1365}{1979}.

\bibitem{GG} H. Georgi and S. L. Glashow, \prl{32}{438}{1974}.

\bibitem{tH} G. 't Hooft, \prl{37}{8}{1976}.

\bibitem{KRS} V. A. Kuzmin, V. A. Rubakov, and M. E. Shaposhnikov, \plb{155}
{36}{1985}; \ibj{191}{171}{1987}.

\bibitem{FS} G. R. Farrar and M. E. Shaposhnikov, \prl{70}{2833}{1993};
\ibj{71}{210(E)}{1993}; CERN report CERN-TH-6732-93, 1993; Rutgers
University report RU-94-40, 1994.

\bibitem{Yana}M. Fukugita and T. Yanagida, \plb{174}{45}{1986};
P. Langacker, R. Peccei, and T. Yanagida, \mpla{1}{541}{1986}.

\bibitem{HS} P. Huet and E. Sather, \prd{51}{279}{1995};
M. B. Gavela \ite, \np{B430}{345}{1994}; \ibj{430B}{382}{1994}.

\bibitem{seesaw} M. Gell-Mann, P. Ramond, and R. Slansky in {\it Supergravity},
edited by P. van Nieuwenhuizen and D. Z. Freedman (Amsterdam, North-Holland,
1979), p.~315; T. Yanagida {\it Proceedings of the Workshop on Unified Theory
and Baryon Number in the Universe}, edited by O. Sawada and A. Sugamoto
(Tsukuba, Japan, National Laboratory for High Energy Physics, 1979).

\bibitem{BKCP} B. Kayser, in \cp89, p.~334.

\bibitem{GQW} H. Georgi, H. Quinn, and S. Weinberg, \prl{33}{451}{1974}.

\bibitem{Amaldi} U. Amaldi \ite, \prd{36}{1385}{1987};
U. Amaldi, W. de Boer,and H. F\"urstenau, \plb{260}{447}{1991};
P. Langacker and N. Polonsky, \prd{47}{4028}{1993}.

\bibitem{SOten} H. Georgi in {\it Proceedings of the 1974
Williamsburg DPF Meeting}, ed. by C. E. Carlson  (New York, AIP, 1975) p.~575;
H. Fritzsch and P. Minkowski, \apny{93}{193}{1975}.

\bibitem{MD} M. Dine, in \corn, p.~695.

\bibitem{Leut} H. Leutwyler, presented at Workshop on Yukawa Couplings and the
Origins of Mass, Gainesville, FL, February, 1994.

\bibitem{PS} P. Sikivie, \prl{51}{1415}{1983}.

\bibitem{RGS} R. G. Sachs, \prl{73}{377}{1994}.

\bibitem{tHc} G. 't Hooft, in {\it Recent Developments in Gauge Theories},
Proceedings of the Carg\`ese Summer Institute, Carg\`ese, France, 1979,
ed. by G. 't Hooft \ite, NATO Advanced Study Institutes, Series B: Physics,
Vol. 59 (Plenum, New York, 1980), p.~135.

\bibitem{PTP} J. L. Rosner, \ptp{66}{1422}{1981}.

\bibitem{RW} J. L. Rosner and M. P. Worah, \prd{46}{1131}{1992}.

\bibitem{RS} J. L. Rosner and D. E. Soper, \prd{45}{3206}{1992}.
\end{thebibliography}
\end{document}